\documentclass[crcready]{iosart2x}
\makeatletter
\let\numberlines@hook\relax
\makeatother
\usepackage{dcolumn}
\usepackage{xcolor}
\usepackage{graphicx}
\usepackage{enumerate}
\usepackage{verbatim}
\usepackage{listings}
\usepackage{caption}
\usepackage{filecontents}
\usepackage{amssymb}
\usepackage{todonotes}
\usepackage{subfig}
\usepackage{url}
\usepackage{multicol}
\usepackage{mdframed}
\usepackage{lipsum}
\usepackage{wrapfig}
\usepackage{rotating}
\usepackage{booktabs}
\usepackage{multirow}

\usepackage{tcolorbox}

\usepackage{lineno, blindtext}
\usepackage{enumitem}

\usepackage[algo2e,ruled,vlined]{algorithm2e}
\usepackage{xargs} % Use more than one optional parameter in a new commands
\usepackage{amsmath}
\usepackage{amsthm}

\DeclareFontFamily{\encodingdefault}{\ttdefault}{\hyphenchar\font=`\-}

\newenvironment{sftabbing}
{\par
	\begingroup\lccode`~=`' \lowercase{\endgroup\let~}\textquotesingle
	\begingroup\lccode`~=`" \lowercase{\endgroup\def~}{{\fontencoding{T1}\selectfont\textquotedbl}}
	\catcode`'=\active \catcode`"=\active
	\sffamily\footnotesize\tabbing}
{\endtabbing\par}

\newenvironment{sftabbinglist}
{\par
	\begingroup\lccode`~=`' \lowercase{\endgroup\let~}\textquotesingle
	\begingroup\lccode`~=`" \lowercase{\endgroup\def~}{{\fontencoding{T1}\selectfont\textquotedbl}}
	\catcode`'=\active \catcode`"=\active
	\ttfamily\footnotesize\tabbing}
{\endtabbing\par}

\newcolumntype{d}[1]{D{.}{.}{#1}}

\firstpage{1}
\lastpage{5}
\volume{1}
\pubyear{0000}

\begin{document}
\begin{frontmatter} % The preamble begins here.

%
%\pretitle{Pretitle}
\title{Multidimensional Enrichment of Spatial RDF Data for SOLAP - Full Version\thanks{Journal paper is submitted to Semantic Web -- Interoperability, Usability, Applicability (IOS Press Journal) on 2020/02/07.}}

\runningtitle{Multidimensional Enrichment of Spatial RDF Data for SOLAP}
%\subtitle{Subtitle}

\author[A]{\inits{F.}\fnms{Nuref\c san} \snm{ G\" ur}\ead[label=e1]{nurefsan@cs.aau.dk}%
\thanks{Corresponding author. \printead{e1}.}%
%\thanks{Do not use capitals for the author's surname.}
},
\author[A]{\inits{S.}\fnms{Torben Bach} \snm{Pedersen}\ead[label=e2]{tbp@cs.aau.dk}}
,
\author[A]{\inits{T.}\fnms{Katja} \snm{Hose}\ead[label=e3]{khose@cs.aau.dk}}
and
\author[A]{\inits{T.}\fnms{Mikael} \snm{Midtgaard}\ead[label=e4]{mikaelmidt@gmail.com}}
%\runningauthor{F. Author et al.}

\runningauthor{N. G\" ur et al.}
\address[A]{Center for Data Intensive Systems, \orgname{Aalborg University}, Selma Lagerl\"ofsvej 300, DK-9220 Aalborg {\O}, \cny{Denmark}\printead[presep={\\}]{e1,e2,e3,e4}}

%\begin{review}{editor}
%\reviewer{\fnms{First} \snm{Editor}\address{\orgname{University or Company name}, \cny{Country}}}
%\reviewer{\fnms{Second} \snm{Editor}\address{\orgname{University or Company name}, \cny{Country}}}
%\end{review}
%\begin{review}{solicited}
%\reviewer{\fnms{First Solicited} \snm{Reviewer}\address{\orgname{University or Company name}, \cny{Country}}}
%\reviewer{\fnms{Second Solicited} \snm{Reviewer}\address{\orgname{University or Company name}, \cny{Country}}}
%\end{review}
%\begin{review}{open}
%\reviewer{\fnms{First Open} \snm{Reviewer}\address{\orgname{University or Company name}, \cny{Country}}}
%\reviewer{\fnms{Second Open} \snm{Reviewer}\address{\orgname{University or Company name}, \cny{Country}}}
%\end{review}

\begin{abstract}
{Large volumes of spatial data and multidimensional data are being published on the Semantic Web, which has led to new opportunities for advanced analysis, such as Spatial Online Analytical Processing (SOLAP). The RDF Data Cube (QB) and QB4OLAP vocabularies have been widely used for annotating and publishing statistical and multidimensional RDF data. Although such statistical data sets might have spatial information, such as coordinates, the lack of spatial semantics and spatial multidimensional concepts in QB4OLAP and QB prevents users from employing SOLAP queries over spatial data using SPARQL.
The QB4SOLAP vocabulary, on the other hand, fully supports annotating spatial and multidimensional data on the Semantic Web and enables users to query endpoints with SOLAP operators in SPARQL.
To bridge the gap between QB/QB4OLAP  and QB4SOLAP, we propose an RDF2SOLAP enrichment model that automatically annotates spatial multidimensional concepts with QB4SOLAP and in doing so enables SOLAP on existing QB and QB4OLAP data on the Semantic Web. 
Furthermore, we present and evaluate a wide range of enrichment algorithms and apply them on a non-trivial real-world use case involving governmental open data with complex geometry types.} 
\end{abstract}

\end{frontmatter}

%-------------------------------------------------------------------------
\section{Introduction}
\label{sec:introduction}
% !TeX spellcheck = en_US
% !TEX root = ../iosart2x.tex

Data warehouses (DWs) as well as Online Analytical Processing (OLAP) tools and queries are well-established for interactive data analysis. DWs have multidimensional (MD) models and store large volumes of data. MD models locate data in an n-dimensional space and are usually referred to as \textit{data cubes}. The \textit{cells} of a cube represent the topic of the analysis and associate observation \textit{facts} with (numerical) \textit{measures} that can be aggregated. Spatial data cubes can also contain \textit{spatial} measures, which can be aggregated with spatial functions. 
For example, a data cube for \textit{farms} might have a numerical measure `number of animals' as well as the `farm's coordinates' as spatial measure. Facts are linked to \textit{dimensions}, which provide contextual information, e.g., farm production, farm location, and farm livestock. Dimensions are organized into \textit{hierarchies} with \textit{levels}, e.g., parish of the farm or herd type of livestock, which allows users to analyze and aggregate measures at different levels of detail. Levels have a set of \textit{attributes} describing the characteristics of the level members. 

In traditional DWs, the location dimension is generally used as a conventional (non-spatial) dimension with alphanumeric data and thus provided with only a nominal reference to places and areas, e.g., parish name. 
This does not allow for applying spatial operations or truly deriving topological relations between hierarchy levels based on geometric information such as coordinates, which are essential for enabling spatial OLAP (SOLAP) analysis. 

By including the geometric information of locations in MD models, we can significantly improve the analysis process (e.g., proximity analysis of locations) with additional perspectives by revealing dynamic spatial hierarchy levels and new spatial level members in SOLAP operations (details and examples in~\cite{nuref2017qb4solapSWJ, nuref2017GeoSemOLAP}). 
In addition, by using geometric attributes of level members, topological relations between the levels, and levels and facts can be specified implicitly. 
Such topological relations are essential to correctly aggregate measures between levels with many-to-many (N:M) cardinality relations.

The Semantic Web (SW) has evolved, from prominently focusing on data publishing to also supporting complex queries, such as interactive analytical queries. Simultaneously, the data available on the SW has evolved from being simple, mostly alphanumerical data, to include complex data types, such as geospatial data. There are many examples of governmental and statistical Linked Open Data (LOD) sets with geographical attributes. However, such datasets are typically not modeled with multidimensional (MD) concepts. Thus, they cannot be queried with interactive analytical queries (OLAP). 
Although in recent years several platforms and tools for Business Intelligence (BI) and data warehouses have emerged~\cite{zimanyi2008advanced}, there is still a lack of common standards to model and publish (geo)semantic cubes on the SW~\cite{nuref2017qb4solapSWJ}. 

More and more statistical datasets using the RDF Data Cube Vocabulary (QB)~\cite{cyganiak2013rdf}, the current W3C standard,  are published on the SW. These datasets have observations and measures, which are well-suited for analytical queries. However, QB lacks the underlying structural metadata for multidimensional models and OLAP operations (Section~\ref{sec:relatedwork}). Well-defined structural metadata is required to translate OLAP queries into SPARQL 1.1~\cite{sparql,nuref2017GeoSemOLAP}. QB4ST~\cite{qb4st} is a recent attempt to define extensions for spatio-temporal components to QB. However, it inherits the limitations of multidimensional modeling from QB.

To address the MD modeling challenges of the QB vocabulary, QB4OLAP~\cite{etcheverry2014modeling} has been proposed, which reuses QB definitions by adding the required MD schema semantics. A significant number of data sets have already been published using the QB vocabulary. QB4OLAP descriptions of a QB data cube can be generated semi-automatically by adding the necessary MD semantics (e.g., the hierarchical structure of the dimensions) and the corresponding instances to populate the dimension levels. However, existing QB4OLAP annotation techniques~\cite{varga2016dimensional} only cover \emph{non-spatial} MD data cube concepts and its operations. Even though such statistical data sets have spatial information, not annotating the spatial MD concepts (e.g., spatial hierarchy levels such as administrative regions) hinders querying the data with interesting spatial OLAP operations. To emerge this need the QB4SOLAP vocabulary was proposed~\cite{nuref2015solap}, which allows modeling the data cubes fully with both multidimensional and spatial concepts on the SW.

\begin{figure}[h!tb]%
	\vspace{-0.19in}
	\centering
	\includegraphics[width=1\columnwidth]{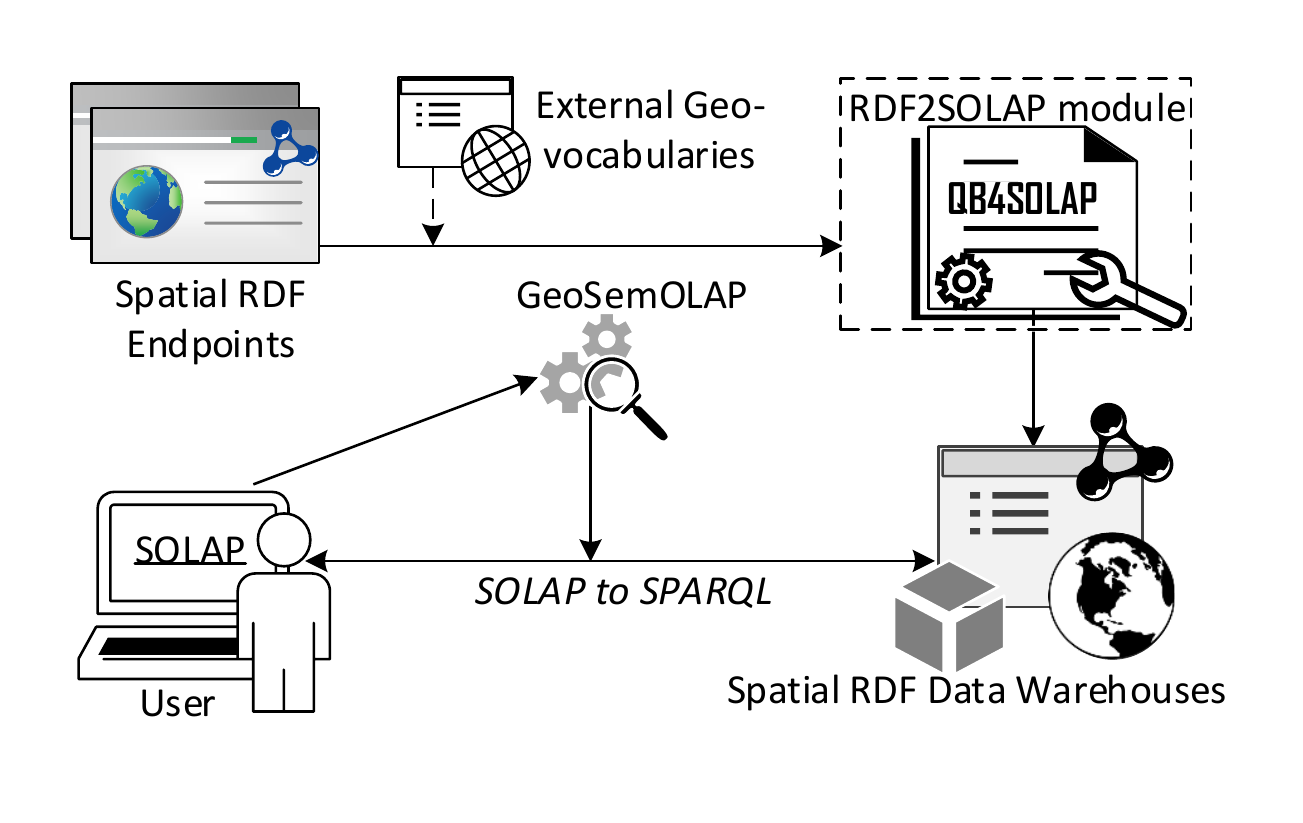}%
	\vspace*{-0.3in}
	\caption{Future vision of SOLAP on the SW }%
	\label{fig:motivation}%
	%\vspace{-0.2in}
\end{figure}

\paragraph{Problem Definition.} In the current state of the SW, spatial OLAP (SOLAP) queries are not supported by existing spatial RDF stores and endpoints. If a (spatial) data warehouse user would like to query spatial RDF data from the SW with SOLAP operations, the user needs to download the RDF data, map it to a relational data model (e.g., with a snowflake schema), and then import it into a traditional spatial data warehouse, which is slow, labor-intensive, and stores the data in a non-open format.

There are existing tools and vocabularies for (spatial) data warehouses on the SW: the QB4SOLAP vocabulary~\cite{nuref2015solap}, for instance, allows publishing data with spatial multidimensional concepts on the SW and high-level SOLAP operators can be translated into SPARQL~\cite{nuref2017qb4solapSWJ}. Based on these algorithms, GeoSemOLAP~\cite{nuref2017GeoSemOLAP} enables users to issue queries involving SOLAP operations on the SW without detailed knowledge of SPARQL or RDF. However, GeoSemOLAP is restricted to RDF data sets that are annotated with QB4SOLAP. 

To minimize user effort for querying existing spatial RDF datasets and endpoints (which are already published in other vocabularies, e.g., QB  or QB4OLAP) with spatial analytical queries (SOLAP), an automated way of annotating spatial metadata with QB4SOLAP from existing endpoints is necessary. Therefore, this paper proposes an \emph{RDF2SOLAP enrichment module} that operates at the back-end of GeoSemOLAP.

\paragraph{Contributions.} In summary, the main contributions of this paper are:

\begin{itemize}
\item [*] We illustrate the need for QB4SOLAP, i.e., the need to enable fully-fledged data warehouse concepts, and introduce running examples from real world governmental open data on environment and farming with complex geometry types.

\item [*] A detailed explanation and comparison of RDF data examples, which are depicted as graphs, and annotated both with QB4OLAP and QB4SOLAP vocabularies, then identifying the required spatial MD metadata and concepts (e.g., spatial hierarchies and topological relations) for SOLAP analysis based on the given comparison. 
\item [*] Hierarchical enrichment algorithms for (1) detecting topological relations at \emph{explicit} hierarchy steps with direct links between the level members; and (2) discovering topological relations at \emph{implicit} hierarchy steps (without direct links between the level members). 
\item [*] Factual enrichment algorithms for both implicit and explicit fact-level relations between fact and level members.
\item [*] An automated way of re-defining a fact schema after factual enrichment, and association of spatial aggregate functions with spatial measures. 
\item [*] General implementation of our approach for both hierarchical enrichment and factual enrichment processes.
\item [*] Evaluation of our approach in terms of accuracy and coverage in comparison to two standard environments (RDBMS and GIS tool). 
\end{itemize}

\paragraph{Paper organization.} The remainder of this paper is organized as follows. Section~\ref{sec:background} defines the preliminary concepts used throughout the paper with a running use case example. Section~\ref{sec:sysarch} presents the system architecture for the MD enrichment process. Section~\ref{sec:rdf2solap} defines the RDF2SOLAP enrichment algorithms with necessary helper functions and formalization of (spatial) RDF data. Section~\ref{sec:implementation} presents the implementation details along with interesting examples and discusses the challenges and implemented solutions. Section~\ref{sec:evaluation} presents the qualitative and performance evaluation with comparison baselines. Finally, Section~\ref{sec:relatedwork} discusses related work and Section~\ref{sec:conclusion} concludes the paper with an outlook to future work.

%-------------------------------------------------------------------------

\section{Preliminaries}
\label{sec:background}
% !TeX spellcheck = en_US
% !TEX root = ../iosart2x.tex

In this section, we explain the preliminary concepts of spatial data warehouses and spatial OLAP (SOLAP) (Section~\ref{subsec:presdw}) and how to deploy them on the Semantic Web (Section~\ref{subsec:preRDF}) using the QB4SOLAP vocabulary.

\subsection{Spatial Data Warehouses and SOLAP}
\label{subsec:presdw}

\paragraph{\textbf{Data cubes and spatially extended cube concepts}}
Data warehouses (DW) are based on a multidimensional (MD) model that models data in an $n$-dimensional space -- often referred to as a data cube. A cube \emph{schema} defines the structure of a cube with MD concepts. The cells of the cube represent \emph{(observation) facts} with a set of attributes called \emph{measures}. Facts are linked to \emph{dimensions}, which are the axes of an MD space and provide perspectives to analyze the data. Dimensions are organized into \emph{hierarchies}, which allow users to aggregate measures at different granularities along the levels of a hierarchy. Hierarchies are composed of \emph{levels}, which have a set of \emph{attributes} describing the characteristics of the level members. Each \emph{level member} is defined by its attributes and attribute values.

Cube members are MD concepts that are defined at the \emph{instance} level and composed of level members, attributes of level members, \emph{partial order} on level members, and fact members. A \emph{hierarchy step} between levels (a child level and a parent level) defines a set of \emph{roll-up} relations, where each relation relates a child level member to a parent level member. These roll-up relations define a partial order between level members with a \emph{cardinality} relation. The cardinality (1:1, 1:N, N:1, N:M) describes the number of members in one level that can be related to a member in the other level for both child and parent levels.        

\begin{figure}[t]
	%\vspace{-0.19in}
	\centering
	\includegraphics[width=1\columnwidth]{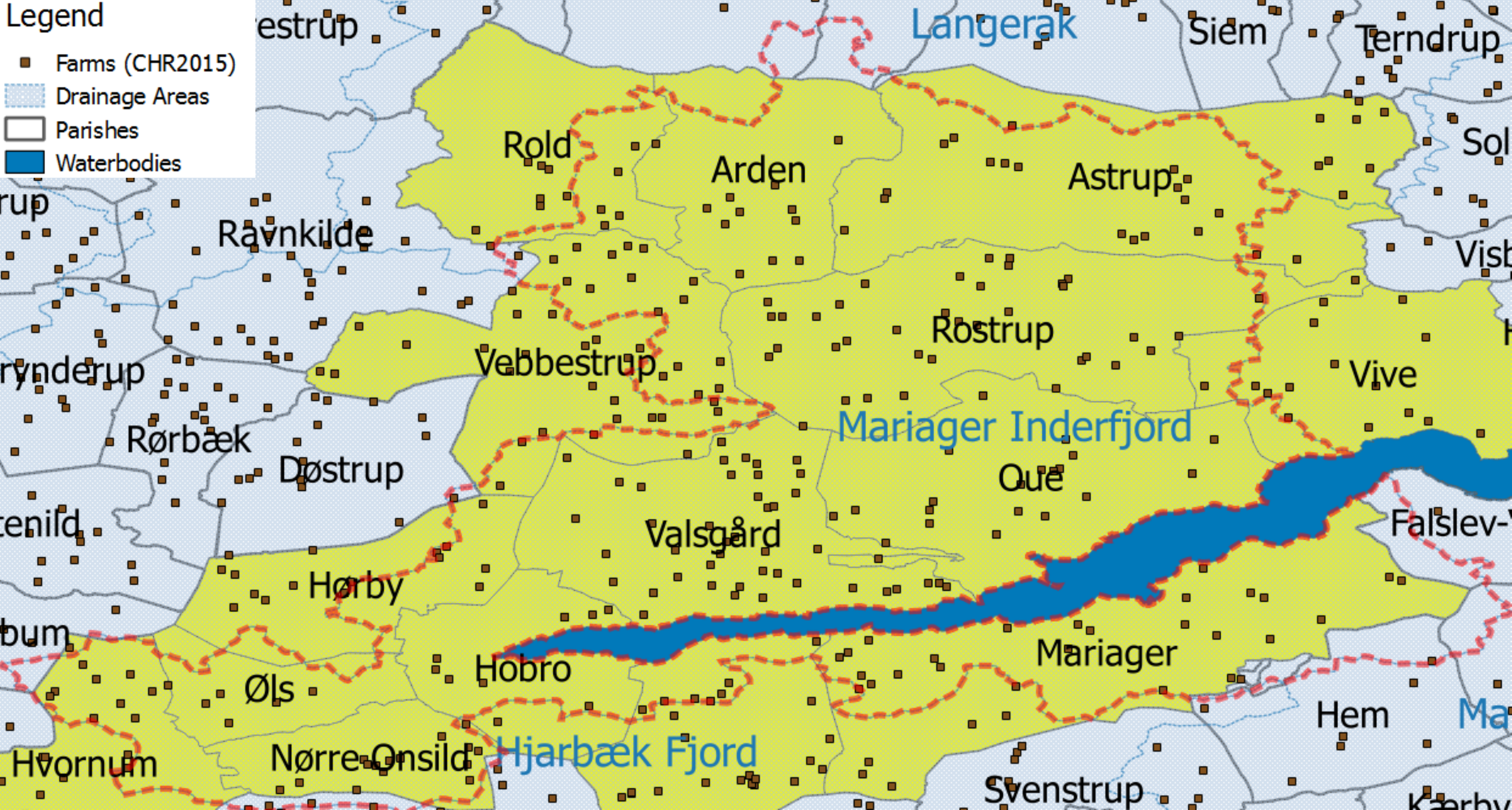}%
	\vspace*{-2ex}
	\caption{GeoFarmHerdState -- Parish, Farm, and Drainage area instances }%
	\label{fig:exmap}%
%	\vspace{-0.1in}
\end{figure}

 Spatial data warehouses (SDW) extend a DW by storing geometries such as \emph{point, line,} and \emph{polygon} in the values of spatial measures and values of level attributes for spatial dimensions. The spatially extended MD schema of an SDW has spatial dimensions, spatial hierarchies, spatial levels~\cite{Malinowski:2004:RSC:1032222.1032226}, spatial hierarchy steps, and topological relations\footnote{Topological relations are Boolean spatial predicates that specify how two spatial objects are related to each other, e.g., \textit {within, intersects, touches, crosses} and etc.~\cite{DE9DIM}.} (in addition to cardinality relations) between spatial levels for each spatial hierarchy step~\cite{nuref2015solap}. Similar to conventional DWs, facts of an SDW can be associated with numeric measures, which are using aggregation functions such as \texttt{SUM, AVG,} etc.~A fully extended spatial MD schema of an SDW should also define spatial measures, which have geometries and spatial aggregate functions such as \texttt{UNION, CONVEX HULL,} etc. For a detailed explanation of SDW concepts we refer the reader to~\cite{vaisman2014spatial}.

\begin{figure}[t]%
	\vspace{-0.19in}
	\centering
	\includegraphics[width=1\columnwidth]{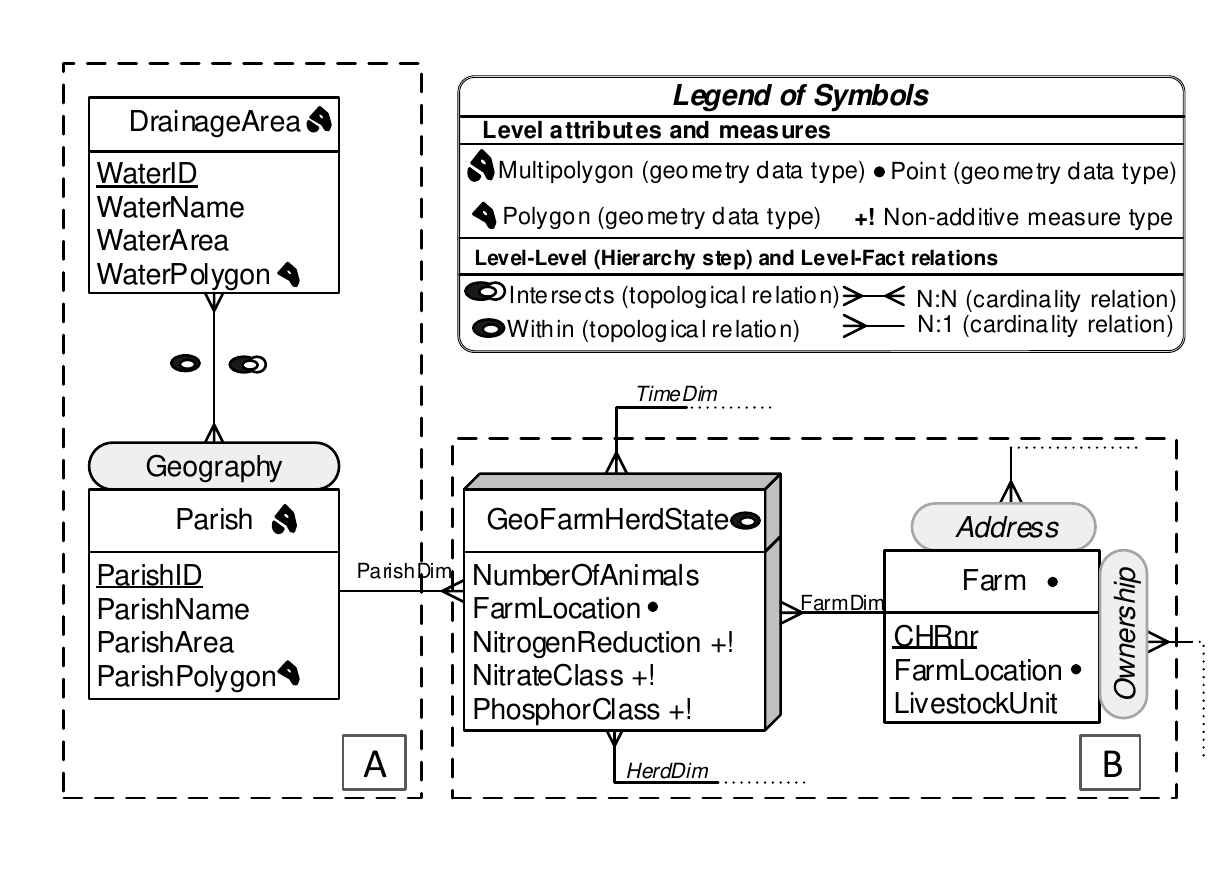}%
	\vspace*{-0.2in}
	\caption{GeoFarmHerdState -- Conceptual MD schema of livestock holdings data (Spatial concepts) }%
	\label{fig:geodyrdw}%
	\vspace{-0.1in}
\end{figure}

\paragraph{\textbf{OLAP and spatial OLAP operations}}
DWs are commonly used to store large volumes of data for decision support with On-Line Analytical Processing (OLAP) operations. Spatial OLAP (SOLAP) integrates the features of OLAP tools and geographical information systems (GIS)~\cite{rivest2005solap}. SOLAP enables advanced analytical processing by taking the spatial information in the cube into account. 

For example, a spatial data cube of livestock holdings in farms (referred to as \emph{GeoFarmHerdState} in the rest of this paper) defines the farm location as a spatial measure, which is linked to the observation facts. In order to derive perspectives and relations on the state of the farms' livestock holdings (herds), spatial levels are defined: parishes and drainage areas. A sample set of the corresponding spatial data cube members are given in Figure~\ref{fig:exmap}. The spatial MD concepts of the data cube are defined in the conceptual schema in Figure~\ref{fig:geodyrdw}, which depicts a simplified version of the GeoFarmHerdState spatial data cube without its non-spatial dimensions (see~\cite{nuref2016solap} for further details the GeoFarmHerdState cube). The cube has two spatial dimensions: \emph{FarmDim} and \emph{ParishDim}. The latter has a spatial hierarchy (\emph{Geography}) with two spatial levels: \emph{Parish} and \emph{DrainageArea}. \emph{FarmDim} on the other hand does not have a spatial hierarchy, despite its spatial (base) level: \emph{Farm}.

The GeoFarmHerdState cube has spatial fact members for farms within a time frame and different kinds of measures, i.e., numeric measures: \emph{NumberofAnimals} in the farm and \emph{NitrogenReduction} potential of the farm land/soil, spatial measures: \emph{FarmLocation} (Figure~\ref{fig:geodyrdw})\footnote{Non-additive measures are also numeric measures, which are given in percentages or classified in numbers, therefore they cannot be meaningfully summarized by all aggregate functions i.e., \texttt{SUM}. However, depending on the semantics, other aggregate functions can be associated with them, e.g., \texttt{AVG} NitrogenReduction potential, \texttt{MAX} NitrateClass.}. 

To evaluate SOLAP operations, spatial levels such as \emph{Parish} and \emph{DrainageArea} are used to aggregate measures at different levels of detail. Due to the \emph{polygon} geometry of the spatial level members, there are two different roll-up relations for the hierarchy step between the Parish and DrainageArea levels, where a parish can be completely contained \emph{within} a drainage area or a parish and a drainage area can \emph{intersect}.

\begin{figure}[t]
	\vspace{-0.25in}
	\centering
	\hspace*{-0.8 cm} %{-0.97cm}%  
	\includegraphics[width=1.1\columnwidth]{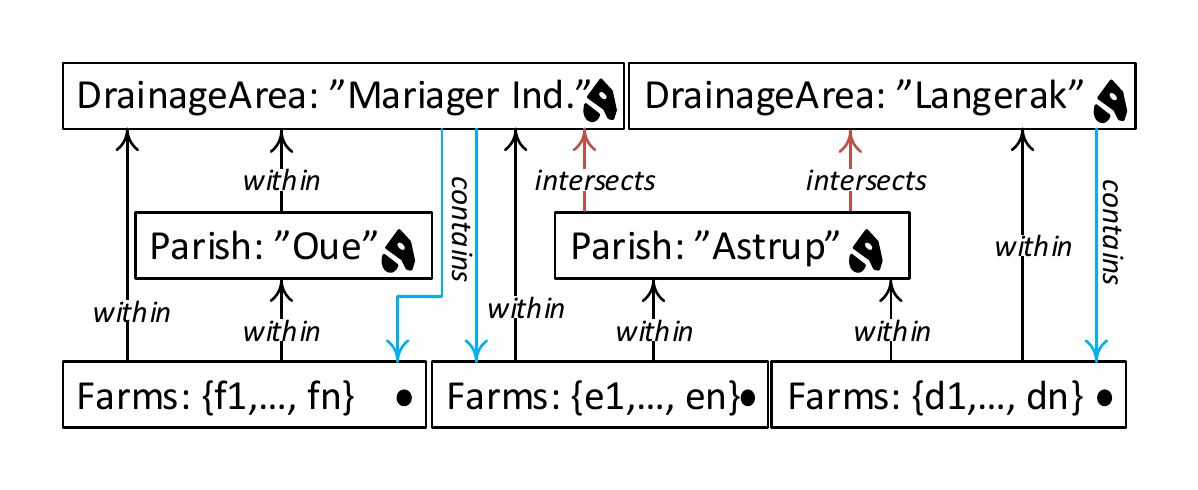}%
	\vspace{-0.15in}
	\caption{Hierarchy example for SOLAP}%
	\label{fig:exlevels}%
	%	\vspace{-0.2in}
\end{figure}

For example, parish ``Oue" is \textit{within} drainage area ``Mariager Inderfjord". Thus all the farms that are \textit{within} ``Oue" are also \textit{within} ``Mariager Inderfjord". Whereas, parish ``Astrup" \textit{intersects} with drainage areas ``Mariager Inderfjord" and ``Langerak". Therefore, some farms that are \textit{within} ``Astrup" are \textit{within} ``Mariager Inderfjord", while the rest of the farms are \textit{within} ``Langerak". Figure~\ref{fig:exmap} displays a sample set of Parish and DrainageArea level members.

The possible roll-up relations for the example above are depicted in Figure~\ref{fig:exlevels} with black and red arrows representing the topological relations \textit{within} and \textit{intersects}. Blue arrows show the topological relation \textit{contains}, which are drill-down (inverse operation of roll-up) relations from DrainageArea level to Farm level.

Topological relations between levels and facts can be implicitly specified through the geometry attributes of their instances (level members and fact members). The relations between spatial levels enable processing spatial roll-up and drill-down through range queries with spatial predicates~\cite{gaede1998multidimensional}. In terms of cardinality, there is an N:M relationship between level members since a parish may intersect with more than one drainage area and vice versa. This induces the problem of computing measures incorrectly when a roll-up operation goes through an N:M relationship, which actually is the case between the Parish level and the DrainageArea level. For example, we would like to aggregate the measure NumberOfAnimals, from Parish level to the DrainageArea level with a roll-up query. In such a roll-up query, we might falsely aggregate the number of animals in farms that are contained \textit{within} the parish, but not contained \textit{within} the drainage area, since the parish \textit{intersects} with another drainage area. In order to refine such an analysis, SOLAP operations are required, where a (spatial) drill-down should be applied to the lowest granularity - from Parish level members to GeoFarmHerdState fact members, and then a spatial roll-up (with within predicate) can be applied from fact members (Farm instances) to DrainageArea level members. This would prevent falsely aggregating the number of animals from the farms that are (spatially) disjoint to the corresponding drainage area.

\begin{figure}[b]
	%\vspace{-0.05in}
	\centering
	\includegraphics[width=0.99\columnwidth]{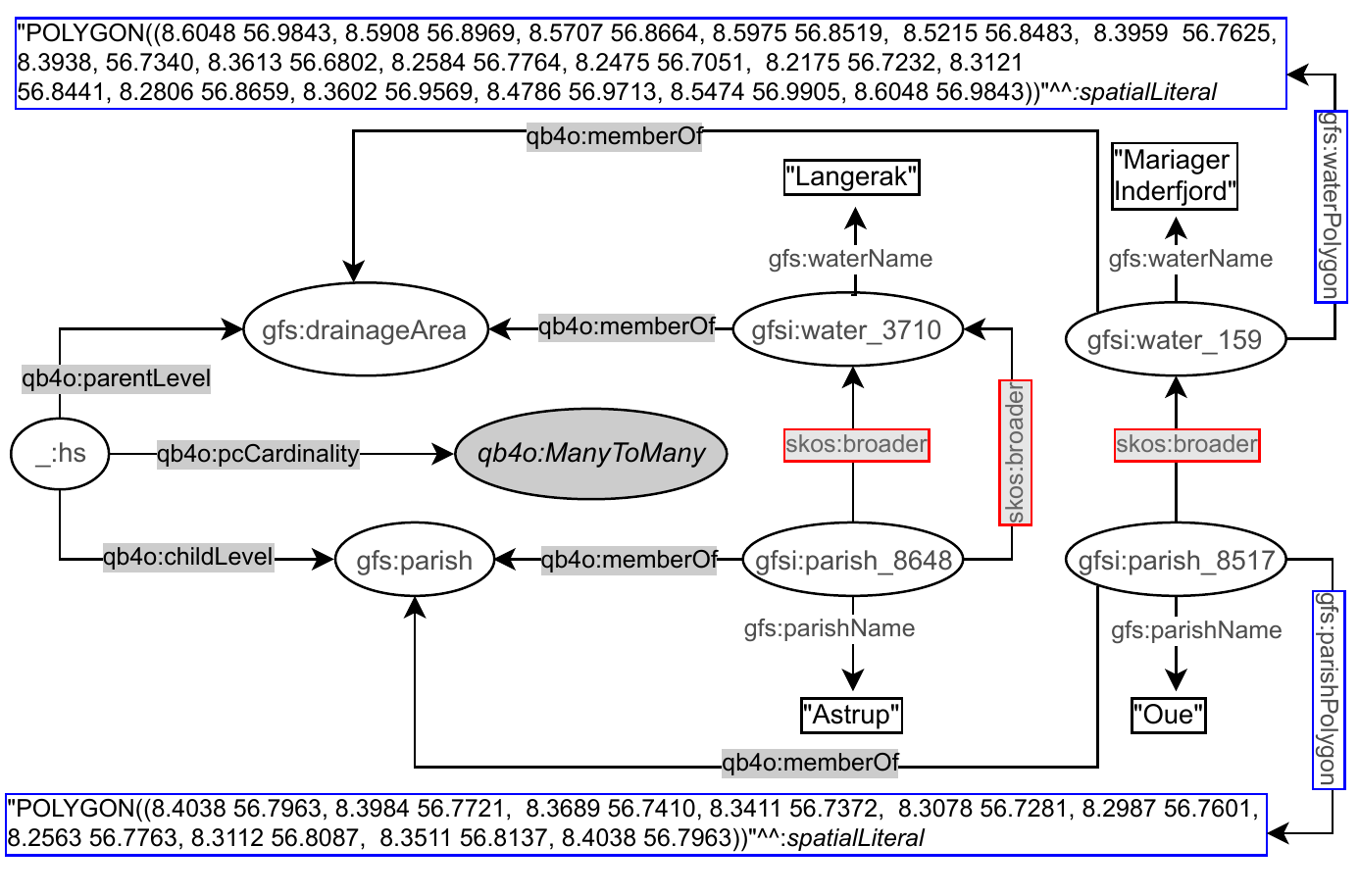}%
	%\vspace*{-2ex}
	\caption{Hierarchy steps in QB4OLAP before multidimensional enrichment}%
	\label{fig:hsqb4olap}%
	\vspace{-0.1in}
\end{figure}

\subsection{QB4SOLAP: Spatial RDF Data Cube Vocabulary for SOLAP operations}\label{subsec:preRDF}

There is an increasing amount of Linked Open Data (LOD) on the Semantic Web containing spatial information and numerical (statistical) data. This led to new opportunities for OLAP over spatial data using semantic web technologies and standards. Datasets on the SW use a standardized format: RDF (Resource Description Framework)\footnote{\url{https://www.w3.org/TR/rdf11-primer/}}. 

In order to enable SOLAP operations on the Semantic Web, a comprehensive vocabulary is needed, i.e., annotation of the spatial hierarchy steps with topological relations. QB4SOLAP~\cite{nuref2017qb4solapSWJ} is a vocabulary that allows the definition of \emph{cube schemas} and \emph{cube instances} in RDF. 
The QB4SOLAP vocabulary is an extension of QB4OLAP~\cite{etcheverry2014modeling} capturing the semantics of spatial MD concepts (i.e., spatial hierarchy steps) that are essential for SOLAP operations. The QB4SOLAP vocabulary V1.3 is available on our project website\footnote{\url{https://extbi.cs.aau.dk/QB4SOLAP}} as well as via a persistent URL\footnote{\url{https://w3id.org/qb4solap\#}}. 

A comprehensive foundation of spatial data warehouses on the Semantic Web can be found in~\cite{nuref2017qb4solapSWJ}, which includes detailed definitions with semantics of spatial MD concepts both at the schema level and instance level using QB4SOLAP.

\begin{figure}[b]
	%\vspace{-0.05in}
	\centering
	\includegraphics[width=0.99\columnwidth]{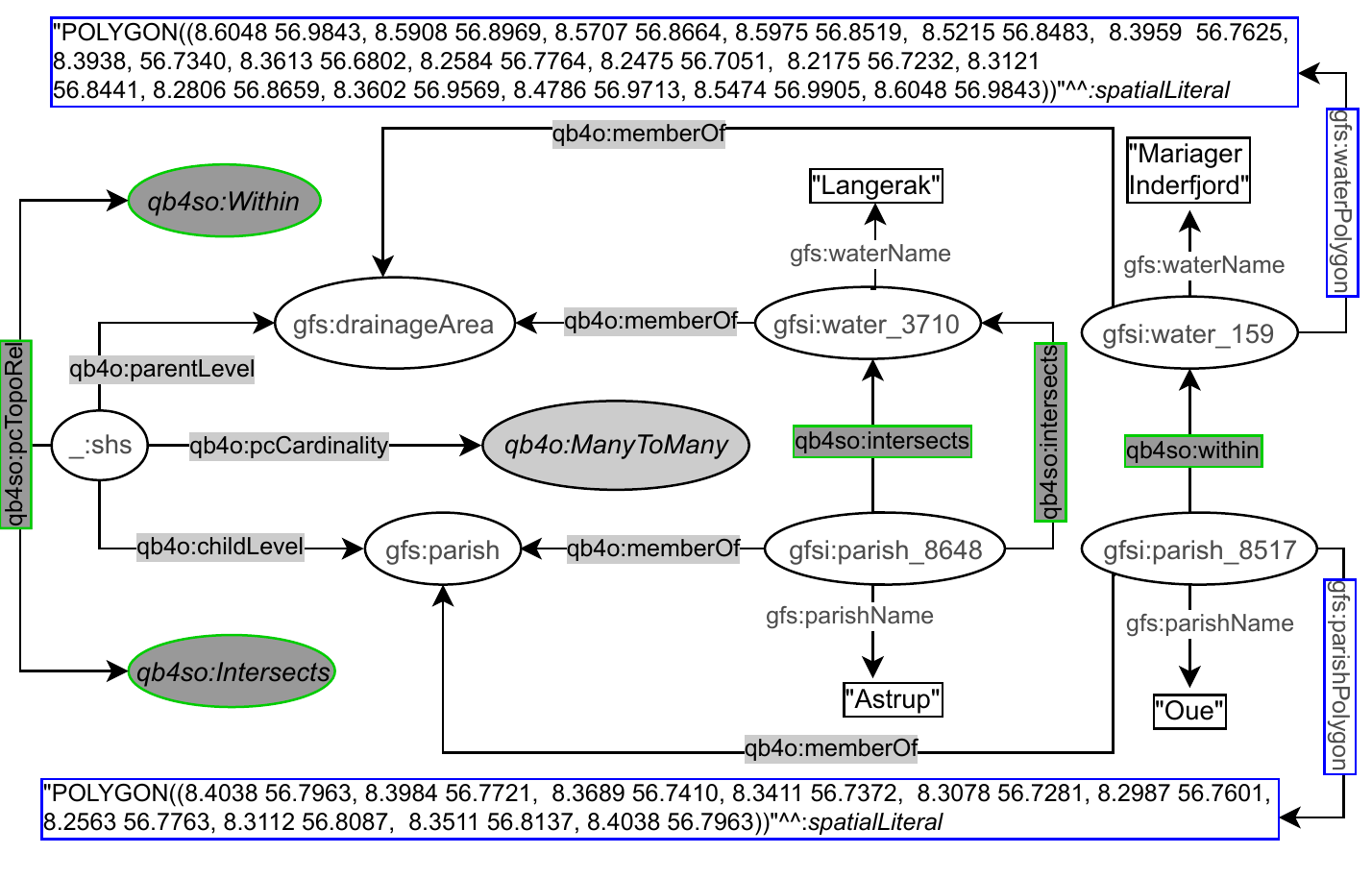}%
	%\vspace*{-2ex}
	\caption{Spatial hierarchy steps in QB4SOLAP after multidimensional enrichment}%
	\label{fig:hsqb4solap}%
	\vspace{-0.1in}
\end{figure}

In the following, we depict an example of a hierarchy step from \texttt{gfs:Parish} child level to \texttt{gfs:}\texttt{drainageArea} parent level (Figure~\ref{fig:hsqb4olap}). In the figure, we prefix the schema elements (attributes, levels, etc.) of the (GeoFarmHerdState) cube with \texttt{gfs:} and instance data from the cube with \texttt{gfsi:}. The left-center part of Figure~\ref{fig:hsqb4olap} shows the hierarchy structure \texttt{\_:hs}, between \texttt{gfs:parish} and \texttt{gfs:drainageArea} levels at the schema level with the QB4OLAP vocabulary. QB4OLAP objects, classes, and properties are prefixed with \texttt{qb4o:}. The levels (\texttt{gfs:parish} and \texttt{gfs:drainageArea}) are linked to the instances of level members (e.g., \texttt{gfsi:parish\_8648}, \texttt{gfsi:water\_3710} and etc.) by \texttt{qb4o:memberOf} property. The polygon geometry attributes are highlighted in blue boxes, on the top and the bottom of the figure. The coordinates recorded in the geometry attributes can be used to derive the topological relation between the level members by applying spatial boolean predicates (e.g., \emph{instersects?}, \emph{within?}) on the \emph{polygon} geometries of the parish and drainage area level members.

However, QB4OLAP does not support annotating the topological relations that might exist between the level members at a hierarchy step. QB4OLAP uses only \texttt{skos:broader} property from SKOS (Simple Knowledge Organization System)~\cite{alistair2009skos} semantic relations for capturing the roll-up relations at hierarchy steps. The roll-up relations with \texttt{skos:broader} property are highlighted in red boxes in Figure~\ref{fig:hsqb4olap}. The \texttt{skos:broader} property does not describe the nature of the roll-up relation with topological relations for spatial hierarchies. Therefore, QB4OLAP cannot capture the topological relations in a hierarchy step from Parish level to DrainageArea level or between these levels' members.

%%%%%%%%%%%%%%%%%%%%%%%%%%%%%%%%%%%%%%%%%%%%%%%%%%%%%%%%%%%%%%%%%%%%%%%%%%%%%%%%

On the other hand, QB4SOLAP can define topological relations both at the schema level and the instance level. In Figure~\ref{fig:hsqb4solap}, we prefix QB4SOLAP objects, classes, and properties with \texttt{qb4so:} and highlight them in green lines. The left-center part of the figure shows the spatial hierarchy structure \texttt{:\_shs}, which has a QB4SOLAP property \texttt{qb4so:pcTopoRel} with two QB4SOLAP class instances \texttt{qb4so:Within} and \texttt{qb4so:Intersects}. This means that when we compare the geometry attributes of parish level members and drainage area level members, we discover two different topological relations (\emph{within} and \emph{intersects}) for all the (spatial) hierarchy steps between the parish and drainage area levels. And these relations are annotated at the schema level on the left-center part of Figure~\ref{fig:hsqb4solap}.

Similarly, {gfs:parish} and \texttt{gfs:drainageArea} levels are linked to the instances of level members (e.g., \texttt{gfsi:parish\_8648}) by \texttt{qb4o:memberOf} property. The explicit topological relations between each level member along a spatial hierarchy step are depicted in the figure with \texttt{qb4so:intersects} or \texttt{qb4so:within} predicates, which are highlighted in green boxes (e.g., \texttt{gfsi:parish\_8648} \textit{intersects} with \texttt{gfsi:water\_159} and \texttt{gfsi:water\_3170} etc.).

In conclusion, QB4SOLAP enables SOLAP operations by defining the semantics of spatial MD concepts both at the schema level and instance level. These semantics are essential for SOLAP operations and they are defined as extensions to the QB4OLAP vocabulary.

%-------------------------------------------------------------------------

\section{System Architecture}
\label{sec:sysarch}
% !TeX spellcheck = en_US
% !TEX root = ../iosart2x.tex

The importance of SOLAP to get accurate results in operations over spatial data warehouses is explained in Section~\ref{subsec:presdw}. However, the RDF data cubes (with spatial attributes) on the Semantic Web are not always annotated with vocabularies that allow users to formulate SOLAP queries. In this section we present an overview of the MD enrichment flow from RDF QB to QB4OLAP data cubes and QB4OLAP to QB4SOLAP data cubes. Thus, users can query the RDF data cubes with SOLAP queries. 

\begin{figure}[htb]
	%\vspace{-0.19in}
	\centering
	\hspace{-1cm}
	\includegraphics[width=1.1\columnwidth]{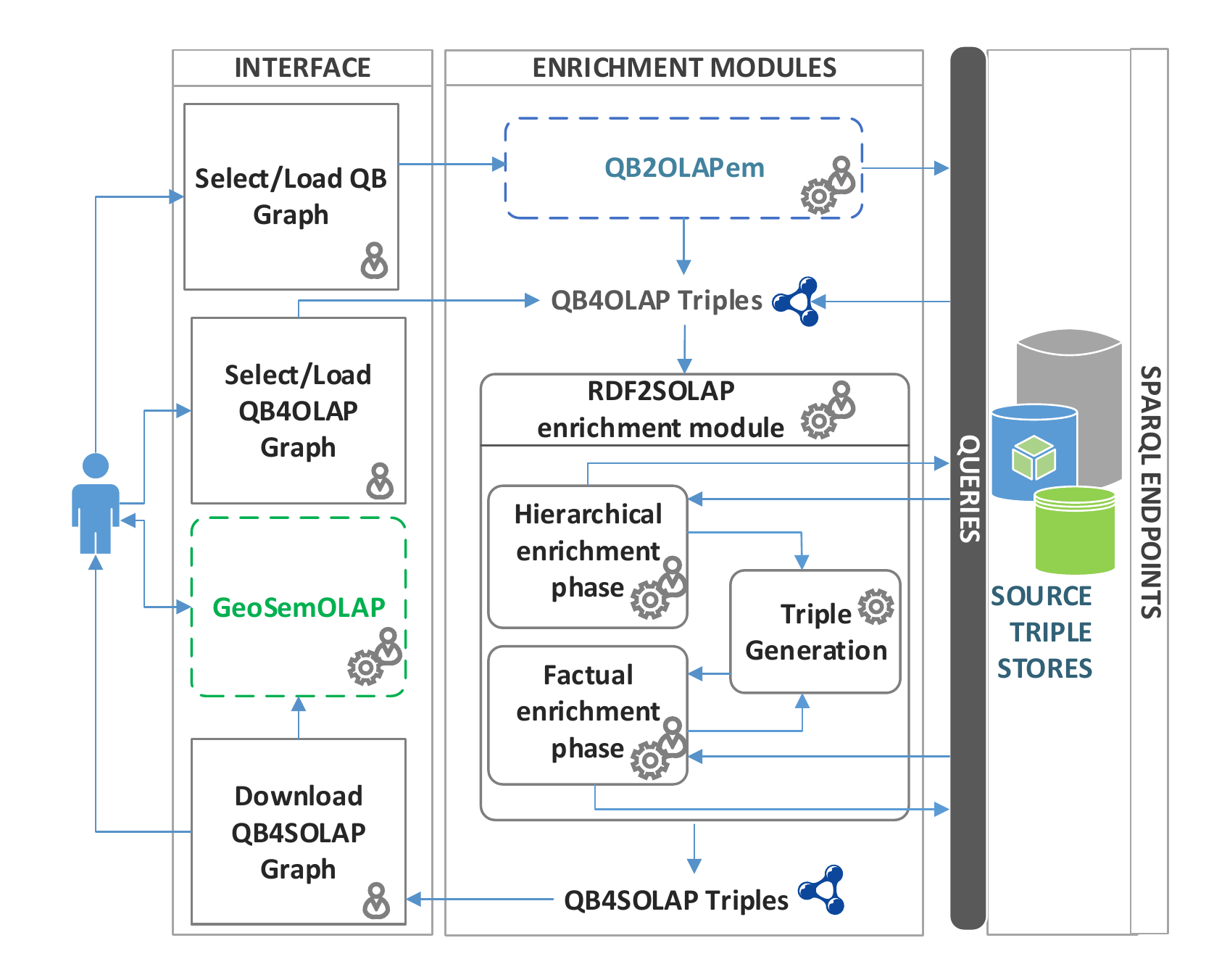}%
	\vspace*{-2ex}
	\caption{Multidimensional~Enrichment~Process}%
	%\caption{RDF2SOLAP Process Flow}%
	\label{fig:processflow}%
	\vspace{-0.1in}
\end{figure}

A multidimensional enrichment process flow is illustrated in Figure~\ref{fig:processflow} with three main architectural layers: Interface, Enrichment Modules, and SPARQL Endpoints. The first layer facilitates user interaction with the enrichment modules (i.e., QB2OLAPem)  and third party tools (i.e., GeoSemOLAP). Our main contribution in this paper is the RDF2SOLAP enrichment module, which is the core of the second layer. The RDF2SOLAP enrichment module operates on QB4OLAP triples that either already exist in the original data or have been generated by the QB2OLAPem enrichment module~\cite{varga2016dimensional}. QB2OLAPem allows users to enrich an RDF QB dataset with QB4OLAP concepts and returns a graph of QB4OLAP triples. 

The internal process flow of the RDF2SOLAP enrichment module consists of three phases: hierarchical enrichment, factual enrichment, and triple generation. The hierarchical and factual enrichment phases iteratively perform the enrichment algorithms explained in Section~\ref{sec:rdf2solap}. Both of these enrichment phases allow interaction with external SPARQL endpoints to enhance the enrichment process via potential spatial and multidimensional concepts that could be retrieved externally. The third phase is the triple generation, which creates QB4SOLAP triples that can be used in third party tools such as GeoSemOLAP. GeoSemOLAP allows users without knowledge of RDF and SPARQL to query with SOLAP operations by interactively formulating the queries using a GUI with interactive maps~\cite{nuref2017GeoSemOLAP}.

The third layer (SPARQL endpoints) allows interaction between user and SPARQL endpoint for retrieving QB or QB4OLAP graphs as well as interaction between system and SPARQL endpoints, where the RDF2SOLAP enrichment module queries external triple stores for hierarchical enrichment and factual enrichment.

RDF2SOLAP is implemented in Javascript on the Node.js platform using the N3.js library for parsing the RDF triples in Javascript and the Turfjs library for spatial analysis\footnote{\textbf{N3.js:} \url{https://github.com/rdfjs/N3.js}  \textbf{Turfjs:} \url{http://turfjs.org/}}.

%-------------------------------------------------------------------------

\section{RDF2SOLAP Enrichment Algorithms}
\label{sec:rdf2solap}
% !TeX spellcheck = en_US
% !TEX root = ../iosart2x.tex

This section presents the core algorithms of our RDF2SOLAP enrichment module. Our MD enrichment approach builds upon QB4OLAP triples that either already exist in the original data or have been generated by the QB2OLAPem enrichment module~\cite{varga2016dimensional} as depicted in Figure~\ref{fig:processflow}.  QB4OLAP defines only the non-spatial multidimensional semantics of RDF data, whereas QB4SOLAP enriches the MD semantics of RDF data with spatial concepts (formalizations and further details can be found in~\cite{nuref2017qb4solapSWJ}). 
Nevertheless, in the following we briefly introduce basic notations.

The basic construct of RDF is a triple $t = (s, p ,o)$ consisting of three components; $s$ is the subject, $p$ is the predicate, and $o$ is the object. RDF triples are defined over {\small ${\mathcal T = (\mathcal I \cup \mathcal B) \times \mathcal I \times (\mathcal I \cup \mathcal B \cup \mathcal L)}$}, where $\mathcal{I}$ is the set of \emph{IRIs} (Internationalized Resource Identifiers), $\mathcal{B}$ is the set of \emph{blank nodes}, and $\mathcal{L}$ is the set of \emph{literals}. An object value can be a \emph{literal} (i.e., string, spatial literal\footnote{Spatial literals are represented as $\mathcal{L}_s$.}, integer etc.). Subjects and objects can be represented by a \emph{blank node} for anonymous resources. Predicates are always represented by IRIs. A set of RDF triples is referred to as an RDF \emph{graph} $\mathcal{G}$. We use superscript notation to represent the type of a graph: schema graph $\mathcal{G}^S$ and instance graph $\mathcal{G}^I$. An instance graph has entities from a use-case dataset as a set of RDF triples. The schema graph describes the structure (schema) of the dataset recorded in the instance graph. We use subscript notation to represent the MD concepts in RDF terms as a graph. For example, $\mathcal{G}^{I}_{A(lm)}$ is the RDF instance graph for attributes of level members -- in the use case example this graph corresponds to the set of triples in Listing~2, Lines~3-6 or Lines~9-13 and Lines~17-22. $\mathcal{G}^{S}_{HS(h)}$ is the RDF schema graph for hierarchy steps -- in the use case example this graph corresponds to the set of triples in Listing~1.

We define function $id(x):\mathcal{G} \rightarrow \mathcal{I}$, which given an MD element $x$ returns its identifier $\mathcal{I}$ from graph $\mathcal{G}$. Similarly, we use superscript notation to indicate the type of the identifier from the schema graph ($\mathcal{G}^{S}$) and instance graph ($\mathcal{G}^{I}$), e.g.,~$id^{S}(a)$ for a schema identifier of a level (\texttt{gfs:parish} in Listing~2, Line~2 or in Listing~1, Line~2) and $id^{I}(lm)$ for an instance identifier of a level member (\texttt{gfsi:parish\_8648} in Listing~2, Line~1 or Line~8). 

The MD enrichment process in RDF2SOLAP runs in two phases (\emph{hierarchical enrichment phase} and \emph{factual enrichment phase}), which are explained in the following.

\subsection{Hierarchical enrichment phase}
\label{subsec:hierenrichment}
% !TeX spellcheck = en_US
% !TEX root = ../iosart2x.tex

The hierarchical enrichment phase is built around spatial levels and their level members forming the spatial hierarchy of a dimension. Thus, by identifying the spatial relations between spatial levels and their level members, we can find the spatial hierarchy steps and the possible topological relations for these hierarchy steps. 

Each spatial hierarchy corresponds to a path of roll-up relationships between the child level and parent level: each of these roll-up relationships corresponds to a \emph{spatial hierarchy step} (Section~\ref{subsec:presdw}). An example of a (spatial) hierarchy with QB4SOLAP is given in Listing~1. Line~4 extends the QB4OLAP schema definitions by enriching the hierarchy step with the possibility to annotate the spatial hierarchy steps with topological relations (see Section~\ref{sec:background} for details and Section~\ref{subsec:preRDF} for examples). 

\begin{sftabbing} \label{list:member2}
	\setcounter{AlgoLine}{0}
	\LinesNumbered
	XX \= XX \= XX \= XX \= \kill 
	\textbf{\#\# Spatial hierarchies in \textit{QB4SOLAP with topological relations}\#\# }\\
	\nl \_:\_shs rdf:type qb4o:HierarchyStep ; qb4o:inHierarchy gfs:geography ;\\
	\nl \> qb4o:childLevel gfs:parish ; qb4o:parentLevel gfs:drainageArea ;\\
	\nl \> qb4o:pcCardinality qb4o:ManyToMany ;\\
	\nl \> \textcolor{blue}{qb4so:pcTopoRel qb4so:Within , qb4so:Intersects .}
\end{sftabbing}
\noindent\footnotesize{Listing 1: Spatial Hierarchy structure in QB4SOLAP}
\normalsize
\medskip

Listing 2 shows the GeoFarmHerdState spatial level members from Parish and Drainage Area levels. Lines 1-7 (Listing~2) represent the QB4OLAP annotation of a child level member from Parish level before multidimensional enrichment (with \texttt{skos:broader}), which is depicted in Figure~\ref{fig:hsqb4olap}. Lines 8-14 represent the QB4SOLAP annotation of the same Parish level member after the multidimensional enrichment with topological relations (depicted in Figure~\ref{fig:hsqb4solap}). Lines 15-22 represent the annotation of a parent level member from the Drainage area level, which remains the same before and after multidimensional enrichment \textcolor{black}{since the hierarchy steps are defined with bottom-up relationships from child level to parent level and the roll-up relations and thus also the topological relations are annotated at the child level members of the hierarchy step.}

\begin{sftabbing} \label{list:member1}
	\setcounter{AlgoLine}{0}
	\LinesNumbered
	XX \= XX \= XX \= XX \= \kill 
	\textbf{\#\# Parish (child) Level member before hierarchical enrichment\#\#} \\
	\nl	\textcolor{black}{gfsi:parish\_8648 rdf:type qb4o:LevelMember ;} \\
	\nl	\>\textcolor{black}{qb4o:memberOf gfs:parish ;} \\
	\nl	\>\textcolor{black}{gfs:parishID 8648 ; 	gfs:parishName "Astrup" ;} \\
	\nl	\>\textcolor{black}{gfs:parishArea 47,969 ;} \textcolor{teal}{gfs:parishPolygon "POLYGON((8.438 56.796,} \\
	\nl	\>\textcolor{teal}{8.3984 56.7721, 8.3689 56.7410, 8.3411 56.7372, 8.3078 56.7281,} \\
	\nl	\>\textcolor{teal}{8.3112 56.8087, 8.3511 56.8137, 8.438 56.796))"\textasciicircum\textasciicircum geo:spatialLiteral ;}	\\
	\nl	\>\textcolor{red}{skos:broader gfsi:water\_3710 , gfsi:water\_159} . \\
	%	XX \= XX \= XX \= XX \= \kill 
	\textbf{\#\# Parish (child) Level member after hierarchical enrichment\#\#} \\
	\nl	\textcolor{black}{gfsi:parish\_8648 rdf:type qb4o:LevelMember ;} \\
	\nl	\>\textcolor{black}{qb4o:memberOf gfs:parish ;} \\
	%\> \nl	\>\textcolor{orange}{skos:broader gfsi:water\_3710 , gfsi:water\_159} ; \\	
	\nl	\>\textcolor{black}{gfs:parishID 8648 ; 	gfs:parishName "Astrup" ;} \\
	\nl	\>\textcolor{black}{gfs:parishArea 47,969 ;} \textcolor{teal}{gfs:parishPolygon "POLYGON((8.438 56.796,} \\
	\nl	\>\textcolor{teal}{8.3984 56.7721, 8.3689 56.7410, 8.3411 56.7372, 8.3078 56.7281,} \\
	\nl	\>\textcolor{teal}{8.3112 56.8087, 8.3511 56.8137, 8.438 56.796))"\textasciicircum\textasciicircum geo:spatialLiteral ;}	\\
	\nl	\>\textcolor{blue}{qb4so:intersects gfsi:water\_3710 , gfsi:water\_159} . \\
	\textbf{\#\# DrainageArea (parent) Level member\#\# }\\
	\nl	\textcolor{black}{gfsi:water\_159 rdf:type qb4o:LevelMember ;} \\
	\nl	\>\textcolor{black}{ qb4o:memberOf gfs:drainageArea ;}\\		
	\nl	\> \textcolor{black}{gfs:waterName "Mariager Inderfjord" ; gfs:waterArea 267,477 ;} \\
	\nl	\>\textcolor{teal}{ gfs:waterPolygon  "POLYGON((8.6048 56.9843, 8.5908 56.8969,} \\
	\nl	\> \textcolor{teal}{8.5707 56.8664, 8.5975 56.8519, 8.5215 56.8483,} \\
	\nl	\> \textcolor{teal}{8.3959 56.7625, 8.3938, 56.7340, 8.3613 56.6802,} \\
	\nl	\> \textcolor{teal}{8.2584 56.7764, 8.2475 56.7051, 8.2175 56.7232,} \\
	\nl	\> \textcolor{teal}{8.5474 56.9905, 8.6048 56.9843))"\textasciicircum\textasciicircum geo:spatialLiteral .}
 
\end{sftabbing} 
\noindent\footnotesize{Listing 2: GeoFarmHerdState level members, attributes, and \textit{spatial} roll-up relations}

\normalsize
\medskip

We exploit QB4OLAP semantics, such as \emph{non-spatial} hierarchy steps and levels as a starting point to find the \emph{spatial} hierarchy steps. We distinguish two cases: 

\noindent\textit{Case 1:} Finding \textit{explicit} spatial hierarchy steps for QB4OLAP levels, with \texttt{skos:broader} roll-up relations between their child-parent level members by \emph{detecting spatial hierarchy steps} in Section~\ref{subsec:detectSpatial}. For this case we assume that level members have direct \texttt{skos:broader} relations as depicted in Figure~\ref{fig:hsqb4olap} and Listing~2, Line~7 with \texttt{skos:broader} property.

\noindent\textit{Case 2:} Finding \textit{implicit} spatial hierarchy steps from QB4OLAP levels without direct roll-up relations through the \texttt{skos:broader} property. In this case, we assume that the level members are only defined by the \texttt{qb4o:memberOf} property as shown in Listing~2, (Line~2) but \textit{do not} have the  \texttt{skos:broader} roll-up relation as given in Line~7. In this case, it is still possible to \emph{discover spatial hierarchy steps} by finding spatial (topological) relations between level members through their attributes as explained in Section~\ref{subsec:discoverSpatial}.

\subsubsection{Spatial helper functions}\label{subsec:spatialhelper}

To address the cases explained above, we need two spatial helper functions; for retrieving spatial attribute values (Algorithm~\ref{algo:getSpatialValues}, \texttt{getSpatialValues}), and for relating spatial attributes (Algorithm~\ref{algo:relateSpatialValues}, \texttt{relateSpatialValues}). 

\paragraph{Algorithm~\ref{algo:getSpatialValues} \textnormal{(\texttt{getSpatialValues})}.} 
The first helper function gets an input graph of attributes of level members $\mathcal{G}^{I}_{A(lm)}$ and returns a set of spatial attribute values $V_{s(a)}$. For example, the function could receive Lines~3-6 from Listing~2 as input. In the algorithm, Lines~3 and~4 check the values $v_{a_i}$ of each attribute $id^{S}(a_i)$ (e.g., \texttt{gfs:parishName}, \texttt{gfs:ParishArea}, etc.) If the value is a type of \texttt{geo:SpatialLiteral} (e.g., the \texttt{POLYGON} geometry value linked to the \texttt{gfs:parishPolygon} attribute), then the value is incrementally added to the output set $V_{s(a)}$\footnote{Note that a level member might have the polygon geometry type for the parish borders and the point geometry type for the parish center, therefore a set of spatial values is required.} in Line~5. 

\begin{algorithm2e}[t]
\DontPrintSemicolon
\KwIn{$\mathcal{G}^I_{A(lm)}$}
\KwOut{$V_{s(a)}$}
\nl \Begin{
\nl $V_{s(a)} = \emptyset $; \hfill/*initialize output set as empty set*/\\	
\nl \ForEach{$(id^I(lm) \ id^S(a_i) \ v_{a_i}) \in \mathcal{G}^I_{A(lm)}$}{
			\nl \If{$v_{a_i}$ is a \textnormal{\texttt{geo:spatialLiteral}} }{\nl $V_{s(a)} \cup = \{v_{a_i}\}$;}
			%}
		}
		\nl \KwRet{$V_{s(a)}$}	
	}
	\caption{getSpatialValues$(\mathcal{G}^{I}_{A(lm)}):~V_{s(a)}$} \label{algo:getSpatialValues}	
\end{algorithm2e}

\paragraph{Algorithm~\ref{algo:relateSpatialValues}: \textnormal{(\texttt{relateSpatialValues})}.} The next helper function is designed based on Table~\ref{tab:topoRel}, w.r.t. the geometry values of the child-parent level members and based on the structure of a hierarchy step. We prepared Table~\ref{tab:topoRel} with topological relations based on DE-9IM\footnote{DE-9IM (Dimensionally Extended Nine-Intersection Model) is a topological model that describes spatial relations of two geometries in two dimensions~\cite{DE9DIM}.}. We consider only the three simple geometry types, \emph{point, line, and polygon} as the spatial attribute values of child-parent level members in roll-up relations, excluding complex geometry types, such as multi-polygon, multi-point, etc. The possible topological relations that can occur in a spatial hierarchy step with a roll-up relation from child level to parent level are marked with check sign (\checkmark) in the table. Topological relations, such as \emph{contains} and \emph{covers}, are not \emph{hierarchically applicable} since a spatial child level member cannot contain or cover a spatial parent level member. For these relations, we mark the complete rows with  minus sign (--) in the table, since they are not hierarchically applicable. Similarly, we mark the complete columns of \emph{line-point}, \textit{polygon-point}, and \textit{polygon-line} roll-up relations with the minus sign (--) since these are also not hierarchically applicable. This is because we assume that in the instance data, a parent level member should always have a spatial attribute of a geometry type of the same or higher dimensionality of its child level member (a point is 0-dimensional, a line is 1-dimensional and a polygon is 2-dimensional).

\begin{table}[t]
	\centering
	\caption{Topological relations for Hierarchy Steps (\checkmark: hierarchically and topologically applicable, $\times$: topologically not applicable, --: hierarchically not applicable)}
	\label{tab:topoRel}	
	\resizebox{\columnwidth}{0.2\columnwidth}{ %
		\begin{tabular}{@{}c|l|ccc|ccc|ccc@{}}
			\toprule
			\multicolumn{1}{l|}{\multirow{2}{*}{\begin{tabular}[c]{@{}l@{}}Roll-up \\ Relations\end{tabular}}} & child level & \multicolumn{3}{c|}{point (pt.)} & \multicolumn{3}{c|}{line (ln.)} & \multicolumn{3}{c}{polygon (po.)} \\ \cmidrule(l){2-11} 
			\multicolumn{1}{l|}{} & parent level & \multicolumn{1}{l}{pt.} & \multicolumn{1}{l}{ln.} & \multicolumn{1}{l|}{po.} & \multicolumn{1}{l}{pt.} & \multicolumn{1}{l}{ln.} & \multicolumn{1}{l|}{po.} & \multicolumn{1}{l}{pt.} & \multicolumn{1}{l}{ln.} & \multicolumn{1}{l}{po.} \\ \midrule
			\parbox[t]{2mm}{\multirow{9}{*} {\rotatebox[origin=c]{90}{Topological Relations}}} & within & $\times$ & \checkmark & \checkmark & -- & \checkmark & \checkmark & --  & -- & \checkmark \\
			& contains & --  & -- & -- & -- & -- & -- & -- & -- & -- \\
			& intersects & \checkmark & \checkmark & \checkmark & -- & \checkmark & \checkmark & -- & -- & \checkmark \\
			& touches & $\times$ & $\times$ & $\times$ & --  & \checkmark & \checkmark & -- & -- & \checkmark \\
			& overlaps & $\times$ & $\times$ & $\times$ & -- & \checkmark & \checkmark & -- & -- & \checkmark \\
			& crosses & $\times$ & $\times$ & $\times$ & -- & \checkmark & \checkmark & -- & -- & $\times$ \\
			& coveredBy & $\times$ & $\times$ & $\times$ & -- & $\times$ & \checkmark & -- & -- & \checkmark \\
			& covers & --  & -- & -- & -- & -- & -- & -- & -- & -- \\
			& equals & \checkmark & $\times$ & $\times$ & -- & \checkmark & $\times$ & -- &--  & \checkmark
		\end{tabular}
	}
\end{table}

 For example, a child level member with a spatial attribute of line geometry can only have parent level member(s) with spatial attributes of line or polygon geometries but not point geometry. We mark the \emph{topologically not applicable} relations with cross sign ($\times$) according to the DE-9IM model (e.g, a line cannot overlap a polygon). 

\begin{figure}[b]
	\vspace{-0.1in}
	\centering
	\hspace{-1cm}
	\includegraphics[width=1.08\columnwidth]{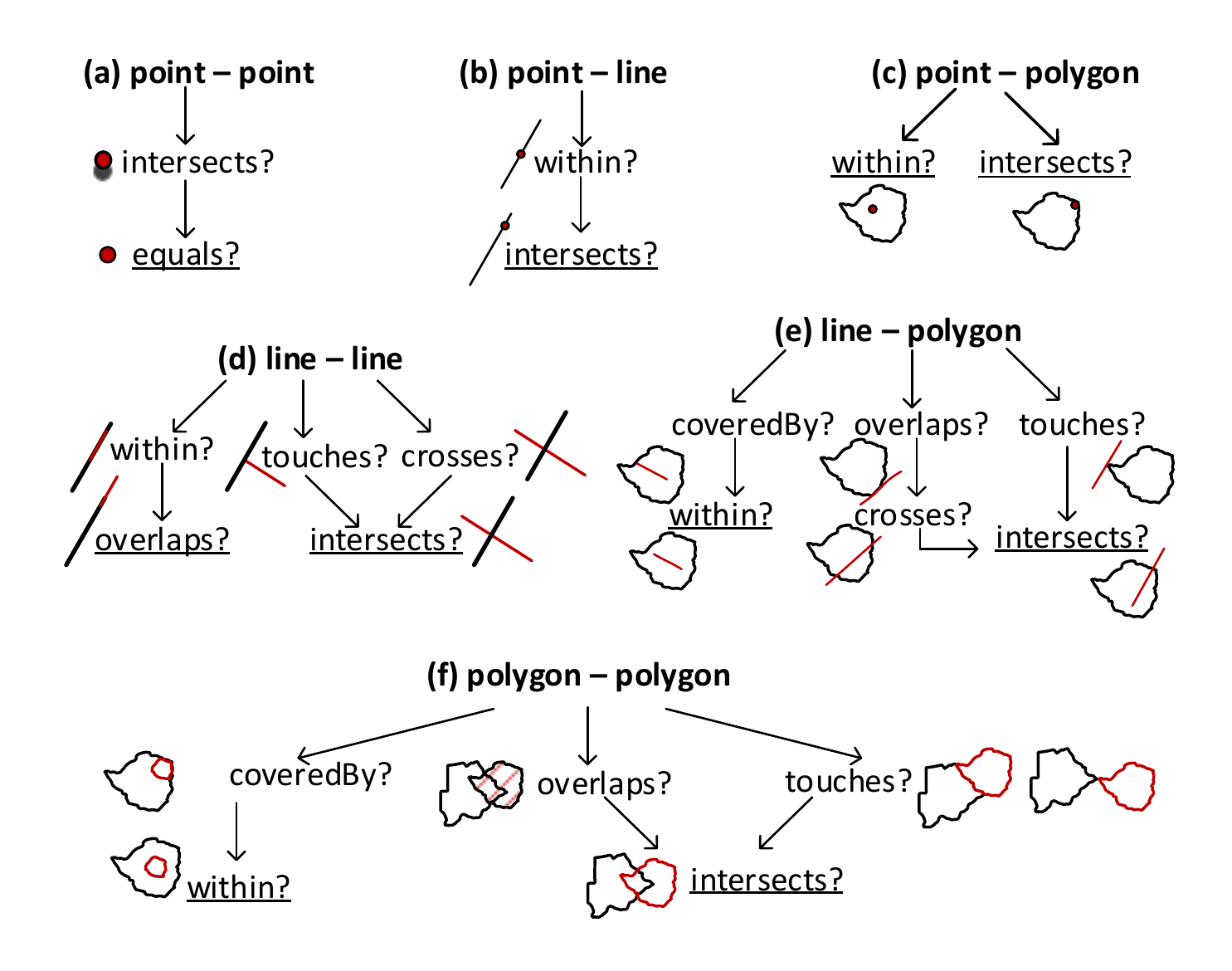}
	\vspace*{-2ex}
	\caption{Simplifying Topological Relations}%
	\label{fig:topoRel}%
	%	\vspace{-0.2in}
\end{figure}

In Figure~\ref{fig:topoRel}, we depict the hierarchically and topologically applicable topological relations from Table~\ref{tab:topoRel}. We simplified them by generalizing the possible relations, e.g., if a line \textit{touches} or \textit{crosses} another line at one point, they are both classified as \textit{intersects} in Fig.~\ref{fig:topoRel}(d). The most general relations are \underline{underlined} in Fig.~\ref{fig:topoRel} for each pair of geometry types (Fig.~\ref{fig:topoRel}(a), (b), (c), (d), (e), and (f)). 

In Algorithm~\ref{algo:relateSpatialValues} \texttt{relateSpatialValues}, we only consider these general topological relations that have a higher probability to satisfy the corresponding spatial predicates. For example, the topological relation \emph{intersects} has the highest probability to satisfy from the DE-9IM matrix~\cite{DE9DIM}. We generalize similar spatial predicates to ones that have higher probability to occur in a 2-dimensional space. For example, relations, such as a line \textit{overlaps} (along the border of) a polygon, can be generalized to the relation - a line \textit{crosses} a polygon at a minimum two points, which can later be generalized to the relation - a line \textit{intersects} a polygon at a (minimum) single point as in Figure~\ref{fig:topoRel}(e). Similarly, a line \textit{touches} a polygon at a single point can be generalized to the relation - a line \textit{intersects} a polygon at a (minimum) single point.
 
The topological relation \textit{coveredBy} requires an area of a geometry, therefore it is applicable only in line-polygon and polygon-polygon relations (Figure~\ref{fig:topoRel}(e) and \ref{fig:topoRel}(f)). For reasons of simplicity, we choose to generalize them as the \textit{within} topological relation. In the algorithm, we also prioritize to check the topological relations based on the compared geometry types. If the spatial attribute values to relate are \emph{point} and \emph{polygon} geometry types, as in Fig.~\ref{fig:topoRel}(c), it is more likely that a \emph{point} is \emph{within} a \emph{polygon} than a \emph{point} \emph{intersects} a \emph{polygon} in the instance data. 

\begin{algorithm2e}[t!]
	\DontPrintSemicolon
	\KwIn {$v_{a_c}, v_{a_p}$}
	\KwOut {\textnormal{\texttt{topoRel$_i$}}}
	\nl \Begin{
		\nl \texttt{topoRel}$_i = null$; /*$geoType(v_{a})$ function returns the geometry type of a given attribute value*/\\ 
		\nl \Switch{(geoType($v_{a_c}$), geoType($v_{a_p}$))}{ 
			\nl	\Case{\textnormal{(\texttt{POINT}}, \textnormal{\texttt{POINT})}}{
				\nl		\If{\textnormal{\texttt{equals?}}$(v_{a_c}, v_{a_p})$}{\nl \textnormal{\texttt{topoRel$_i$}}= \textnormal{\texttt{qb4so:equals}}}}
			\nl \Case{\textnormal{(\texttt{POINT}}, \textnormal{\texttt{LINE})}}{
				\nl	\If{\textnormal{\texttt{intersects?}}$(v_{a_c}, v_{a_p})$ }{\nl \textnormal{\texttt{topoRel$_i$}}= \textnormal{\texttt{qb4so:intersects}}}}	
			\nl \Case {\textnormal{(\texttt{POINT}}, \textnormal{\texttt{POLYGON})}}{
				\nl \If{\textnormal{\texttt{within?}}$(v_{a_c}, v_{a_p})$ }{\nl \textnormal{\texttt{topoRel$_i$}}= \textnormal{\texttt{qb4so:within}}}	
				\nl \ElseIf{\textnormal{\texttt{intersects?}}$(v_{a_c}, v_{a_p})$ }{\nl \textnormal{\texttt{topoRel$_i$}}= \textnormal{\texttt{qb4so:intersects}}}}	
			\nl \Case {\textnormal{(\texttt{LINE}}, \textnormal{\texttt{LINE})}}{
				\nl \If{\textnormal{\texttt{intersects?}}$(v_{a_c}, v_{a_p}) $ }{\nl \textnormal{\texttt{topoRel$_i$}}= \textnormal{\texttt{qb4so:intersects}}}	
				\nl \ElseIf{\textnormal{\texttt{overlaps?}}$(v_{a_c}, v_{a_p})$ }{\nl \textnormal{\texttt{topoRel$_i$}}= \textnormal{\texttt{qb4so:overlaps}}}}	
			\nl \Case {\textnormal{(\texttt{LINE}}, \textnormal{\texttt{POLYGON})}}{
				\nl \If{\textnormal{\texttt{within?}}$(v_{a_c}, v_{a_p})$ }{\nl \textnormal{\texttt{topoRel$_i$}}= \textnormal{\texttt{qb4so:within}}}	
				\nl \ElseIf{\textnormal{\texttt{intersects?}}$(v_{a_c}, v_{a_p})$ }{\nl \textnormal{\texttt{topoRel$_i$}}= \textnormal{\texttt{qb4so:intersects}}}}	
			\nl \Case {\textnormal{(\texttt{POLYGON}}, \textnormal{\texttt{POLYGON})}}{
				\nl \If{\textnormal{\texttt{within?}}$(v_{a_c}, v_{a_p})$ }{\nl \textnormal{\texttt{topoRel$_i$}}= \textnormal{\texttt{qb4so:within}}}	
				\nl \ElseIf{\textnormal{\texttt{intersects?}}$(v_{a_c}, v_{a_p})$ }{\nl \textnormal{\texttt{topoRel$_i$}}= \textnormal{\texttt{qb4so:intersects}}}}				
		} 
		\nl \KwRet{\textnormal{\texttt{topoRel$_i$}}}	
	}
	\caption{relateSpatialValues($v_{a_c}, v_{a_p}$):\texttt{topoRel$_i$}}\label{algo:relateSpatialValues}	
\end{algorithm2e}

Therefore, we initially check for a more probable relation in the algorithm. For example, for the point-polygon relations case in Algorithm~\ref{algo:relateSpatialValues}, Line~10: initially, the \emph{within} spatial predicate is checked in the if statement (Line~11), then the  \emph{intersects} spatial predicate is checked in the else if statement (Line~13). After checking all the possible combinations of spatial attribute values in a \texttt{switch case}, a topological relation is returned from the algorithm (Line~30). 

Now that we have introduced spatial helper functions, we present the main algorithms for finding the spatial hierarchy steps in the following.

\subsubsection{Detecting spatial hierarchy steps}\label{subsec:detectSpatial}
In this section, we present the algorithm for \emph{Case 1}, given in the beginning of Section~\ref{subsec:hierenrichment}, to find the \textit{explicit} spatial hierarchy steps for QB4OLAP levels with \texttt{skos:broader} roll-up relations between their child-parent level members.

\paragraph{Algorithm~\ref{algo:detectspatialHS} \textnormal{(\texttt{detectspatialHS})}.}

The input variables for Algorithm~\ref{algo:detectspatialHS} are the instance graphs of attributes of level members $\mathcal{G}^{I}_{A(lm)}$ and roll-up relations of the hierarchy steps $\mathcal{G}^{I}_{RU(hs)}$ between the level members.~The RDF graph formulation~of~the attributes of the level members $A(lm)$ is: $\mathcal{G}^{I}_{A(lm)} = \bigcup_{i=1}^p \{(id^{I}\!(lm) \ \ id^{S}\!(a_{i}) \ \ v_{a_{i}}) \mid lm\rightsquigarrow v_{a_{i}}\}$. Here, we denote by $lm \rightsquigarrow v_{a_{i}}$ that a level member $lm$ has value $v_{a_{i}}$ for attribute $a_i$ (e.g., Listing~2, Lines~3-6, Lines~9-13, and Lines~17-22). The RDF graph formulation of the roll-up relations $RU(hs)$ is: $\mathcal{G}^I_{RU(hs)} = \bigcup_{i=1}^k \{(id^{I}\!(lm_{c}) \ \texttt{skos:broader} \ id^{I}\!(lm_{p})) \mid lm_{c_{i}} \sqsubseteq lm_{p_{i}} \}$. Here, we denote by $lm_{c_{i}} \sqsubseteq lm_{p_{i}}$ the partial order between level members, where a child level member $lm_{c_{i}}$ rolls up to a parent level member $lm_{p_{i}}$\footnote{We use subscript \emph{c} and \emph{p} to distinguish values for child and parent level members.} (e.g., Listing~2, Line~7).

The output of Algorithm~\ref{algo:detectspatialHS} is the instance graph of roll-up relations for the \emph{detected} spatial hierarchy steps $\mathcal{G}^{I}_{RU(shs)}$ (e.g., Listing~2, Line~14). In Line~2, initially the output graph is initialized as an empty set. Next, in Line~3 we create two temporary graphs: $\mathcal{G}^{I}_{A(lm_c)}$ and $\mathcal{G}^{I}_{A(lm_p)}$ as empty sets\footnote{Remark: a set of RDF triples is referred to as an RDF graph}, to keep triple patterns separately in two graphs for attributes of child and parent level members. We also create two temporary sets: $V_{s(a_c)}$ and $V_{s(a_p)}$ for keeping the spatial attribute values from the child and parent level members, and initialize them as empty sets in Line~3. A set of spatial attribute values is defined over spatial literals $\mathcal{L}_s$ as $V_{s(a)} = \{v_{a_1}, \ldots, v_{a_i},\ldots v_{a_n} \mid 1\leq i \leq n \wedge v_{a_i} \in \mathcal{L}_s \}$. 

\begin{algorithm2e}[t]
	\DontPrintSemicolon
	\SetKwFunction{getSpatialValues}{getSpatialValues}	
	\SetKwFunction{relateSpatialValues}{relateSpatialValues}
	\KwIn{$\mathcal{G}^I_{A(lm)}$, $\mathcal{G}^I_{RU(hs)}$}
	\KwOut{$\mathcal{G}^I_{RU(shs)}$}
	\nl \Begin{
		\nl $\mathcal{G}^{I}_{RU(shs)} = \emptyset$; /*initialize output graph as emptyset*/\\
		\nl  $\mathcal{G}^{I}_{A(lm_{c})} = \emptyset$; $\mathcal{G}^{I}_{A(lm_{p})} = \emptyset$;  $V_{s(a_c)} = \emptyset$; $V_{s(a_p)} = \emptyset$; \texttt{topoRel}$_i = null$; \hfill /*temporary variable and sets*/\\	
		\nl \ForEach{$((id^{I}(lm_c) \ id^S(a_c) \ v_{a_c}), (id^{I}(lm_p) \ id^S(a_p) \ v_{a_p}))  \mid  (id^I(lm_c) \ id^S(a_c) \ v_{a_c}), (id^I(lm_p) \ id^S(a_p) \ v_{a_p}) \in \mathcal{G}^I_{A(lm)} \wedge  (id^{I}(lm_c) \ \textnormal{\texttt{skos:broader}} \ id^{I}(lm_p))  \in \mathcal{G}^I_{RU(hs)} \ \wedge$ 
			$lm_c \rightsquigarrow v_{a_c} \wedge \ lm_p \rightsquigarrow v_{a_p} \wedge lm_c \sqsubseteq lm_p$} 
		{
			\nl $\mathcal{G}^{I}_{A(lm_{c})} = \{(id^I(lm_c) \ id^S(a_c) \ v_{a_c})\}$; \\
			\nl $V_{s(a_c)} = $\getSpatialValues{$\mathcal{G}^I_{A(lm_c)}$};\\
			\nl \If{$V_{s(a_c)} \neq \emptyset$}{ 
				\nl $\mathcal{G}^{I}_{A(lm_{p})} = \{ (id^I(lm_p) \ id^S(a_p) \ v_{a_p}) \}$; \\
				\nl $V_{s(a_p)} = $ \getSpatialValues{$\mathcal{G}^I_{A(lm_p)}$}; \\
				\nl \If{$V_{s(a_p)} \neq \emptyset$}{
					\nl \ForEach{$(v_{a_{c}}, v_{a_{p}}) \in V_{s(a_c)} \times V_{s(a_p)}$}
					{\nl \texttt{topoRel$_i$} = \relateSpatialValues{$v_{a_c},v_{a_p}$}; \\
						\nl \If{\textnormal{\texttt{topoRel}}$_i \neq null$}{
							\nl $\mathcal{G}^{I}_{RU(shs)}\cup=\{(id^I(lm_c)$~\texttt{topoRel$_i$}~$id^I(lm_p))\}$;}}}}	
		}
		\nl \KwRet  $\mathcal{G}^{I}_{RU(shs)}$	
	}
	\caption{detectSpatialHS$(\mathcal{G}^{I}_{RU(hs)},\mathcal{G}^{I}_{A(lm)}):~\mathcal{G}^{I}_{RU(shs)}$} \label{algo:detectspatialHS}	
\end{algorithm2e}

In the \texttt{foreach} loop in Line~4, we go through the elements of the input graphs $\mathcal{G}^{I}_{A(lm)}$ and $\mathcal{G}^{I}_{RU(hs)}$ that are fulfilling a specific criteria, which is having an \emph{explicit} \texttt{skos:broader} relation between child and parent level members.  

In Line~5, while iterating through the \texttt{foreach} loop, we assign the set of triples of child level members and their attributes to the temporary graph $\mathcal{G}^{I}_{A(lm_c)}$. This temporary graph is given in Line~6 as an input to the helper function \texttt{getSpatialValues} (Algorithm~\ref{algo:getSpatialValues}), which finds the spatial attribute values from the given graph, and returns a set of spatial attribute values (i.e., $V_{s(a_c)}$) that are found in the input graph. The output of the helper function ($V_{s(a_c)}$) keeps the spatial attribute values of the child level member $id^I(lm_c)$. 

Next in Line~7, if $V_{s(a_c)}$ is not empty and has some spatial values of $id^I(lm_c)$, we populate the next temporary graph $\mathcal{G}^{I}_{A(lm_p)}$ with its parent level $id^I(lm_p)$ and attributes of the parent level in Line~8. 

Similar to Line~6, Line~9 calls the helper function \texttt{getSpatialValues} with the input graph $\mathcal{G}^{I}_{A(lm_p)}$ and the output of the function is assigned to the temporary set $V_{s(a_p)}$. If this set is also not empty (Line~10), we go through the pairs of values ($v_{a_c}, v_{a_p}$) of the child-parent level members (Line~11), which are selected from the temporary graphs $\mathcal{G}^{I}_{A(lm_c)}$ and $\mathcal{G}^{I}_{A(lm_p)}$.

In this loop, we call the next helper function \texttt{relateSpatialValues} (Algorithm~\ref{algo:relateSpatialValues}), where the input is the spatial value pairs. The output value of this function is the topological relation between the corresponding child and parent level members, and it is assigned to the initially created temporary variable \texttt{topoRel}$_i$ (Line~12). If this value is not null (checked in Line~13), \texttt{relateSpatialValues} function returns a topological relation (Line~12) that is satisfied as shown with a check-mark ($\checkmark$) from  Table~\ref{tab:topoRel}. 

Finally, the output graph for \emph{spatial} hierarchy steps $\mathcal{G}^{I}_{RU(shs)}$ is incrementally generated by adding the triple pattern with the topological relation (Line~14) and the output graph for the detected spatial hierarchy steps is returned (Line~15).

\begin{algorithm2e*}[htb]
	\DontPrintSemicolon
	\SetKwFunction{getSpatialValues}{getSpatialValues}	
	\SetKwFunction{relateSpatialValues}{relateSpatialValues}
	\SetKwFunction{countNo}{count}	
	\SetKwFunction{insertToList}{insertToList}
	\KwIn{$\mathcal{G}^S_D,$ $\mathcal{G}^S_{H(d)},$ $\mathcal{G}^S_{L(h)},$ $\mathcal{G}^I_{LM(l)},$ $\mathcal{G}^I_{A(lm)}$}
	\KwOut{$\mathcal{G}^I_{RU(shs)}$}
	\nl \Begin{
		\nl $\mathcal{G}^{I}_{RU(shs)} =\emptyset$;~\texttt{topoRel}$_i=null$ \hfill/*initialize the output graph as an empty set and a temporary variable as null*/\\
		\nl $V_{s(a_n)} =\emptyset$;~$V_{s(a_k)}=\emptyset$; \hfill/*initialize temporary sets as empty sets for keeping spatial attribute values*/ \\
		\nl $\mathcal{G}^{I}_{A(lm_n)}=\emptyset$; $\mathcal{G}^{I}_{A(lm_k)}=\emptyset$;\hfill/*initialize empty sets to keep triple patterns for attributes of level members*/\\	
		\nl \ForEach{$(id^{S}(d) \ \textnormal{\texttt{qb4o:hasHierarchy}} \ id^{S}(h))~\in~\mathcal{G}^{S}_D$  \hfill\textnormal{/*iterate through the dimensions*/}}
		{
			\nl	 \ForEach{$(id^{S}(h) \ \textnormal{\texttt{qb4o:inDimension}} \ id^{S}(d)) \in \mathcal{G}^{S}_H(d)$ \hfill\textnormal{/*iterate through the hierarchies*/}} 
			{
				\nl \ForEach{$ (id^{S}(h) \ \textnormal{\texttt{qb4o:hasLevel}} \ id^{S}(l)) \in \mathcal{G}^{S}_H(d)$ \hfill\textnormal{/*while iterating through the levels in the hiearchy*/}}
				{
					\nl \ForEach{$(id^S(l_i), id^S(l_j)) \in \mathcal{G}^S_{L(h)} \times\mathcal{G}^S_{L(h)}~\mid~id^S(l_i)~\neq~id^S(l_j) \wedge$ \hfill\textnormal{/*\ldots get level pairs} $(id^S(l_i), id^S(l_j))*/ \linebreak$ 		
						\nl $\bigcup_{lm \in LM(l)} ((id^{I}(lm) \ \textnormal{\texttt{qb4o:memberOf}} \ id^{S}(l_i)), (id^{I}(lm) \ \textnormal{\texttt{qb4o:memberOf}} \ id^{S}(l_j))) \in \mathcal{G}^I_{LM(l)}$ \hfill\textnormal{/*in each level pair, while iterating through their level members, get a pair of level members $(id^I(lm_n), id^I(lm_k)$, where each level member comes from different levels*/}}
				%foreach block begin
				{
									\nl \ForEach{$(id^I(lm_n), id^I(lm_k)) \in \mathcal{G}^I_{LM(l)} \times \mathcal{G}^I_{LM(l)} \mid id^I(lm_n) \neq id^I(lm_k) \wedge id^I(lm_n) \in \mathcal{G}^I_{LM(l_i)} \implies id^I(lm_k) \in \mathcal{G}^I_{LM(l_j)} \mid  \mathcal{G}^I_{LM(l_i)} \subset  \mathcal{G}^I_{LM(l)} \wedge  \mathcal{G}^I_{LM(l_j)} \subset  \mathcal{G}^I_{LM(l)} \wedge  \mathcal{G}^I_{LM(l_i)} \neq  \mathcal{G}^I_{LM(l_j)} $ \hfill\textnormal{/*iterate through the pairs of level members*/} } 
				{
					\nl \ForEach{$((id^I(lm_n) \ id^S(a_i) \ v_{a_i}),((id^I(lm_k) \ id^S(a_j) \ v_{a_j})) \in \mathcal{G}^I_{A(lm)} \times \mathcal{G}^I_{A(lm)} $ \hfill\textnormal{/*iterate through the pairs of level members' attributes*/}}
					{\nl $\mathcal{G}^{I}_{A(lm_n)}~=~\{(id^I(lm_n) \ id^S(a_i) \ v_{a_i})\}; ~\mathcal{G}^{I}_{A(lm_k)}~=~\{(id^I(lm_k) \ id^S(a_j) \ v_{a_j})\}; \linebreak  
						\nl V_{s(a_n)}~=~\getSpatialValues(\mathcal{G}^{I}_{A(lm_n)}); V_{s(a_k)} = \getSpatialValues(\mathcal{G}^{I}_{A(lm_k)});$ \linebreak
						\nl \If{$V_{s(a_n)} \neq \emptyset \wedge V_{s(a_k)} \neq \emptyset $ \hfill\textnormal{/*make sure there are spatial values in the temporary sets*/}}
						{\nl \ForEach {$(v_{a_i}, v_{a_j}) \in V_{s(a_n)} \times V_{s(a_k)} $} { \nl \texttt{topoRel}$_i$ = \relateSpatialValues{$v_{a_i}, v_{a_j}$}; \linebreak
								\nl \If {\textnormal{\texttt{topoRel}}$_i \neq$ null \hfill\textnormal{/*make sure there is a topological relation assigned to the variable*/}}{\nl $\mathcal{G}^{I}_{RU(shs)} \cup = \{(id^{I}(lm_{n}) \ \textnormal{\texttt{topoRel}}_i \ id^{I}(lm_{k}) ) \};$}		
						}}
				}}
				}
				}
					}}
					\nl  \KwRet  $\mathcal{G}^{I}_{RU(shs)}$	
				}
				\caption{discoverSpatialHS($\mathcal{G}^S_D,\mathcal{G}^S_{H(d)},\mathcal{G}^S_{L(h)},\mathcal{G}^I_{LM(l)},\mathcal{G}^I_{A(lm)}$): $\mathcal{G}^I_{RU(shs)}$}\label{algo:discoverspatialHS}
			\end{algorithm2e*}

\subsubsection{Discovering spatial hierarchy steps}\label{subsec:discoverSpatial}
In this section, we present the algorithm for \emph{Case 2}, given in the beginning of Section~\ref{subsec:hierenrichment}, to find the \textit{implicit} spatial hierarchy steps from QB4OLAP levels that do not have direct (\texttt{skos:broader}) roll-up relations. In this algorithm, we have to handle the situations where there are no explicit hierarchy steps between the level members. Therefore, we benefit from schema graphs of dimensions, hierarchies, and levels for iterating through the RDF triples and compare the spatial attribute values of the level members to find the topological relations within the same dimension. 
  
\paragraph{Algorithm~\ref{algo:discoverspatialHS} \textnormal{(\texttt{discoverSpatialHS})}.} 
 The input variables for Algorithm~\ref{algo:discoverspatialHS} are the schema graphs of dimensions $\mathcal{G}^S_{D}$, hierarchies of the dimensions $\mathcal{G}^S_{H(d)}$, levels of the hierarchies $\mathcal{G}^S_{L(h)}$, the instance graphs of level members of levels $\mathcal{G}^I_{LM(l)}$, and attributes of level members $\mathcal{G}^I_{A(lm)}$. Each dimension $d \in D$ has a set of hierarchies $H(d)$, which is shown in the RDF graph formulation for a dimension $d \in D$ as: $\mathcal{G}^S_{d} = \bigcup_{h\in H(d)} \{(id^S(d) \ \texttt{qb4o:hasHierarchy} \ id^S(h)) \}$. Each hierarchy $h \in H(d)$ belongs to a dimension $d$ and has a set of levels $L(h)$, which is shown in the RDF graph formulation for a hierarchy $h \in H(d)$ as: $\mathcal{G}^S_{h} = \{(id^S(h) \ \texttt{qb4o:inDimension} \ id^S(d)\} \cup \bigcup_{l\in L(h)} \{(id^S(h) \ \texttt{qb4o:hasLevel} \ id^S(l)) \}$. Each level $l$ has a set of level members $LM(l) = \{lm_1,\ldots,lm_y\}$, which is shown in the RDF graph formulation for a level member $lm \in LM(l)$ as:
 
  $\mathcal{G}^I_{lm}=\{(id^I(lm) \ \texttt{qb4o:memberOf} \ id^S(l)\} $. 
  
  \noindent Each level member $lm$ has a set of attributes $A(lm)$. The RDF graph formulation of attributes of level members $\mathcal{G}^I_{A(lm)}$ is already given in Section~\ref{subsec:detectSpatial}. In Listing~2, examples of a triple pattern for level members and attributes of level members are given in Lines~1-6, Lines~8-13 and Lines~15-22, without explicit roll-up relations (Line~7).

The output of Algorithm~\ref{algo:discoverspatialHS} is the instance graph of roll-up relations for the \emph{discovered} spatial hierarchy steps $\mathcal{G}^{I}_{RU(shs)}$ (e.g., Listing~2, Line~14). In Line~2, the output graph is initialized as an empty set. And a temporary variable (\texttt{topoRel}$_i$) for keeping the discovered topological relations is initialized as \emph{null}. In Line~4, we create two temporary graphs: $\mathcal{G}^{I}_{A(lm_n)}$ and $\mathcal{G}^{I}_{A(lm_k)}$ as empty sets similar to Algorithm~\ref{algo:detectspatialHS}.  
~We also create two temporary sets: $V_{s(a_n)}$ and $V_{s(a_k)}$ for storing spatial attribute values and initialize them as empty sets in Line~3.

To discover the spatial hierarchy steps, we need to get the attributes of all the level members from the instance graph ($\mathcal{G}^I_{A(lm)}$) and compare their spatial attribute values in pairs, where the pairs of level member attributes should be coming from two different levels in the same dimension hierarchy. Therefore, before getting the attributes of the level members, we need to classify the level members as they are grouped in different levels of a dimension hierarchy.

To achieve that, we use the schema definitions readily available in QB4OLAP, by looping through in Algorithm~\ref{algo:discoverspatialHS}, in nested loops of dimensions in Line~5,  hierarchies in the dimension (Line~6), levels in the hierarchy (Line~7). This helps us to determine the levels in a dimension hierarchy, where we can get level pairs from the same hierarchy (Line~8).

Now, while looping through the level pairs, we can identify the level members via the \texttt{qb4o:memberOf} property (Line~9). We get a pair of level members, where each level member should come from a different level, then we iterate through that pair of level members (Line~10).

Then, we get the triple patterns for the attributes of the level members from the each of the level member in the pair, and iterate through those pairs of the triple patterns (Line~11). While iterating through the triple patterns, we insert them to the temporary graphs $\mathcal{G}^I_{A(lm_n)}$ and $\mathcal{G}^I_{A(lm_k)}$ (Line~12), which are created earlier as empty sets in Line~4. So, we can filter the spatial values from the triple patterns kept in the temporary graphs by calling the helper function \texttt{getSpatialValues} (Algorithm~\ref{algo:getSpatialValues}), with those input graphs $\mathcal{G}^I_{A(lm_n)}$ and $\mathcal{G}^I_{A(lm_k)}$ (Line~13).

 Next, we call the helper function \texttt{getSpatialValues} (Algorithm~\ref{algo:getSpatialValues}) twice, with the input graphs $\mathcal{G}^I_{A(lm_n)}$ and $\mathcal{G}^I_{A(lm_k)}$. The outputs of the each (helper) function call are assigned to the temporary sets $V_{s(a_n)}$ and $V_{s(a_k)}$ correspondingly (Line~13). If these sets are not empty (Line~14), it means that \texttt{getSpatialValues} identified spatial values in the triple patterns of the input graphs.

 Then, we iterate through the spatial value pairs retrieved from the each of the sets (Line~15). In this loop, we call the next helper function \texttt{relateSpatialValues} (Algorithm~\ref{algo:relateSpatialValues}), where the input is the spatial value pairs. The output value of this function is the topological relation between the corresponding level members, and it is assigned to the initially created temporary variable \texttt{topoRel}$_i$ (Line~16). 
 
 Finally, if this \texttt{topoRel}$_i$ value is not null (Line~17), the output graph for the \emph{spatial} hierarchy steps $\mathcal{G}^I_{RU(shs)}$ is incrementally generated by adding the triple pattern with the topological relation (Line~18) and the output graph for the \emph{discovered} spatial hierarchy steps is returned in Line~19.

\subsection{Factual enrichment phase}
\label{subsec:factenrichment}
% !TeX spellcheck = en_US
% !TEX root = ../iosart2x.tex

The factual enrichment phase is built around the observation facts and their spatial attributes a.k.a spatial measures and fact-dimension relations (Section~\ref{subsec:presdw}). 

In QB4OLAP facts are linked to the dimensions at the lowest granularity level, which is the base level of the dimensions. For example, the GeoFarmHerdState cube has two spatial base levels linked to the cube: Parish level and Farm level. The GeoFarmHerdState cube also has a spatial measure listed in the cube: FarmLocation (Figure~\ref{fig:geodyrdw}). In QB4OLAP, a fact schema defines the structure of a cube with the \texttt{qb:DataStructureDefinition} property (Listing~3, Line~1).  Base levels (Lines~2 and~4) and measures (Line~6) are given as \texttt{qb:component}s of the fact (Listing~3). The cardinality relationship between the base level and the fact can also be represented with \texttt{qb4o:cardinality} in QB4OLAP as given in Lines~2 and~4 in Listing~3.

On the other hand, with QB4SOLAP we can also represent fact-level topological relations that are similar to the topological relations between the child-parent levels at the hierarchy steps. Fact-level topological relations are given in spatial fact schema with blue in Lines~3 and 5 (Listing~3). QB4SOLAP also extends the (cube) schema with spatial aggregate functions, which are defined over spatial measures as highlighted in blue (Listing~3, Line~7).

\begin{sftabbing}
	\setcounter{AlgoLine}{0}
	\LinesNumbered
	XX \= XX \= XX \= XX \= \kill 
	\textbf{	\#\#Spatial Fact Schema in QB4SOLAP\#\# }\\
	\nl gfs:GeoFarmHerdState a qb:DataStructureDefinition ;\\
	\textbf{\#Lowest spatial level for each dimension in the cube\#}\\
	\nl qb:component [qb4o:level gfs:farm ; qb4o:cardinality qb4o:ManyToOne ;\\ 
	\nl\> \textcolor{blue}{qb4so:topologicalRelation qb4so:Equals}] ;\\
	\nl qb:component [qb4o:level gfs:parish ; qb4o:cardinality qb4o:ManyToOne ;\\ 
	\nl\> \textcolor{blue}{qb4so:topologicalRelation qb4so:Within}] ;\\
	\textbf{ \#Example of a spatial measure in the cube\#}\\
	\nl qb:component [qb:measure gfs:farmLocation ; \\
	\nl\> qb4o:aggregateFunction \textcolor{blue}{qb4so:ConvexHull}] .
\end{sftabbing} 
\vspace{-0.05in}
\noindent\footnotesize{Listing 3: GeoFarmHerdState fact schema definition in QB4SOLAP}
\normalsize

\medskip
 
An example of an observation fact (fact member) at the instance level is given in Listing~4. A fact member is a \texttt{qb:Observation} (Line~1), which is related to the base levels (Line~2) with respect to the data structure definition (DSD) of the fact schema, and has a set of measures (Lines~3, 4) where some measures (Line~4) might have spatial values (Listing~4). To define a QB4OLAP fact schema, first, we need to enrich the fact members by annotating with topological relations as highlighted with blue in Line~5. We can derive topological relations between fact members and the (base) level members by comparing the spatial measures of the fact members and spatial attributes of the (base) level members with Boolean spatial predicates. The links between fact members and base level members are already given explicitly in Line~2 (Listing~4). However, these links are simple references between the fact and base level members, which do not describe the nature of the topological relation. By applying Boolean spatial predicates on fact and level members, we can find the exact topological relations, i.e., if a fact member \emph{intersects} with the level member or if a fact member is \emph{within} the level member. 
We explain how to detect these \emph{explicit} fact-level (topological) relations in Section~\ref{subsec:detectFactLevel}.

Moreover, there might also be some missing links between the (observartions) fact members and the corresponding base level members. For this case we need to find all the base level members that are spatial and derive the links between the spatial measure values and spatial attribute values (of the base level members) by using Boolean spatial predicates. We explain how to discover fact-level (topological) relations, which are not explicitly linked between observation fact and base level members in Section~\ref{subsec:discoverFactLevel}.

There are also cases where we would like to establish a direct (topological) relation between the fact members and higher granularity (parent) level members, which are not at the base level of the dimension. Using the example depicted in Figure~\ref{fig:exlevels} we explained that wrongly aggregating the measures (i.e., double counting) becomes a problem when we roll-up between the levels that have many-to-many (N:M) cardinality relations (as in Parish and Drainage Area levels). Therefore, it is necessary to drill-down to the lowest granularity (fact members) and find the direct relation between the observation fact members and the corresponding level members of the higher level in many-to-many cardinality relations.

In order to prevent this problem, we address the issue in our algorithm to discover and annotate the fact-level (topological) relations that are between the observation fact members and level members of a higher level in an N:M cardinality relation in Section~\ref{subsec:discoverFactLevel}. For example, such a relation is given in green in Line~6 (Listing~4) that shows a topological relation between an observation fact member (farm state) and a higher level -- not a base level -- member (drainage area).

\begin{sftabbing} %[htb]
	\setcounter{AlgoLine}{0}
	\LinesNumbered
	XX \= XX \= XX \= XX \= \kill 
	\textbf{	\#\#GeoFarmHerdState cube: observation fact example\#\# }\\
	\nl gfsi:farmState\_103850\_12\_2015 a qb:Observation ;\\
	\nl\> gfs:farm gfsi:farm\_103850 ; gfs:parish gfsi:parish\_8648 ; \\
	\nl\> gfs:livestockUnit "4.2699999999999996"\textasciicircum\textasciicircum xsd:double ;\\
	\nl\> gfs:farmLocation "POINT (8.31941 56.75822)"\textasciicircum\textasciicircum geo:spatialLiteral ;  \\
	\nl\> \textcolor{blue}{qb4so:equals gfsi:farm\_103850 ; qb4so:within gfsi:parish\_8648 ;} \\
	\nl\> \textcolor{teal}{qb4so:within gfsi:water\_3770 .} \\	
\end{sftabbing} 
\vspace{-0.03in}
\noindent\footnotesize{Listing 4: GeoFarmHerdState fact member with base levels and measures}
\normalsize

\medskip

Finally, in Section~\ref{subsec:defineSpatialFdsd} we explain how to define a data structure definition (DSD) of spatial fact schema using a QB4OLAP fact schema and the spatial fact member instances derived in the previous two algorithms.  

\subsubsection{Detecting explicit fact-level relations}\label{subsec:detectFactLevel}

In this section, we present an algorithm for detecting explicit fact-level topological relations between observation fact members and base level members where there is a direct reference between the fact member and the base level member. To derive these topological relations we need to get the spatial attributes of fact members (spatial measures) and base level members.

\paragraph{Algorithm~\ref{algo:detectFactLevel} \textnormal{(\texttt{detectFactLevelRelations})}.}

The input variables for Algorithm~\ref{algo:detectFactLevel} are the instance graphs of fact members $\mathcal{G}^{I}_{FM(F)}$, level members $\mathcal{G}^{I}_{LM(l)}$, and attributes of level members $\mathcal{G}^{I}_{A(lm)}$. 

Every fact member $f_i \in FM$ has an IRI $id^I(f_i)$ and defined as a \texttt{qb:Observation}. The RDF graph formulation of a fact member $f_i$ is: 

$\mathcal{G}^{I}_{f_i} = \bigcup_{l_j \in L(f_i)} \{(id^{I}\!(f_i) \  id^{S}\!(l_{j})  \ id^{I}\!(lm_{j}) \mid f_i\rightsquigarrow lm_j\} \cup \bigcup_{m_k \in M(f_i)} \{(id^{I}\!(f_i) \ \ id^{S}\!(m_{k}) \ \ v_{m_k} \mid f_i\rightsquigarrow v_{m_k}\}$. 

\noindent Here, we denote by $f_i\rightsquigarrow lm_j$ that a fact member $f_i$ has an explicit link to a level member $lm_j$ (e.g., Listing~4, Line~3). Note that we denote by $lm \rightsquigarrow v_{a_{i}}$ that a level member $lm$ has value $v_{a_{i}}$ for attribute $a_i$ (Section~\ref{subsec:detectSpatial}), which is used in Algorithm~\ref{algo:detectFactLevel} (Line~12) to get the attribute values of the linked level members. Moreover, we denote here by  $f_i \rightsquigarrow v_{m_{k}}$ that a fact member $lm$ has value $v_{m_{k}}$ for measure $m_k$ (e.g., Listing~4, Lines 5 and 6). The RDF graph formulation of the other input variables are: attributes of level members $\mathcal{G}^{I}_{A(lm)}$ and level members $\mathcal{G}^{I}_{LM(l)}$ are already given, respectively, in Sections~\ref{subsec:detectSpatial} and~\ref{subsec:discoverSpatial}.  

The output of Algorithm~\ref{algo:detectFactLevel} is the enriched instance graph of fact members with topological relations $\mathcal{G}^{I}_{FM(F_s)}$. 
In Line~2, we initialize the output graph as the input graph of fact members (without topological relations) so that we can gradually enrich it with the detected topological relations (Line~22). Initially, the topological relation variable \texttt{topoRel}$_i$ is set to \emph{null}. We also create two temporary graphs: $\mathcal{G}^{I}_{A(lm_j)}$ and $\mathcal{G}^{I}_{A(f_{i}m_{k})}$  as empty sets to keep triple patterns separately in two graphs for attributes of level members and (measures of) fact members. We also create two temporary sets: $V_{s(m_k)}$ and $V_{s(a_i)}$ for keeping the spatial values from the fact and level members, and initialize them also as empty sets in Line~3.

\begin{algorithm2e}[h!]
\DontPrintSemicolon
\SetKwFunction{getSpatialValues}{getSpatialValues}	
\SetKwFunction{relateSpatialValues}{relateSpatialValues}
\KwIn{$\mathcal{G}^I_{FM(F)}$, $\mathcal{G}^I_{A(lm)}$}
\KwOut{$\mathcal{G}^I_{FM(F_s)}$}
\nl\Begin{
\nl $\mathcal{G}^{I}_{FM(F_s)} = \mathcal{G}^I_{FM(F)}$;  \texttt{topoRel}$_i = null$; 
$\mathcal{G}^{I}_{A(f_{i}m_{k})} = \emptyset$; \\
\nl $\mathcal{G}^{I}_{A(lm_{j})} = \emptyset$;  $V_{s(m_k)} = \emptyset$; $V_{s(a_i)} = 
\emptyset$;  \hfill /*initialize the ouput graph, temporary variable and sets*/ 	

\nl \ForEach{\textnormal{\hfill /*get each observation fact (fact member)*/} \linebreak
	\nl $(id^{I}(f_i) \ \textnormal{\texttt{rdf:type qb:Observation}})\in\mathcal{G}^{I}_{FM(F)}$}
 {
	 \nl \ForEach{\hfill \textnormal{/*get measure-level member pairs*/} 
	 \nl $((id^I(f_i) \ id^S(m_k) \ v_{m_k}), (id^I(f_i) \ id^S(l_j) \ id^I(lm_j)))$ \linebreak
	 \nl $\in \mathcal{G}^{I}_{FM(F)} \times \mathcal{G}^{I}_{FM(F)} \mid f_i \rightsquigarrow v_{m_k} \wedge  lm_j \rightsquigarrow v_{a_i} \wedge $ \linebreak 
	 \nl $(id^I(lm_j) \ id^S(a_i) \ v_{a_i}) \in \mathcal{G}^{I}_{A(lm)}$ %\linebreak 
	 \textnormal{/*get measure and attribute values of level members*/} }{ 
	 \nl $\mathcal{G}^{I}_{A(f_{i}m_{k})} = \{ (id^I(f_i) \ id^S(m_k) \ v_{m_k}) \}$; \linebreak 
	 \nl $V_{s(m_k)} = $ \getSpatialValues{$\mathcal{G}^I_{A(f_{i}m_{k})}$};
	 \linebreak
	 \nl \If{$V_{s(m_k)} \neq \emptyset $}{\nl $\mathcal{G}^{I}_{A(lm_{j})} = \{ (id^I(lm_j) \ id^S(a_i) \ v_{a_i}) \}$; \linebreak
	 	\nl $V_{s(a_i)} = $ \getSpatialValues{$\mathcal{G}^I_{A(lm_{j})}$}; \linebreak
		\nl \If {$V_{s(a_i)} \neq \emptyset$}{\nl \ForEach {$(v_{m_k},v_{a_i}) \in V_{s(m_k)} \times V_{s(a_i)}$  \textnormal{/*foreach spatial value pairs*/} } {\nl \texttt{topoRel}$_i$ = \relateSpatialValues{v$_{m_k}$,v$_{a_i}$}; \linebreak \nl \If{\textnormal{\texttt{topoRel}}$_i \neq null$}
				{\nl $\mathcal{G}^{I}_{FM(F_s)}\cup=
					\{(id^I(f_i)$~\texttt{topoRel$_i$}~$id^I(lm_j))\}$;  } }}}	 
	 
	 }
	}
	\nl \KwRet  $\mathcal{G}^{I}_{FM(F_s)}$	 	
}	
\caption{detectFactLevelRelations$(\mathcal{G}^{I}_{FM(F)},  \mathcal{G}^{I}_{A(lm)}):\mathcal{G}^{I}_{FM(F_s)}$} \label{algo:detectFactLevel}		
\end{algorithm2e}

In the first \texttt{foreach} loop (Line~4 and 5) we retrieve the observation fact members from the input graph of fact members, which corresponds to Line~1 in Listing~4. Getting the fact members allows us to access each of their measures in Line~6 and level members in Line~7 (Algorithm~\ref{algo:detectFactLevel}). In the next \texttt{foreach} loop (Line~9) we match each measure-level member pair, where we can already retrieve the measure values from the input graph of fact members $\mathcal{G}^I_{FM(F)}$ (Line~10) and through the input graph for attributes of the level members $\mathcal{G}^I_{A(lm)}$ (Line~11 and~12), we can retrieve the attribute values. In Line~13, we assign the set of triples for measure attributes of fact members to a temporary graph $\mathcal{G}^{I}_{A(f_{i}m_{k})}$ created earlier in Line~2. This temporary graph is given as an input to the helper function \texttt{getSpatialValues} (Algorithm~\ref{algo:getSpatialValues}) in Line~14 (Algorithm~\ref{algo:detectFactLevel}). The helper function returns the spatial attribute (measure) values of the fact members, which are kept in the temporary set $V_{s(m_k)}$. If this set is not empty (checked in Line~15) and has some spatial measures of fact member $id^I(f_i)$, we repeat the same procedure for retrieving the spatial attribute values of level member $id^I(lm_j)$ in Lines~16 and~17. If the output set for spatial attribute values $V_{s(a_i)}$ is also not empty (Line~18), then we go through the pairs of spatial values $(v_{m_k}, v_{a_i})$ in Line~19. In this loop, we call the next helper function \texttt{relateSpatialValues} (Algorithm~\ref{algo:relateSpatialValues}), where the input is the spatial value pairs. The output value of this function is the topological relation between the corresponding fact and level members, which is assigned to the variable \texttt{topoRel}$_i$ (Line~20).

\begin{algorithm2e*}[htb]
	\DontPrintSemicolon
	\SetKwFunction{getSpatialValues}{getSpatialValues}	
	\SetKwFunction{relateSpatialValues}{relateSpatialValues}
	\KwIn{$\mathcal{G}^I_{FM(F)}$, $\mathcal{G}^I_{LM(l)}$, $\mathcal{G}^I_{A(lm)}, \mathcal{G}^{S}_{D}, \mathcal{G}^{S}_{H(d)},  \mathcal{G}^{S}_{HS(h)}$}
	\KwOut{$\mathcal{G}^I_{FM(F_s)}$}
	\nl\Begin
	{
		\nl $\mathcal{G}^{I}_{FM(F_s)} = \mathcal{G}^I_{FM(F)}$;  \texttt{topoRel}$_i = null$; \hfill /*initialize the output graph and temporary variable*/ 	\linebreak 
		\nl $\mathcal{G}^{I}_{A(f_{i}m_{k})} = \emptyset$; $\mathcal{G}^{I}_{A(lm_{j})} = \emptyset$;  $V_{s(m_k)} = \emptyset$; $V_{s(a_i)} = 
		\emptyset$;  \hfill /*initialize temporary graphs and sets as empty set*/ \\
		\nl \ForEach {$(id^{S}(d) \ \textnormal{\texttt{qb4o:hasHierarchy}} \ id^{S}(h))~\in~\mathcal{G}^{S}_D$  \hfill\textnormal{/*iterate through the dimensions*/}}
		{\nl \ForEach{$(id^{S}(h) \ \textnormal{\texttt{qb4o:inDimension}} \ id^{S}(d)) \in \mathcal{G}^{S}_H(d)$ \hfill\textnormal{/*iterate through the hierarchies*/}}
		{\nl \ForEach{$(id^{S}(h) \ \textnormal{\texttt{qb4o:hasLevel}} \ id^{S}(l_n)) \in \mathcal{G}^{S}_H(d)$ \hfill\textnormal{/*iterate through the levels in the hierarchy*/}}
			{\nl \ForEach {$(\textnormal{\texttt{\_:hs}}_i \ \textnormal{\texttt{qb4o:inHierarchy}} \ id^S(h)) \in  \mathcal{G}^{S}_{HS(h)} \mid (\textnormal{\texttt{\_:hs}}_i \ \textnormal{\texttt{qb4o:childLevel}} \ id^S(l_c)) \in  \mathcal{G}^{S}_{HS(h)} \wedge (\textnormal{\texttt{\_:hs}}_i \ \textnormal{\texttt{qb4o:parentLevel}} \ id^S(l_p)) \in  \mathcal{G}^{S}_{HS(h)} \wedge (\textnormal{\texttt{\_:hs}}_i \ \textnormal{\texttt{qb4o:pcCardinality}} \ id^S(card)) \in  \mathcal{G}^{S}_{HS(h)}$ \hfill\textnormal{/*each hierarchy step has a child level ($l_c$), a parent level ($l_p$), and a cardinality relation between these levels*/}} %\blankline
				{\nl \If{ $(id^S(l_n) \neq id^S(l_p)) \vee (id^S(l_n) = id^S(l_p) \wedge id^S(card) = \textnormal{\texttt{qb4o:ManyToMany}}) $ \hfill\textnormal{/*check in each hierarchy step that level $l_n$ should not be annotated as a parent level $l_p$, thus it is a base level OR if it is a parent level, there should be also a N:M cardinality realtion in the hierarchy step*/} }
					{\nl \ForEach {$(id^I(lm_j) \ \textnormal{\texttt{qb4o:memberOf}} \ id^S(l_n )) \in \mathcal{G}^{I}_{LM(l)}$ \hfill\textnormal{/*get level members of the level $l_n$*/}}
						{\nl \ForEach{$((id^I(lm_j) \ \textnormal{\texttt{qb4o:memberOf}} \ id^S(l_n)), (id^I(f_i) \ \textnormal{\texttt{rdf:type qb:Observation}})) \linebreak
							\nl	\in \mathcal{G}^{I}_{LM(l)} \times \mathcal{G}^{I}_{FM(F)} \mid \bigcup_{m_k \in M(f_i)} (id^I(f_i) \ id^S(m_k) \ v_{m_k}) \in \mathcal{G}^{I}_{FM(F)} \wedge \bigcup_{a_i \in A(lm)}$ \linebreak
							\nl $(id^I(lm_j) \ id^S(a_i) \ v_{a_i}) \in \mathcal{G}^{I}_{A(lm)} $ \hfill\textnormal{/*get level member-fact member pairs, where each fact member has some measure values $v_{m_k}$, and each level member has some attribute values $v_{a_i}$ */} 
							} 
							{\nl \ForEach {$((id^I(f_i) \ id^S(m_k) \ v_{m_k}), (id^I(lm_j) \ id^S(a_i) \ v_{a_i})) \in \mathcal{G}^{I}_{FM(F)} \times \mathcal{G}^{I}_{A(lm)} $} {\nl $\mathcal{G}^{I}_{A(f_{i}m_{k})} = \{(id^I(f_i) \ id^S(m_k) \ v_{m_k})\}; \ \mathcal{G}^{I}_{A(lm_{j})} = \{(id^I(lm_j) \ id^S(a_i) \ v_{a_i})\};$  \\  
									\nl $ V_{s(m_k)} =$ \getSpatialValues{$\mathcal{G}^{I}_{A(f_{i}m_{k})}$}; $ V_{s(a_i)} =$ \getSpatialValues{$\mathcal{G}^{I}_{A(lm_{j})}$}; \\
									\nl \If{$ V_{s(m_k)} \neq \emptyset \wedge V_{s(a_i)} \neq \emptyset $}{\nl \ForEach {$(v_{m_k}, v_{a_i}) \in V_{s(m_k)} \times V_{s(a_i)}$}{ \nl $\textnormal{\texttt{topoRel}}_i =$ \relateSpatialValues{$v_{m_k}, v_{a_i}$}; \\ 
											\nl \If{$\textnormal{\texttt{topoRel}}_i \neq null$}{\nl $\mathcal{G}^{I}_{FM(F_s)}\cup=
												\{(id^I(f_i)$~\texttt{topoRel$_i$}~$id^I(lm_j))\}$;} } } } }} } 
					
				} 
			}
		}	}

		\nl \KwRet  $\mathcal{G}^{I}_{FM(F_s)}$	 	
	}	
	\caption{discoverFactLevelRelations$(\mathcal{G}^{I}_{FM(F)},\mathcal{G}^{I}_{LM(l)},$  $\mathcal{G}^{I}_{A(lm)}, \mathcal{G}^{S}_{D}, \mathcal{G}^{S}_{H(d)}, \mathcal{G}^{S}_{HS(h)} ):\mathcal{G}^{I}_{FM(F_s)}$} \label{algo:discoverFactLevel}		
\end{algorithm2e*}

\begin{algorithm2e*}[htb]
	\DontPrintSemicolon
	\KwIn {$\mathcal{G}^{I}_{FM(F_s)},\mathcal{G}^{S}_{F}$}
	\KwOut {$\mathcal{G}^{S}_{F_s}$}
	\nl \Begin{
		\nl $\mathcal{G}^{S}_{F_s} = \mathcal{G}^{S}_{F}$; \texttt{aggFunc}$_i = null$; \hfill \textnormal{/*initalize the output graph and temporary variable*/} \\ 
		\nl \ForEach{$(id^I(f_i) \ \textnormal{\texttt{rdf:type qb:Observation}}) \in \mathcal{G}^{I}_{FM_(F_s)}$}
		{\nl \ForEach {$(id^I(f_i) \ \textnormal{\texttt{topoRel}}_i \ id^I(lm_j)) \in \mathcal{G}^{I}_{FM_(F_s)}\mid \bigcup_{l_n \in L(f_i)} (id^I(f_i) \ id^S(l_n) \ id^I(lm_j)) \in \mathcal{G}^{I}_{FM_(F_s)}    $ \hfill \textnormal{/*each \texttt{topoRel}$_i$ in the fact member triples goes into the DSD with its corresponding level $l_n$*/} }
			{\nl $\mathcal{G}^{S}_{F_(F_s)} \cup = \{ (id^S(F) \ \textnormal{\texttt{qb:component
						[qb4o:level}} \ id^S(l_n), \ \textnormal{\texttt{qb4so:topologicalRelation}} \ id^S(\textnormal{\texttt{topoRel}}_i) \textnormal{\texttt{]}})\}$; }
			\nl \ForEach{$v_{m_k} \in (id^I(f_i) \ id^S(m_k) \ v_{m_k})$ \hfill \textnormal{/*find the spatial measures from the fact triples*/}}
			{\nl \If{$v_{m_k}$ is a \textnormal{\texttt{geo:spatialLiteral}}} 
				{\nl \Switch{(geoType($v_{m_k}$)) \hfill \textnormal{/*$geoType(v_{a})$ function returns the geometry type of a given attribute value*/}}{
						\nl \Case {\textnormal{(\texttt{POINT})} \hfill \textnormal{/*point geometry measures are supported to be aggregated with ConvexHull function*/} }{ \nl $\textnormal{\texttt{aggFunc}}_i = \textnormal{\texttt{qb4so:ConvexHull}}$} 		
						\nl \Case {\textnormal{(\texttt{LINE})} \hfill \textnormal{/*line geometry measures are supported to be aggregated with Union function*/}}{ \nl $\textnormal{\texttt{aggFunc}}_i = \textnormal{\texttt{qb4so:Union}}$}
						\nl \Case {\textnormal{(\texttt{POLYGON})} \hfill \textnormal{/*polygon geometry measures are supported to be aggregated with Union, Centroid,*/}}{ \nl $\textnormal{\texttt{aggFunc}}_i = \textnormal{\texttt{qb4so:Union}} \vee \textnormal{\texttt{qb4so:Centroid}} \vee \textnormal{\texttt{qb4so:MBR}}$  \hfill \textnormal{/*or MBR functions*/}} 				 					
					}
					\nl $\mathcal{G}^{S}_{F_(F_s)} \cup = \{ (id^S(F) \ \textnormal{\texttt{qb:component
							[qb:measure}} \ id^S(m_k), \ \textnormal{\texttt{qb4o:aggregateFunction}} \ id^S(\textnormal{\texttt{aggFunc}}_i) \textnormal{\texttt{]}})\}$;
				}		
		}}
		\nl \KwRet{$\mathcal{G}^S_{F_s}$}	
	}
	\caption{defineSpatialFactDSD$(\mathcal{G}^{I}_{FM(F_s)},\mathcal{G}^{S}_{F} ):\mathcal{G}^{S}_{F_s}$}\label{algo:defineSDSD}	%\vspace{-0.1in}
\end{algorithm2e*}

\subsubsection{Discovering implicit fact-level relations}\label{subsec:discoverFactLevel}
In this section, we present an algorithm for discovering fact-level (topological) relations, where there are no direct links between the fact and level members. This algorithm handles the following situations: 1) Finding the topological relations between observation facts and base level members; 2) Finding the topological relations between observation facts and parent level members in an N:M cardinality relation. In both cases there are no direct links between the observation facts and level members. Therefore, we benefit from (QB4OLAP) schema graphs of dimensions, hierarchies, and levels for iterating through the RDF triples to distinguish the base level members, and find the parent level members, when there is an N:M cardinality relation between the levels of a hierarchy at a hierarchy step. 

\paragraph{Algorithm~\ref{algo:discoverFactLevel} \textnormal{(\texttt{discoverFactLevelRelations})}.}

The input variables at the schema level for Algorithm~\ref{algo:discoverFactLevel} are the schema graphs of dimensions $\mathcal{G}^S_{D}$, hierarchies of the dimensions $\mathcal{G}^S_{H(d)}$, levels of the hierarchies $\mathcal{G}^S_{L(h)}$, and hierarchy steps of the hierarchies $\mathcal{G}^S_{HS(h)}$. The RDF graph formulations of the schema level input variables (dimensions $\mathcal{G}^S_{H(d)}$, hierarchies $\mathcal{G}^S_{H(d)}$, and levels $\mathcal{G}^S_{L(h)}$) are already given in Section~\ref{subsec:discoverSpatial}. Therefore, we only explain the structure of a hierarchy step in the schema graph. 
Each hierarchy step $hs_i$ is defined in the schema graph $\mathcal{G}^S_{HS(h)}$ as a blank node $\_$\texttt{:hs}$_i \in \mathcal{B}$. Each hierarchy step is linked to a hierarchy $id^S(h)$ with the \texttt{qb4o:inHierarchy} predicate and has a child level $id^S(l_c)$, a parent level $id^S(l_p)$, and a cardinality relation $id^S(card)$, which are provided with \texttt{qb4o:childLevel}, \texttt{qb4o:parentLevel}, and \texttt{qb4o:pcCardinality} predicates in Line~6.

The input variables at the instance level are the instance graphs of fact members $\mathcal{G}^I_{FM(F)}$, level members of levels $\mathcal{G}^I_{LM(l)}$, and attributes of level members $\mathcal{G}^I_{A(lm)}$. We have already explained the RDF graph formulations of the instance level input variables (fact members $\mathcal{G}^I_{FM(F)}$, level members $\mathcal{G}^I_{LM(l)}$, and attributes of level members $\mathcal{G}^I_{A(lm)}$) in Section~\ref{subsec:detectFactLevel}. 

The output of Algorithm~\ref{algo:discoverFactLevel} is the enriched instance graph of fact members with the topological relations $\mathcal{G}^{I}_{FM(F_s)}$. 
In Line~2, we initialize the output graph as the input graph of fact members (without topological relations) so that we can gradually enrich it with the detected topological relations (Line~22). Initially, the topological relation variable \texttt{topoRel}$_i$ is set to \emph{null}. We also create two temporary graphs: $\mathcal{G}^{I}_{A(lm_j)}$ and $\mathcal{G}^{I}_{A(f_{i}m_{k})}$  as empty sets to keep triple patterns separately in two graphs for attributes of level members and (measures of) fact members. We also create two temporary sets: $V_{s(m_k)}$ and $V_{s(a_i)}$ for keeping the spatial values from the fact and level members and initialize them also as empty sets in Line~3. 

To find the topological relations between observation facts (with spatial measures) and base level members (with spatial attributes), first, we need to find all the base levels since there is no direct link between the fact and level members. To achieve this in Algorithm~\ref{algo:discoverFactLevel}, we use the schema definitions readily available in QB4OLAP. In Line~4, we iterate through the nested loops of dimensions to get the hierarchies and in Line~5 we iterate the nested loops of hierarchies to get the hierarchy levels. To find the base level of a hierarchy, we have to iterate through the hierarchy steps, where each hierarchy step describes a child level, a parent level and a cardinality relation between the levels (Line~6). If a level $id^S(l_n)$ has never been assigned as a parent level with \texttt{qb4o:parentLevel} predicate in any of the hierarchy steps in a hierarchy $h$ from the schema graph $\mathcal{G}^S_{HS(h)}$, then $l_n$ is the base level of a hierarchy $h$ (Line~7).

Thus, we can retrieve the level members of level $l_n$ from the instance graph level members $\mathcal{G}^I_{LM(l)}$ (Line~8). In the next \texttt{foreach} loop we can pair the level members from the instance graph $\mathcal{G}^I_{LM(l)}$, and observation facts from the instance graph of fact members $\mathcal{G}^I_{FM(F)}$ (Line~9). We can retrieve a set of attributes (measures) for fact members from the fact members graph (Line~10), and a set of attributes for level members from the instance graph $\mathcal{G}^I_{A(lm)}$ (Line~11). 

Then, in the next \texttt{foreach} loop in Line~12, we get the triple patterns with each measure values of the fact member and attribute values of the level member in pairs. While iterating through the (pair of) triple patterns, we insert each member of the pair to the temporary graphs for measures of fact members $\mathcal{G}^I_{A(f_{i}m_{k})}$ and attributes of level members $\mathcal{G}^I_{A(lm_j)}$ (Line~13), which are created earlier as empty sets in Line~3. Then, we can filter the spatial values from the triple patterns kept in the temporary graphs by calling the helper function \texttt{getSpatialValues} (Algorithm~\ref{algo:getSpatialValues}), with those input graphs $\mathcal{G}^I_{A(f_{i}m_{k})}$ and $\mathcal{G}^I_{A(lm_j)}$ (Line~14). We call the helper function \texttt{getSpatialValues} (Algorithm~\ref{algo:getSpatialValues}) twice, with the input graphs $\mathcal{G}^I_{A(f_{i}m_{k})}$ and $\mathcal{G}^I_{A(lm_j)}$, where the outputs of the each (helper) function call are assigned to the temporary sets $V_{s(m_k)}$ and $V_{s(a_i)}$ correspondingly (Line~14). If these sets are not empty (Line~15), it means that \texttt{getSpatialValues} identified spatial values in the triple patterns of the input graphs. 

Then, we iterate through the spatial value pairs retrieved from the each of the sets (Line~16). In this loop, we call the next helper function \texttt{relateSpatialValues} (Algorithm~\ref{algo:relateSpatialValues}), where the input is a spatial value pair. The output value of this function is the topological relation between the corresponding level members, and it is assigned to the initially created temporary variable \texttt{topoRel}$_i$ (Line~17). If this \texttt{topoRel}$_i$ value is not null (Line~18), the output graph for the spatial fact members is incrementally enriched by adding the triple pattern with the topological relation (Line~19).

To find the topological relations between the observation facts and parent level members in an N:M cardinality relation, we check in Line~20 that if level $id^S(l_n)$ is assigned as a parent level in a hierarchy step with \texttt{qb4o:parentLevel} predicate and the hierarchy step entails an N:M relation with \texttt{qb4o:ManyToMany} predicate. If that is the case, we repeat the same steps from Lines~8 to 19.

Finally, the output graph for the spatial fact members with \emph{discovered} fact-level (topological) relations is returned in Line~22.

\subsubsection{Defining spatial fact DSD}\label{subsec:defineSpatialFdsd}
In this section, we present an algorithm for re-defining the fact schema data structure definition (DSD) by enriching the DSD with fact-level topological relations. An example of a fact schema in QB4OLAP is given in the black-colored lines of Listing~3 (for now please ignore Lines~3, 5 and~7). We re-define the spatial fact schema to QB4SOLAP (Listing~3 Lines~1-7)  by using the enriched fact members that are generated via Algorithms~\ref{algo:detectFactLevel} and~\ref{algo:discoverFactLevel}.

\paragraph{Algorithm~\ref{algo:defineSDSD} \textnormal{(\texttt{defineSpatialFactDSD})}.}
The input variables for Algorithm~\ref{algo:defineSDSD} are the instance graph of spatial fact members $\mathcal{G}^{I}_{FM(F_s)}$ and schema graph of QB4OLAP fact schema $\mathcal{G}^{S}_{F}$. Spatial fact members in the instance graph $\mathcal{G}^{I}_{FM(F_s)}$ must be annotated with QB4SOLAP or can be generated by using Algorithms~\ref{algo:detectFactLevel} and~\ref{algo:discoverFactLevel} from QB4OLAP fact members. A QB4OLAP fact schema $\mathcal{G}^{S}_{F}$ has (base) levels and measures of the cube as  \texttt{qb:component}s  and defines the fact-level cardinality relation with \texttt{qb4o:cardinality} predicate, aggregate functions on (numerical) measures with \texttt{qb4o:aggregateFunction} predicate\footnote{In QB4OLAP, \texttt{qb4o:AggregateFunction} class has only instances (e.g., \texttt{qb4o:Avg, qb4o:Sum} functions) for numerical measures. QB4SOLAP extends this class with a subclass \texttt{qb4so:SpatialAggregateFunction}, which has instances of spatial aggregate functions (e.g., \texttt{qb4so:ConvexHull, qb4so:Union}) for spatial measures~\cite{nuref2015solap, nuref2017qb4solapSWJ}.}. 

The output of Algorithm~\ref{algo:defineSDSD} is the enriched fact schema graph $\mathcal{G}^{S}_{F}$ annotating the fact-level relations with QB4SOLAP topological relations and measures with spatial aggregate functions. 

In Line~2, we initialize the output graph as the input schema graph so that we can gradually enrich it with QB4SOLAP schema annotations (Lines~5 and ~15). Initially, an aggregate function variable \texttt{aggFunc}$_i$ is created and set to $null$ (Line~2).

The first \texttt{foreach} loop iterates through the fact members graph $\mathcal{G}^{I}_{FM(F_s)}$ and finds each fact member $f_i$ by using the triple pattern $(id^I(f_i) \ \texttt{rdf:type} \ \texttt{qb:Observation})$. The second \texttt{foreach} loop gets every \emph{distinct} topological relation \texttt{topoRel}$_i$ of the fact member $f_i$ (Line~4). Then the output schema is annotated with the identifier of these topological relations (Line~5). Next, we get every measure $v_{m_k}$ of the fact member $f_i$ (Line~6), and check if it is a spatial measure (Line~7). If it is a spatial measure, we find the geometry type with \emph{geoType} function (Line~8). We have appointed the corresponding spatial aggregate functions (Lines~~10, 12, and~14) with regard to the geometry type of the spatial measure (Lines~9, 11, and~13). Finally, the output schema $\mathcal{G}^{S}_{F_s}$ is annotated with the identifier of these spatial aggregate functions (Line~15) and returned (Line~16).

%-------------------------------------------------------------------------

\section{Implementation}
\label{sec:implementation}
% !TeX spellcheck = en_US
% !TEX root = ../iosart2x.tex

In this section, first we provide the details on how the algorithms from  Section~\ref{sec:rdf2solap} are implemented to generate spatially enriched RDF triples with QB4SOLAP (Sections~\ref{subsec:triplegen},~\ref{subsec:impDetectTopoRel},~\ref{subsec:impDiscoverTopoRel}, and\ref{subsec:impFactSchema}). Afterwards, we discuss our implementation choices in Section~\ref{subsec:choices} and present the results of applying the algorithms on the use case data (GeoFarmHerdState) in Section~\ref{sec:evaluation} (Table~\ref{tab:results}).
    
\subsection{QB4SOLAP triples generation}\label{subsec:triplegen}

To implement the algorithms from Section~\ref{sec:rdf2solap}, we have chosen a use case data set that can be annotated with multi-dimensional concepts in QB4OLAP and has the required spatial properties to be enriched as a fully spatial multidimensional cube with QB4SOLAP. The required spatial properties are: 1) Level members in a (spatial) hierarchy must have spatial attributes, where the geometry of the attributes should be different than only a simple \emph{point} geometry type, e.g., \emph{polygon, line}, etc. Thus we can implement the hierarchical enrichment (Section~\ref{subsec:hierenrichment}). 2) Fact members should have spatial measures, thus we can implement the factual enrichment (Section~\ref{subsec:factenrichment}).

Therefore, we have chosen GeoFarmHerdState as use case, which we have already used as running example throughout the paper. In Section~\ref{sec:background}, we discussed the spatial multi-dimensional concepts of the GeoFarmHerdState data cube and in Section~\ref{sec:rdf2solap} we provided RDF triple snippet examples of those concepts: \textit{(a)}~spatial hierarchy structure with QB4SOLAP (Listing~1), \textit{(b)}~level members annotated with QB4OLAP and with QB4SOLAP after hierarchical enrichment (Listing~2), \textit{(c)}~spatial fact schema (Listing~3), and \textit{\textit{(d)}}~spatial fact members with spatial measures (Listing~4). A full overview of the GeoFarmHerdState cube with spatial and non-spatial dimensions can be found in our previous work \cite{nuref2016solap} and on our project website \href{http://extbi.cs.aau.dk/GeoFarmHerdState/}{http://extbi.cs.aau.dk/GeoFarmHerdState/}.

Note that we use the non-spatial annotation of the GeoFarmHerdState data cube with QB4OLAP as an input to our algorithms, which is publicly available from our SPARQL endpoint\footnote{SPARQL Endpoint: \href{http://lod.cs.aau.dk:8890/sparql}{http://lod.cs.aau.dk:8890/sparql}} with corresponding namespaces for schema data triples\footnote{QB4OLAP schema: \href{http://extbi.cs.aau.dk/geofarm/qb4olap/farm-qb4olap-schema.ttl}{http://extbi.cs.aau.dk/geofarm/qb4olap/farm-qb4olap-schema.ttl}} and instance data triples\footnote{QB4OLAP instances: \href{http://extbi.cs.aau.dk/geofarm/qb4olap/farm-qb4olap-input.tar.gz}{http://extbi.cs.aau.dk/geofarm/qb4olap/farm-qb4olap-input.tar.gz}}.  

We query the endpoint and extract RDF data in JSON format as an input to our implementation of the four main enrichment algorithms; Algorithm~\ref{algo:detectspatialHS} - \texttt{detectSpatialHS}, Algorithm~\ref{algo:discoverspatialHS} - \texttt{discoverSpatialHS}, Algorithm~\ref{algo:detectFactLevel} - \texttt{detectFactLevel}, and Algorithm~\ref{algo:discoverFactLevel} - \texttt{discoverFactLevel}. 

In the following, we show the implementation highlights of each algorithm and helper function along with code snippets. 

\subsection{Detecting explicit topological relations} \label{subsec:impDetectTopoRel}

Detecting explicit topological relations are addressed in the following algorithms: Algorithm~\ref{algo:detectspatialHS} - \texttt{detectSpatialHS} and  Algorithm~\ref{algo:detectFactLevel} - \texttt{detectFactLevel}. In both cases the source data has explicitly defined roll-up relations, which means there is a direct relation between level members with \texttt{skos:broader} for hierarchy steps (e.g., Listing~2, Line~7) and there is a direct relation between a fact member and a base level member's foreign key URI (e.g., Listing~4, Line~2)

The input variables for Algorithm~\ref{algo:detectspatialHS} - \texttt{detectSpatialHS} are the triples with roll-up relations of the hierarchy steps ($\mathcal{G}^I_RU(hs)$) and the attributes of level members ($\mathcal{G}^I_A(lm)$) from the instance data graph. Explicit \texttt{skos:broader} relations are annotated in the instance graph of hierarchy steps ($\mathcal{G}^I_RU(hs)$). Therefore, we query the endpoint by filtering with the explicit \texttt{skos:broader} relations between all the level members. We fetch the results of the query in Node.js in JSON format.

The input variables for Algorithm~\ref{algo:detectFactLevel} - \texttt{detectFactLevel} are the triples with fact members  ($\mathcal{G}^I_FM(F)$) and the attributes of level members ($\mathcal{G}^I_A(lm)$) from the instance data graph. Explicit fact-level relations (by referring to the foreign key URI of level members) are annotated in the instance graph of fact members ($\mathcal{G}^I_FM(F)$). Therefore, we query the endpoint with all the fact members and the corresponding attributes of level members. We fetch the results of the query in Node.js in JSON format. 

Initially, we need to provide the explicit (roll-up) relations between the level members and fact-level members to implement Algorithms~\ref{algo:detectspatialHS} and~\ref{algo:detectFactLevel} for detecting the (explicit) topological relations. As mentioned above, we provide these relations from the data set by querying the endpoint and fetching the results of the query in Node.js in JSON format.

The next step is to retrieve the spatial attribute and measure values from the attributes of the level members and fact members.

\paragraph{Retrieving attribute and measure values.}

In this step, we retrieve the (spatial) attribute values and measure values of level members and fact members by accessing object (\emph{o}) of the each triple pattern  $t= (s, p, o) $ from the instance graphs of attributes of level members ($\mathcal{G}^I_A(lm)$) and fact members ($\mathcal{G}^I_FM(F)$) (Listing~5). This is followed by passing the \texttt{getLevelMemberAttributes} and \texttt{getMeasures} constants to \texttt{getSpatialValues} constant\footnote{We differentiate measure and level attribute values in seperate constants since a measure is annotated as \texttt{qb:MeasureProperty} and a level attribute is annotated as \texttt{qb4o:LevelAttribute} in the schema graph.} as explained below (\emph{filtering spatial values}) and given in Listing~6.  

\begin{tcolorbox}
\begin{sftabbinglist}
	\setcounter{AlgoLine}{0}
	\LinesNumbered
	XX \= XX \= XX \= XX \= \kill 
\nl const getLevelMemberAttributes = val => \\
\nl \> val.substring (val.indexOf("(") +1, \\
\nl \> val.indexOf(")"));\\
\nl const getMeasures = mval => \\
\nl \> mval.substring (val.indexOf("(") +1, \\
\nl \> mval.indexOf(")"));
\end{sftabbinglist}
\end{tcolorbox}
\noindent\footnotesize{Listing 5: Get level member attributes and fact member measures}

\normalsize

\paragraph{Filtering spatial values.} Before employing spatial analysis functions, we have to filter the spatial attributes of level members and spatial measures of fact members. 
Spatial values are always an object (\emph{o}) value in a triple pattern $t= (s, p, o) $, which is defined as spatial literals $\mathcal{L}_s$ (Section~\ref{sec:rdf2solap}). 
Therefore, we have retrieved the attribute and measure values as objects as mentioned above. 

We have shown the helper function Algorithm~\ref{algo:getSpatialValues} - \texttt{getSpatialValues}, which is used in the main algorithms. We have implemented this helper function on Node.js by filtering the WKT geometries from the input JSON data as exemplified in Listing~6. We create a \texttt{locationString} constant that takes a string value from \texttt{getLevelMemberAttributes} (Line~2). The string value is the last index location of a triple pattern constructed in \texttt{getLevelMemberAttributes}\footnote{Similarly, we create a second \texttt{locationString(2)} for spatial measure values that takes the string value from \texttt{getMeasures}, which is not repeated in Listing~6.}. 

\begin{tcolorbox}
\begin{sftabbinglist}
	\setcounter{AlgoLine}{0}
	\LinesNumbered
	XX \= XX \= XX \= XX \= \kill 
	\nl const getSpatialValues = value => \{ \\
	\nl	const locationString = \\
	\nl \> getLevelMemberAttributes (value); \\
	\nl	if (value.startsWith("POLYGON")) \{ \\
	\nl \>		const polygons = \\
	\nl \> \> generatePolygonPoints(locationString); \\
	\nl \>		return turf.polygon(coordinates:[polygons]); \}\\
	\nl	if (value.startsWith("LINE")) \{ \\
	\nl \>		const lines = locationString;\\
	\nl \>		return turf.lineString(coordinates:[lines]); \}\\
	\nl	if (value.startsWith("POINT"))\{ \\
	\nl \>		const points = locationString; \\
	\nl \>		return turf.point(coordinates:[points]); \}\\
	\nl	return null; \};
\end{sftabbinglist} 
\end{tcolorbox}
\noindent\footnotesize{Listing 6: Filtering spatial data types}

\normalsize

\paragraph{Finding topological relations.} Each of the four main enrichment algorithms (Algorithms~\ref{algo:detectspatialHS}, \ref{algo:discoverspatialHS}, \ref{algo:detectFactLevel}, and~\ref{algo:discoverFactLevel}) returns an instance graph of level members or fact members with topological relations annotated in QB4SOLAP. To find these topological relations we have introduced a helper function in Algorithm~\ref{algo:relateSpatialValues} - \texttt{relateSpatialValues}. 
This algorithm is implemented by using \emph{boolean} functions (spatial predicates) from the Turf.js library for relating spatial values and finding the appropriate topological relations. The library supports the following topological relations with corresponding predicates between certain spatial data types (Table~\ref{tab:turfFunctions}). A complete list of functions and details can be found online at \href{http://turfjs.org/docs}{http://turfjs.org/docs}.

\begin{table}[b]
	\caption{Turf.js Spatial Boolean Functions}
	\label{tab:turfFunctions}
	\resizebox{1.02\columnwidth}{0.3\columnwidth}{
		\begin{tabular}{|l|l|l|}
			\hline
			\multicolumn{1}{|c|}{\textbf{EQUALS}} & \multicolumn{1}{c|}{\textbf{WITHIN}} & \multicolumn{1}{c|}{\textbf{INTERSECTS}} \\ \hline
			\multirow{3}{*}{\begin{tabular}[c]{@{}l@{}}\#booleanEqual: (\texttt{equals}) \\ \textit{between} \\POINT-POINT\\ LINE-LINE\\ POLYGON-POLYGON\end{tabular}} & \begin{tabular}[c]{@{}l@{}}\#booleanWithin: (\texttt{within})\\ \textit{between}\\LINE-POLYGON\\ POLYGON-POLYGON\end{tabular} & \begin{tabular}[c]{@{}l@{}}\#booleanCrosses:\\ (\texttt{crosses}) \textit{between}\\ LINE-POLYGON\end{tabular} \\ \cline{2-3} 
			& \multirow{2}{*}{\begin{tabular}[c]{@{}l@{}}\#booleanPointInPolygon:\\    (\texttt{within}) \textit{between}\\ POINT-POLYGON\end{tabular}} & \begin{tabular}[c]{@{}l@{}}\#booleanOverlap:\\ (\texttt{overlaps}) \textit{between}\\ POLYGON-POLYGON\end{tabular} \\ \cline{3-3} 
			&  & \begin{tabular}[c]{@{}l@{}}\#booleanPointOnLine:\\ (\texttt{intersects}) \textit{between}\\ POINT-POLYGON\end{tabular} \\ \hline
	\end{tabular}}
\end{table}

We grouped the available Turf.js spatial boolean functions in Table~\ref{tab:turfFunctions} under three main topological relations (EQUALS, WITHIN, INTERSECTS), with respect to the simplification rules for grouping topological relations (Section~\ref{subsec:spatialhelper}) and explained along with Figure~\ref{fig:topoRel} and Table~\ref{tab:topoRel}. In Table~\ref{tab:turfFunctions}, Turf.js built-in functions (predicates) are shown with \#boolean prefix. In parentheses, we show how we have named them in our implementation by using the corresponding built-in functions.

Listing~7 provides an overview of the implementation of the boolean functions from Table~\ref{tab:turfFunctions} that are called in the main function for relating spatial values (relateSpatialValues) shown in Listing~8. We provide examples for each of the main topological relations (EQUALS, WITHIN, INTERSECTS). 
  
This \textit{first} spatial boolean function in Listing~7 is \texttt{equals} (Lines~1-8), which can be between any pair of the same spatial data type (Table~\ref{tab:turfFunctions}). We have grouped child level spatial (attribute) values and parent level spatial (attribute) values by their unique id (URI) for each spatial level attribute. This allows us to use \textit{javascript array prototype (instance) methods}, e.g., \texttt{every} or \texttt{some}, where we can create our own spatial predicate \texttt{equals} with condition to satisfy that \texttt{every} (grouped) child level attribute values should be equal to \texttt{every} (grouped) parent level attribute values. This ensures the multi-point, multi-line, and multi-polygon data types can be covered in our implementation.

\begin{tcolorbox}
\begin{sftabbinglist}
\setcounter{AlgoLine}{0}
\LinesNumbered
XX \= XX \= XX \= XX \= \kill 
// equals function \\
\nl const equals = (childLevelSpatialValues, \\
\nl parentLevelSpatialValues) => \\
\nl \> childLevelSpatialValues.every(\\
\nl \> childLevelSpatialValue =>\\
\nl \>\> parentLevelSpatialValues.every(\\
\nl \>\> parentLevelSpatialValue =>\\
\nl \>\>\ \ turf.booleanEqual(childLevelSpatialValue, \\
\nl \>\>\ \ parentLevelSpatialValue)));\\
// within function (POLYGON-POLYGON)\\
\nl const within = (childLevelSpatialValues, \\
\nl parentLevelSpatialValues) => \{\\
\nl \> const parentLevelMultipolygonBoundingBox \\
\nl \> = turf.bboxPolygon( \\
\nl \> \ turf.bbox(\\
\nl \> \ \ turf.multiPolygon(coordinates: [ \\
\nl \> \ \ \ parentLevelSpatialValues.map(\\
\nl \> \ \ \ parentLevelSpatialValue => \\
\nl \> \ \ \ \ pathOr([], [0], turf.getCoords(\\
\nl \> \ \ \ \ parentLevelSpatialValue)))])));\\
// all child level values are within the parent level\\
// polygon (simplified with bounding box)\\
\nl return childLevelSpatialValues.every(\\
\nl childLevelSpatialValue => \{\\
\nl \> return turf.booleanWithin(\\
\nl \> \ childLevelSpatialValue,\\
\nl \> \ parentLevelMultipolygonBoundingBox);\});\};\\
// crosses function (LINE-POLYGON)\\
\nl const crosses = (childLevelSpatialValues, \\
\nl parentLevelSpatialValues) => \\
\nl \> childLevelSpatialValues.some(\\
\nl \> childLevelSpatialValue =>\\
\nl \>\> parentLevelSpatialValues.some(\\
\nl \>\> parentLevelSpatialValue =>\\
\nl \>\>\ \ turf.booleanCrosses(childLevelSpatialValue, \\
\nl \>\>\ \ parentLevelSpatialValue)));
\end{sftabbinglist} 
\end{tcolorbox}
\noindent\footnotesize{Listing 7: Spatial Boolean Functions}
\normalsize
\medskip

%This \textit{first} spatial boolean function in Listing~7 is \texttt{equals} (Lines~1-8), which can be between any pair of the same spatial data type (Table~\ref{tab:turfFunctions}). We have grouped child level spatial (attribute) values and parent level spatial (attribute) values by their unique id (URI) for each spatial level attribute. This allows us to use \textit{javascript array prototype (instance) methods}, e.g., \texttt{every} or \texttt{some}, where we can create our own spatial predicate \texttt{equals} with condition to satisfy that \texttt{every} (grouped) child level attribute values should be equal to \texttt{every} (grouped) parent level attribute values. This ensures the multi-point, multi-line, and multi-polygon data types can be covered in our implementation.

For example, in the source data, we had multi-polygons for drainage areas, where each unique drainage areas is a multi-polygon that is composed of several polygons. To simplify we did not store multi-polygon data in RDF. Instead, we have annotated each unique drainage area as several polygons (of the multi-polygon), where each polygon of the drainage area is bound to its drainage area via unique id - URI of the drainage area. This means in the instance graph of parent level members $\mathcal{G}^I_{A(lm_p)}$ (drainage areas), there will be triple patterns $t= (s, p, o)$, where many different polygons - objects (\emph{o})  have the same subject (\emph{s}) - URI of a unique drainage area to represent the multi-polygon. 

To handle these multi-polygons, we gather them in a bounding box by using \texttt{turf.bboxPolygon} and \texttt{turf.bbox} functions in Listing~7 (Lines~13-14). In Listing~7 (Lines 10-18) depicts how several polygons of the same parent level can be put into a bounding box, which is passed as a parameter to our \textit{second} spatial boolean function \texttt{within}. Finally, the function returns in Lines~19-23 with condition to satisfy that \texttt{every} (grouped) child level attribute value should be within the simplified parent level polygon - \texttt{parentLevelMultipolygonBoundinxBox} (Line~23).

The \textit{third} spatial boolean function in Listing~7 is \texttt{crosses} (Lines~24-31), where we re-use the Turf.js spatial predicate \texttt{booleanCrosses}. This function is very similar to \texttt{overlaps} in implementation. The only difference is \texttt{crosses} occurs between LINE-POLYGON, \texttt{overlaps} occurs between POLYGON-POLYGON. For both cases, the condition to satisfy is that \texttt{some} of the (grouped) child level attribute values should cross/overlap \texttt{some} of the (grouped) parent level attribute values.

Listing~8) uses our own spatial predicates (explained above) to implement the helper function Algorithm~\ref{algo:relateSpatialValues} - \texttt{relateSpatialValues}. Note that we have followed the simplification rules for grouping topological relations (Figure~\ref{fig:topoRel}), aligned with switch cases for spatial data type pairs from Algorithm~\ref{algo:relateSpatialValues} in our implementation.

In our implementation illustrated in Listing~8, we create two functions \texttt{childLevelGeoType} (Line~3) and \texttt{parentLevelGeoType} (Line~6), which returns the geometry type of a given attribute value. This way we can implement \texttt{switch}($geoType(v_{a_c}), geoType(v_{a_p})$) \texttt{case}s from Algorithm~\ref{algo:relateSpatialValues} - \texttt{relateSpatialValues}.

\begin{tcolorbox}
\begin{sftabbinglist}
\setcounter{AlgoLine}{0}
\LinesNumbered
XX \= XX \= XX \= XX \= \kill 
\nl const relateSpatialValues = (childLevelSpatialValues,\\
\nl \> \ \ parentLevelSpatialValues) => \{ \\
\nl \> const childLevelGeoType = pathOr(\\
\nl \> \ \ null, [0, "geometry", "type"],\\
\nl \> \ \ childLevelSpatialValues); \\
\nl \>  const parentLevelGeoType = pathOr(\\
\nl \> \ \	null, [0, "geometry", "type"],\\
\nl \> \ \	parentLevelSpatialValues);\\
\nl \> if (childLevelGeoType === "Point" \&\& \\
\nl \>  parentLevelGeoType === "Point") \{\\
\nl \> \ \ if (equals(childLevelSpatialValues, \\
\nl \> \ \ parentLevelSpatialValues)) \{ \\
\nl \> \> return "qb4so:equals";\} \\
\nl \> \} else if (childLevelGeoType === "Point" \&\& \\
\nl \> \ \	parentLevelGeoType === "LineString") \{ \\
\nl \> \ \ if (intersects(childLevelSpatialValues, \\
\nl \> \ \ parentLevelSpatialValues)) \{ \\
\nl \> \> return "qb4so:intersects";\} \\
\nl \> \} else if (childLevelGeoType === "Point" \&\& \\
\nl \> \ \ parentLevelGeoType === "Polygon") \{ \\
\nl \> \ \ if (pointWithin(childLevelSpatialValues, \\
\nl \> \ \ parentLevelSpatialValues)) \{\\
\nl \> \> return "qb4so:within";\} \\
\nl \> \} else if (childLevelGeoType === "LineString" \\
\nl \> \ \ \&\& parentLevelGeoType === "LineString") \{ \\
\nl \> \ \ if (crosses(childLevelSpatialValues, \\
\nl \> \ \ parentLevelSpatialValues)) \{\\
\nl \> \> return "qb4so:intersects";\}\\
\nl \> \ \	if (overlaps(childLevelSpatialValues, \\
\nl \> \ \ parentLevelSpatialValues)) \{ \\
\nl \> \> return "qb4so:overlaps";\} \\
\nl \> \} else if (childLevelGeoType === "LineString" \\
\nl \> \ \ \&\&	parentLevelGeoType === "Polygon") \{ \\
\nl \> \ \	if (within(childLevelSpatialValues, \\
\nl \> \ \ parentLevelSpatialValues)) \{\\
\nl \> \> return "qb4so:within";\}\\
\nl \> \ \ if (crosses(childLevelSpatialValues, \\
\nl \> \ \ parentLevelSpatialValues)) \{\\
\nl \> \> return "qb4so:overlaps";\}\\
\nl \> \} else if (childLevelGeoType === "Polygon" \\
\nl \> \ \ \&\& parentLevelGeoType === "Polygon") \{ \\
\nl \>	\ const isWithin = within( \\
\nl \>	\ \ childLevelSpatialValues,\\
\nl \>	\ \  parentLevelSpatialValues);\\
\nl \>	\ const isOverlaps = overlaps(\\
\nl \>	\ \	childLevelSpatialValues, \\
\nl \>	\ \ parentLevelSpatialValues);\\
\nl \>  \ \	\ if (isWithin) \{ \\
\nl \> \> return "qb4so:within";\}\\
\nl \>  \ \	\ if (isOverlaps) \{\\
\nl \> \> return "qb4so:overlaps";\}\}\\
\nl return null;\};
\end{sftabbinglist} 
\end{tcolorbox}
\noindent\footnotesize{Listing 8: Relating spatial values}

\normalsize
\medskip

%In our implementation illustrated in Listing~8, we create two functions \texttt{childLevelGeoType} (Line~3) and \texttt{parentLevelGeoType} (Line~6), which returns the geometry type of a given attribute value. This way we can implement \texttt{switch}($geoType(v_{a_c}), geoType(v_{a_p})$) \texttt{case}s from Algorithm~\ref{algo:relateSpatialValues} - \texttt{relateSpatialValues}. 

\paragraph{Detecting topological relations.} 

Finally, we have implemented detecting topological relations algorithms (Algorithms~\ref{algo:detectspatialHS} and~\ref{algo:detectFactLevel}) with a bottom-up approach after implementing the core helper functions. In the following, we show the function implemented on Node.js for detecting topological relations (Listing~9) between level members, which is covered in Algorithm~\ref{algo:detectspatialHS}. The same approach with minor differences (in parameter passing) is used in our implementation for detecting topological relations between fact-level members (Algorithm~\ref{algo:detectFactLevel}). 

\begin{tcolorbox}
\begin{sftabbinglist}
\setcounter{AlgoLine}{0}
\LinesNumbered
XX \= XX \= XX \= XX \= \kill 
\nl const detectSpatialHierarchySteps = (\\
\nl \ \ parentLevelMembers,\\
\nl \ \ childLevelMembers,\\
\nl \ \ explicitRelations) => \{\\
\nl \  const spatialHierarchySteps = \\ 
\nl \ \ explicitRelations.results.bindings.map(\\
\nl \ \ binding => \{\\
\nl \> const childLevelMemberId = binding.s.value;\\
\nl \> const parentLevelMemberId = binding.o.value;\\
\nl \> const childLevelSpatialValues = pathOr([ ],\\
\nl \> \ \ [childLevelMemberId],childLevelMembers\\
\nl \> \ \ ).map(childLevelMember => \\
\nl \> \> utils.getSpatialValues(\\
\nl \> \> childLevelMember.value));\\
\nl \> const parentLevelSpatialValues = pathOr([],\\
\nl \> \ \ [parentLevelMemberId], parentLevelMembers\\
\nl \> \ \ ).map(parentLevelMember =>\\
\nl \> \> utils.getSpatialValues(\\
\nl \> \> parentLevelMember.value));\\
\nl \> const topoRel = utils.relateSpatialValues(\\
\nl \> \ \ childLevelSpatialValues,\\
\nl \> \ \ parentLevelSpatialValues);\\
\nl \> return \{\\
\nl \> \> ...binding,\\
\nl \> \>	p: \{type: "uri", value: topoRel ||\\
\nl \> \> "skos:broader"\}\};\\
\nl \});\\
\nl \> return \{\\
\nl \> \> ...explicitRelations,\\
\nl \> \> results: \{ ...explicitRelations.results,\\
\nl \> \> bindings: spatialHierarchySteps\}\\
\nl \> \};\\
\nl \};
\end{sftabbinglist} 
\end{tcolorbox}
\noindent\footnotesize{Listing 9: Detecting topological relations (between level members)}

\normalsize
\medskip

Listing~9 is constructed with the main function \texttt{detectSpatialHierarchySteps} with parameters of \texttt{parentLevelMembers}, \texttt{childLevelMembers}, and \texttt{explicitRelations}\footnote{We do not repeat a similar listing in the paper for detecting topological relations between fact-level members (Algorithm~\ref{algo:detectFactLevel}) where the parameter \texttt{childLevelMembers} from Listing~9 corresponds to fact members and \texttt{parentLevelMembers} corresponds to base level members in the implementation of detecting topological relations between fact-level members.}. In Line~5, the contant \texttt{spatialHierachySteps} takes the \texttt{explicitRelations} between child level and parent level members, and creates constants for those in Lines~8 and~9. The next step is to get the spatial values of the level members (child level members Lines~10-14 and parent level members Lines~15-19), where we utilize the helper function \texttt{getSpatialValues}, which is described in Listing~6. In Line~20, we create a constant \texttt{topoRel}, which takes the helper function \texttt{relateSpatialValues} (Listing~8) with two parameters \texttt{childLevelSpatialValues} and \texttt{parentLevelSpatialValues} that are created, in Lines~10 and~15, respectively. Next, we return the topological relations (\texttt{topoRel}) as predicates ($p$) between Lines~24-26. If a topological relation is not found, we keep the explicit relation as \texttt{skos:broader} (Line~26). Finally, we return the new results by replacing the \texttt{explicitRelations} with \texttt{spatialHierarchySteps} (Lines~28-32). 

We discuss our implementation in Section~\ref{subsec:qualeval}, Table~\ref{tab:results}, for both cases covered in Algorithms~\ref{algo:detectspatialHS} and~\ref{algo:detectFactLevel}, together with a number of input level members and fact members.

\subsection{Discovering implicit topological relations} 
\label{subsec:impDiscoverTopoRel}

Discovering implicit topological relations is addressed in the following algorithms: Algorithm~\ref{algo:discoverspatialHS} - \texttt{discoverSpatialHS} and Algorithm~\ref{algo:discoverFactLevel} - \texttt{discoverFactLevel}. In both cases the source data has not any defined roll-up relations (with \texttt{skos:broader}), or has missing spatial hierarchy steps between level members. Similarly, a fact level member has no defined relation link to any spatial level member of its dimensions.

The input variables for Algorithm~\ref{algo:discoverspatialHS} - \texttt{discoverSpatialHS}
are the triples with dimensions ($\mathcal{G}^S_D$), hierarchies in dimensions ($\mathcal{G}^S_H(d)$), levels in hierarchies ($\mathcal{G}^S_L(h)$) from the schema graph, and level members of levels ($\mathcal{G}^I_LM(l)$) and the attributes of level members ($\mathcal{G}^I_A(lm)$) from the instance data graph. Therefore, we query the endpoint by filtering with the schema elements \texttt{qb4o:hasHierarchy}, \texttt{qb4o:inDimension}, and \texttt{qb4o:hasLevel}. We fetch the results of the query in Node.js JSON format.

The input variables for Algorithm~\ref{algo:discoverFactLevel} - \texttt{discoverFactLevel} are 
the triples with dimensions ($\mathcal{G}^S_D$), hierarchies in dimensions ($\mathcal{G}^S_H(d)$), levels in hierarchies ($\mathcal{G}^S_L(h)$) from the schema graph, and fact members ($\mathcal{G}^I_FM(F)$), level members of levels ($\mathcal{G}^I_LM(l)$) and the attributes of level members ($\mathcal{G}^I_A(lm)$) from the instance data graph. Therefore, we query the endpoint by filtering with the schema elements \texttt{qb4o:hasHierarchy}, \texttt{qb4o:inDimension}, and \texttt{qb4o:hasLevel}. We fetch the results of the query in Node.js JSON format. 

The following listing (Listing~10) shows how we implement a schema wrapper by filtering the schema graph at our endpoint with predicates for schema elements (Lines~3, 7, 11, and 14). Once we get to the levels, we filter the level members in each level with \texttt{qb4o:memberOf} predicate (Line~11). Afterwards, we group level members by level that are in the same hierarchy and pass these grouped level members as inputs to a similar function as in Listing~9, which is called \texttt{detectSpatialHierarchyStepsExpensive}. This function takes only two parameters without explicit relations (two sets of level members grouped by level: \texttt{parentLevelMembers} and \texttt{childLevelMembers}). We run this algorithm several times for each pair of grouped level members (by level) within the same hierarchy as our approach is discovering implicit relations between level members and fact-level members. For fact members we similarly use one parameter (i.e., \texttt{parentLevelMembers}) as the  grouped level members (by level), and the other parameter is fact members (i.e., \texttt{childLevelMembers}), which are annotated as \texttt{qb:Observation}. In the \texttt{detectSpatialHierarchyStepsExpensive} function we utilize the same helper functions that are implemented with child-parent topological relations and simplification rules defined in Section~\ref{subsec:spatialhelper} along with Figure~\ref{fig:topoRel} and Table~\ref{tab:topoRel}. This ensures to apply spatial boolean predicates (on geometries of level members and fact members)  with \texttt{relateSpatialValues} helper function only between the appropriate spatial data types given in Tables~\ref{tab:topoRel} and~\ref{tab:turfFunctions}. Since there are no explicit relations in \texttt{detectSpatialHierarchyStepsExpensive} function, \texttt{relateSpatialValues} helper function is called $NumberOf_{childLevelMembers} \times NumberOf_{parentLevelMembers}$ in one iteration, where with \texttt{detectSpatialHierarchySteps} function, the helper function is called only $NumberOf_{explicitRelations}$ times. 

We discuss the implementation in Section~\ref{subsec:qualeval}, Table~\ref{tab:results}, for both cases covered in Algorithms~\ref{algo:discoverspatialHS} and~\ref{algo:discoverFactLevel}, together with a number of input level members and fact members.

\begin{tcolorbox}
	\begin{sftabbinglist}
		\setcounter{AlgoLine}{0}
		\LinesNumbered
		XX \= XX \= XX \= XX \= \kill 
		\nl const discoverSpatialHierarchySteps = schema =>\\
		\nl \> schema.results.bindings.filter(binding =>\\
		\nl \> \ \ binding.p.value ==="qb4o:hasHierarchy")\\
		\nl \> .map(hierachyBinding =>\\
		\nl \> \ \ schema.results.bindings.filter(binding =>\\
		\nl \> \ \ hierarchyBinding.o.value === binding.s.value \\
		\nl \> \ \ \&\& binding.p.value ==="qb4o:hasLevel"));\\
		\nl \> .map(levelBinding =>\\
		\nl \> \ \ schema.results.bindings.filter(binding =>\\
		\nl \> \ \ levelBinding.o.value === binding.s.value \\
		\nl \> \ \ \&\& binding.p.value ==="qb4o:memberOf"));\\		
		\nl const inDimension = \\
		\nl \> schema.results.bindings.filter(binding => \\
		\nl \> binding.p.value === "qb4o:inDimension");\\
		\nl module.exports =\{\\
		\nl \> wrapper: discoverSpatialHierarchySteps\};
	\end{sftabbinglist} 
\end{tcolorbox}
\noindent\footnotesize{Listing 10: Discovering topological relations (schema wrapper)}

\normalsize
\medskip

\subsection{Generating the fact schema} \label{subsec:impFactSchema}

Finally, we implement the enrichment of the fact schema based on spatially enriched fact instances (members). To extract the input variables for Algorithm~\ref{algo:defineSDSD} - \texttt{defineSpatialFactDSD}, we use the spatially enriched fact members (by Algorithms~\ref{algo:detectFactLevel} and \ref{algo:discoverFactLevel}) and non-spatial fact schema. 

The first step of generating the fact schema is to look for detected and discovered topological relations between the fact and level members and then annotate each of them with \texttt{qb4so:topologicalRelation} in the fact schema as given in Listing~3. The next step is to identify the spatial data types with helper functions \texttt{getMeasures} and \texttt{getSpatialValues} (Listings~5 and~6). Finally, for each of the identified spatial data types we annotate the fact schema with the corresponding spatial aggregate function, e.g., spatial data type POINT can have \textit{ConvexHull} aggregate function, LINE can have \textit{Union} etc.  

In our implementation of detecting and discovering topological relations between fact members and level members, we have only encountered the \texttt{qb4so:within} topological relation. Thus, the fact schema enrichment implementation generates Lines~4 and~5 as exemplified in Listing~3. As spatial measures in fact members, we have found the POINT spatial data type. Therefore, the fact schema enrichment implementation generates Lines~6 and~7, annotating that the spatial measure has \texttt{qb4so:ConvexHull} aggregate function, as exemplified in Listing~3.

After the spatial enrichment is fully completed, both schema\footnote{\href{http://extbi.cs.aau.dk/geofarm/qb4solap/geofarm-qb4solap-schema.ttl}{http://extbi.cs.aau.dk/geofarm/qb4solap/geofarm-qb4solap-schema.ttl}} and instance\footnote{\href{http://extbi.cs.aau.dk/geofarm/qb4solap/geofarm-qb4solap-output.tar.gz}{http://extbi.cs.aau.dk/geofarm/qb4solap/geofarm-qb4solap-output.tar.gz}} data has been published via the same SPARQL endpoint with QB4SOLAP.

Table~\ref{tab:results} shows the results of our implementation and we discuss them in detail in Section~\ref{subsec:qualeval}.

\subsection{Implementation choices} \label{subsec:choices}

After thoroughly describing the necessary steps and enrichment algorithms, we briefly present our implementation choices both in technical and strategical terms for implementing our approach. %the algorithms given in this section. materializing

To answer the question: "\textit{Can this approach be reasonably implemented on top of triple stores by directly using Web and Semantic Web technologies?}", we have come across a number of challenges, where specific choices had to be made. These will be discussed next. 

We chose to store RDF data in a well-established triple store (Virtuoso Open Source) that supports many geometry data types (i.e., POLYGON, MULTIPOLYGON). Even though Virtuoso supports several shape types (e.g., POLYGON, MULTIPOLYGON, etc.), it has a limited number of spatial Boolean functions available as built-in functions from the DE9DIM model from Table~\ref{tab:topoRel}. Therefore, we have also decided to use a third party \textit{Javascript 
library} for spatial analysis, which is called \textit{Turfjs}. This way, we can ensure that RDF2SOLAP can be used on top of
any triple store since the Javascript library provides us with the spatial analysis
capabilities and a flexible development environment, independent from the 
choice of the triple store.

It is mentioned earlier in Section~\ref{subsec:impDetectTopoRel} that we have multi-part POLYGON data (for drainage areas and parishes), which means that, when several polygons are grouped by unique (parish or water) URI they can compose a MULTIPOLYGON for a single parish or drainage area instance. From the implementation point of view, we had to implement a bounding box function for multi-part POLYGON data, in order to call the spatial Boolean functions (within and intersects) between the correct parish and drainage area instances, then annotate the topological relations between their unique URIs. If triple stores already provided overall support of complex spatial data types, spatial indices, and a complete support of built-in spatial functions, decoupling the triple stores during development of RDF2SOLAP would not have been necessary. We could then directly use the spatial capabilities of the triple stores that were required for developing RDF2SOLAP. However, to the best of our knowledge, a third party spatial analysis library was needed to fully implement our RDF2SOLAP (spatial) multi-dimensional enrichment algorithms given in Section~\ref{sec:rdf2solap}.

The details of our approach, endpoints, and data sets can be found on our project page\footnote{Project Page: \href{http://extbi.cs.aau.dk/RDF2SOLAP}{http://extbi.cs.aau.dk/RDF2SOLAP}}.
The code repository for the whole implementation can be found on GitHub\footnote{RDF2SOLAP Repository: \href{https://github.com/lopno/rdf2solap}{https://github.com/lopno/rdf2solap}}.

%-------------------------------------------------------------------------

\section{Experimental Evaluation}
\label{sec:evaluation}
% !TeX spellcheck = en_US
% !TEX root = ../iosart2x.tex

\begin{table*}[htb]
	\caption{Implementation Results of Detecting and Discovering Topological Relations}
	\label{tab:results}
	\begin{tabular}{ll|l|l|l|l|l|c|}
		\cline{3-8}
		&  & \multicolumn{3}{c|}{\multirow{2}{*}{INPUT}} & \multicolumn{3}{c|}{\multirow{2}{*}{OUTPUT}} \\
		&  & \multicolumn{3}{c|}{} & \multicolumn{3}{c|}{} \\ \cline{3-8} 
		& \multicolumn{1}{c|}{} & \multicolumn{1}{c|}{\begin{tabular}[c]{@{}c@{}}NumberOf\\ Child Members\end{tabular}} & \multicolumn{1}{c|}{\begin{tabular}[c]{@{}c@{}}NumberOf\\ Parent Members\end{tabular}} & \multicolumn{1}{c|}{\begin{tabular}[c]{@{}c@{}}NumberOf\\ Explicit Relations\end{tabular}} & \multicolumn{2}{c|}{\begin{tabular}[c]{@{}c@{}}NumberOf\\ Topological Relations\end{tabular}} & \begin{tabular}[c]{@{}c@{}}Run times\\ (in seconds)\end{tabular} \\ \hline
		\multicolumn{1}{|l|}{\multirow{3}{*}{\begin{tabular}[c]{@{}l@{}}Section\\ 5.2\end{tabular}}} & \multirow{2}{*}{Alg. 3} & \multirow{2}{*}{parishes: 2,180} & \multirow{2}{*}{drainageAreas: 134} & \multirow{2}{*}{2,683} & intersects & 636 & \multirow{2}{*}{29 s} \\ \cline{6-7}
		\multicolumn{1}{|l|}{} &  &  &  &  & within & 2,046 &  \\ \cline{2-8} 
		\multicolumn{1}{|l|}{} & Alg. 5 & farmStates: 40,039 & parishes: 2,180 & 39,800 & within & 39,334 & 7 s \\ \hline
		\multicolumn{1}{|l|}{\multirow{4}{*}{\begin{tabular}[c]{@{}l@{}}Section \\ 5.3\end{tabular}}} & \multirow{2}{*}{Alg. 4} & \multirow{2}{*}{parishes: 2,180} & \multirow{2}{*}{drainageAreas: 134} & \multirow{2}{*}{NONE} & intersects & 1,088 & \multirow{2}{*}{2,622 s} \\ \cline{6-7}
		\multicolumn{1}{|l|}{} &  &  &  &  & within & 3,392 &  \\ \cline{2-8} 
		\multicolumn{1}{|l|}{} & \multicolumn{1}{c|}{\multirow{2}{*}{Alg. 6}} & farmStates: 40,039 & parishes: 2,180 & NONE & within & 39,998 & 1,920 s \\ \cline{3-8} 
		\multicolumn{1}{|l|}{} & \multicolumn{1}{c|}{} & farmStates: 40,039 & drainageAreas: 134 & NONE & within & 39,845 & 525 s \\ \hline
	\end{tabular}
\end{table*}

The rationale of developing the RDF2SOLAP enrichment module is to enrich and re-annotate existing RDF data on the Semantic Web with spatial and multi-dimensional data warehouse metadata. After this, the spatial RDF data becomes available for querying with SOLAP operations directly in SPARQL without losing its triple (RDF) format. We do not expect superior performance of our implementation due to the limited spatial and multidimensional technologies available in the RDF/SW stack. Instead, as long as we achieve reasonable performance and results, our proposal will give much more flexibility and analytical power without needlessly spending large amounts of time on hand-crafting specialized software (e.g., RDBMS tool or GIS) for annotation. 

%extracting and converting the data from RDF to GIS and back. 

% however we aim to reach comparable performance and results with respect to the existing (offline) GIS and RDBMS tools. 
First, we briefly introduce the experimental settings in Section~\ref{subsec:expset}. Then, we present the run-times of the algorithms given in the previous section, for assessing the performance our approach (Section~\ref{subsec:quanteval}). To evaluate the performance of our approach, we present the total time for getting similar results over RDF data in two different (non-SW) environments. Next, we give the comparison baselines in Section~\ref{subsec:compbase}, for describing those two different environments (GIS, RDBMS) that we are comparing our results against in the experimental set-up. Then, in Section~\ref{subsec:qualeval} we compare our results with those two environment in terms of accuracy and coverage. Finally, we share the technical lessons learned in Section~\ref{subsec:techless} and summarize our findings in Section~\ref{subsec:expsum}.

%TODO: Already make it clear in the beginning that we do not expect superior performace of  our implementation due to RDF/SW stack but only aim comperable performance. 
%DONE

%TODO: Also we should clearly and explicitly state what the cost of using GIS tool and RDBMS etc. is, such that we need to convert all the RDF data into their native formats. Conversion times and load time 

\subsection{Experimental Settings}\label{subsec:expset}

As triple store we used: \textit{Virtuoso version 07.20.3217 on Linux (x86\_64-ubuntu-linux-gnu), Single Server Edition}.
We implemented RDF2SOLAP on the \textit{Node.js} platform. Hardware set-up of the Node.js machine is given in Table~\ref{tab:hardware}.

\begin{table}[htb]
	\caption{Hardware Setup (Node.js Machine)}
	\label{tab:hardware}
	\begin{tabular}{|l|l|}
		\hline
		Processor Name: & Intel Core i7 \\ \hline
		Processor Speed: & 2,8 GHz \\ \hline
		Num. of Processors: & 1 \\ \hline
		Num. of Cores: & 4 \\ \hline
		L2 Cache (per Core): & 256 KB \\ \hline
		L3 Cache: & 6 MB \\ \hline
		Memory: & 16 GB \\ \hline
	\end{tabular}
\end{table}
For all the algorithms that we have implemented, we provided the test cases in the GitHub repository, where the results can be re-generated. Each experiment given in Table~\ref{tab:results} was run (on Node.js running) on a MacBook Pro 14,3 in a single process. The hardware details of the machine are given in Table~\ref{tab:hardware}. 
%For each test case  we have queried the triple store's SPARQL endpoint and extracted the triple data in JSON format to Node.js, which is kept in memory. Using Node.js and a Javascript library for spatial analysis provides us a flexible development environment, independent from any choice of triple store. 
To elaborate the performance and then accuracy of our approach with choice of technologies to implement RDF2SOLAP, we compare our results (Table~\ref{tab:results}) against two different environments: a leading GIS tool and a leading RDBMS. The software versions of the tools and hardware of the machine running these tools are shown in Table~\ref{tab:hardware2}\footnote{We cannot disclose the names of the GIS tool and RDBMS tool due to license restrictions}.  

\begin{table}[b]
	\caption{Hardware and Software Setup (RDBMS Server and GIS Platform)}
	\label{tab:hardware2}
	\begin{tabular}{|l|l|}
		\hline
		\multicolumn{2}{|c|}{Hardware} \\ \hline
		Processor Name: & Intel Core i7 \\ \hline
		Processor Speed: & 2,7 - 2,9 GHz \\ \hline
		Num. of Processors: & 4 \\ \hline
		Memory: & 32 GB \\ \hline
		\multicolumn{2}{|c|}{Software} \\ \hline
		Operating System: & Windows 10 Enterprise (10.0) 64-bit \\ \hline
		%SQL Server: & Microsoft SQL Server Developer (64-bit) \\ \hline
		%SQL Server Version: & 14.0.2001.14 \\ \hline
		RDBMS Server Memory: & 64-bit  \\ \hline
		RDBMS Server Memory: & 16287 MB \\ \hline
		%SQL Server Memory: & 16287 MB \\ \hline
		GIS Tool Version: &  64 bit 2.18.21 \\ \hline
		%QGIS Version: & Las Palmas 64 bit 2.18.21 \\ \hline
		%QGIS code revision: & 9fba24a3f2 \\ \hline
	\end{tabular}
\end{table}

\subsection{Performance Evaluation}\label{subsec:quanteval}
The results of applying our algorithms on the running use case are summarized in Table~\ref{tab:results}. 
The results show the number of topological relationships found between the level members in spatial hierarchies and between the base level members and fact members. We distinguish the results for explicit and implicit relations as implemented in the algorithms for  spatial hierarchies (Alg.~\ref{algo:detectspatialHS} and~\ref{algo:discoverspatialHS}) and fact-level relations (Alg.~\ref{algo:detectFactLevel} and~\ref{algo:discoverFactLevel}).

The input parameters and figures for each algorithm are shown in Table~\ref{tab:results} under the INPUT column(s). The input datasets to the algorithms are 2,180 parish members, 40,039 farm state members, and 134 drainage area members. The OUTPUT columns show the number of topological relations found and run times of the algorithms. In this section, we only focus on evaluating the performance of our implementation and discuss the output coverage for the number of found topological relations in the qualitative evaluation section. 

%For all the algorithms that we have implemented, we provided the test cases in the GitHub repository, where the results can be re-generated. Each experiment given in Table~\ref{tab:results} was run (on Node.js running) on a MacBook Pro 14,3 in a single process. The hardware details of the machine are given in Table~\ref{tab:hardware}. 

In Table~\ref{tab:results}, we can see that most expensive algorithm is Alg.~\ref{algo:discoverspatialHS} (\texttt{discoverSpatialHS}), which runs in 2,622 seconds. The algorithm takes input instances of parish and drainage area with POLYGON data type, without explicit relations as in Alg.~\ref{algo:detectspatialHS} (\texttt{detecSpatialHS}). In Alg.~\ref{algo:detectspatialHS}, with distinct explicit relations (given 2,683), the algorithm checks for the designated spatial Boolean functions (\textit{within} and \textit{intersects}) just 2,683 times for each Boolean function. However, in Alg.~\ref{algo:discoverspatialHS}, the algorithm calls the spatial Boolean functions (\textit{within} and \textit{intersects}) 134 $\times$ 2,180 $=$ 292,120 times for each function. Similarly, compared to Alg.~\ref{algo:detectFactLevel}., Alg.~\ref{algo:discoverFactLevel} is more expensive because of running without explicit relations, although it is much faster than Alg.~\ref{algo:discoverspatialHS} since Alg.~\ref{algo:discoverspatialHS} calls the spatial Boolean functions between (farm states) POINT data type and POLYGON data type (for parishes and drainage areas).

In Table~\ref{tab:quant}, we compare the run times of our algorithms with two different query platforms (RDBMS and GIS) for detecting and discovering topological relations. From these platforms, Alg.~\ref{algo:detectspatialHS} and~\ref{algo:detectFactLevel} (to detect explicit topological relations) are only implemented on the RDBMS since the GIS tool employs spatial joins instead of joining through referential integrity of explicit relations. The RDBMS demonstrates great performance for both Alg.~\ref{algo:detectspatialHS} and~\ref{algo:detectFactLevel} by processing the queries in less than 1 second for each. However, these query processing times do  not include development time for extracting the needed input data sets from our RDF endpoint, loading the data into RDBMS, writing the SQL queries, etc., which roughly takes 1-1.5 days (Table~\ref{tab:quant}). To discover implicit topological relations in Alg.~\ref{algo:discoverspatialHS} and~\ref{algo:discoverFactLevel}, the GIS tool outperformed the RDBMS in terms of processing time excluding the development cost (preparation and load times), which is about 2 days. We see that while RDF2SOLAP has slower query performance than RDBMS and GIS, it still has acceptable performance as even the longest queries can finish over a lunch break. The main benefit of RDF2SOLAP is that it requires orders of magnitude less development time than RDBMS and GIS and requires similarly less technical knowledge: around 5 minutes of configuring the endpoint versus literally {\em days} of writing complex code and SQL/SPARQL queries to achieve the same task.

%RDF2SOLAP performs slower in query processing time compared to RDBMS and GIS, although, to implement the enrichment approach on native SW/RDF data, RDF2SOLAP is still the most efficient, since development cost of RDF2SOLAP is only 5 min. for fetching the data sets from the endpoint and the users don't have to write SPARQL/SQL queries or know how to use a GIS tool.

% preparation and load time is max. 1/2 hr in total and the users don't have to write SPARQL/SQL queries or know how to use a GIS tool. 

% \usepackage{multirow}
\begin{table}[t]
	\caption{Performance Evaluation Results (f.s.= farm states, p.= parishes, d.a.= drainage areas)}
	\label{tab:quant}
	\begin{tabular}{c|l|l|l|}
		\cline{2-4}
		& \multicolumn{1}{c|}{\multirow{2}{*}{\begin{tabular}[c]{@{}c@{}}Query\\ Platform\end{tabular}}} & \multicolumn{2}{c|}{Performance Results} \\ \cline{3-4} 
		& \multicolumn{1}{c|}{} & \multicolumn{1}{c|}{\begin{tabular}[c]{@{}c@{}}Run times\end{tabular}} & \multicolumn{1}{c|}{\begin{tabular}[c]{@{}c@{}}Development cost \end{tabular}} \\ \hline
		\multicolumn{1}{|c|}{\multirow{2}{*}{\begin{tabular}[c]{@{}c@{}}Alg. 3\\ (p.-- d.a.)\end{tabular}}} & RDF2SOLAP & 29 s & 5 min. \\ \cline{2-4} 
		\multicolumn{1}{|c|}{} & RDBMS & < 1 s & 1-1.5 days \\ \hline
		\multicolumn{1}{|c|}{\multirow{2}{*}{\begin{tabular}[c]{@{}c@{}}Alg. 5\\ (f.s. -- p.)\end{tabular}}} & RDF2SOLAP & 7 s & 5 min. \\ \cline{2-4} 
		\multicolumn{1}{|c|}{} & RDBMS & < 1 s & 1-1.5 days \\ \hline
		\multicolumn{1}{|c|}{\multirow{3}{*}{\begin{tabular}[c]{@{}c@{}}Alg. 4\\ (p. -- d.a.)\end{tabular}}} & RDF2SOLAP & 2,622 & 5 min. \\ \cline{2-4} 
		\multicolumn{1}{|c|}{} & RDBMS & 43 s & 1-1.5 days \\ \cline{2-4} 
		\multicolumn{1}{|c|}{} & GIS & 45 s & 2 days \\ \hline
		\multicolumn{1}{|c|}{\multirow{3}{*}{\begin{tabular}[c]{@{}c@{}}Alg. 6\\ (f.s. -- p.)\end{tabular}}} & RDF2SOLAP & 1,920 s & 5 min. \\ \cline{2-4} 
		\multicolumn{1}{|c|}{} & RDBMS & 95 s & 1-1.5 days \\ \cline{2-4} 
		\multicolumn{1}{|c|}{} & GIS & 72 s & 2 days \\ \hline
		\multicolumn{1}{|c|}{\multirow{3}{*}{\begin{tabular}[c]{@{}c@{}}Alg. 6\\ (f.s. - d.a.)\end{tabular}}} & RDF2SOLAP & 525 s & 5 min. \\ \cline{2-4} 
		\multicolumn{1}{|c|}{} & RDBMS & 48 s & 1-1.5 days \\ \cline{2-4} 
		\multicolumn{1}{|c|}{} & GIS & 41 s & 2 days \\ \hline
	\end{tabular}
\end{table}

Even excluding the preparation and load times, RDF2SOLAP demonstrated a reasonable performance at a very low development cost (configuration) compared to the other non-SW query platforms, considering that the run times (in Table~\ref{tab:results} and~\ref{tab:quant}) for RDF2SOLAP cover also the query processing times, parsing the RDF data in JSON, calling the helper functions when necessary, and returning bounding box objects for multi-part POLYGON data. RDF2SOLAP configuration can be done within 5 minutes by specifying the SPARQL endpoint, where the RDF cube schema namespace URI is located. Then, the enrichment process is automatically initiated by retrieving the input parameters to the enrichment algorithms from the endpoint. RDF2SOLAP demonstrates a significant advantage and ease-of-use by cutting down the development costs on data extraction and preparation times by 2 to 3 orders of
magnitude, where 1.5-2 days of development cost is reduced to 5 minutes. RDF2SOLAP operates natively over RDF data without the need for third party tools and software.

\subsection{Comparison Baselines}\label{subsec:compbase}

To prepare the experimental set-up for the RDBMS and GIS platforms, we load the WKT data (spatial attributes of level and fact members) used in our experiments with the same decimal precision of the coordinates to the GIS tool, and to a geo-database on RDBMS from CSV files. We extract topological relations between the child and parent members by using spatial joins in the GIS tool and built-in spatial functions of the RDBMS. 
%two non-SW query platforms, where we can get similar results of the enrichment process (i.e., finding topological relations between child-parent members), we load the WKT data (spatial attributes of level and fact members) used in our experiments with the same decimal precision of the coordinates to the GIS tool, and to a geo-database on RDBMS from CSV files. We extract topological relations between the child and parent members by using spatial joins in the GIS tool and using built-in spatial functions of the RDBMS. 

%The results for detecting and discovering topological relations between child-parent members on an RDBMS and GIS tool are given in Table~\ref{tab:eval}. Run times include only looking for topological relations (with spatial Boolean functions in structured query language in RDBMS and table joins in GIS tool) and returning the results. The preparation and load times of the data is not included. 

Since both the GIS tool and the RDBMS cannot process RDF data in native format, we have to prepare the data to load into these environments. The preparation and load times of the data is given as development cost in Table~\ref{tab:quant}. The preparation and load time is calculated assuming that the developer has basic knowledge of the domain, extraction of RDF data with SPARQL queries, can write SQL queries, and knows how to use the RDBMS and GIS tools.   
We extracted the spatial level members (farms states, parishes, and drainage areas) used in the algorithms from our RDF endpoint in CSV format. To prepare the data to be loaded into GIS tool and RDBMS we have to also use the relational schema defined by QB4SOLAP.    

On the GIS tool we saved CSV data layers (for each level; farm states, parishes and drainage ares) and converted these into native GIS format, which are shape files. Then, we run the \emph{Join Attributes By Location} function, which is a built-in data management process. We run this function as a batch process, for parishes-drainage areas (Alg.~\ref{algo:discoverspatialHS}), farm states-parishes, and farm states-drainage areas (Alg.~\ref{algo:discoverFactLevel}) as given in Table~\ref{tab:quant}.

% We did not measure the separate run times of each algorithm, since we run all possible topological relations as a batch process on the GIS tool. The total run time for these two algorithms (Alg.~\ref{algo:discoverspatialHS} and Alg.~\ref{algo:discoverFactLevel}) as a batch process for three test cases (parishes-drainage areas, farm states-parishes, and farm states-drainage areas) took a little more than \textit{2 minutes }(\textit{127 s}).

%(between Tables~\ref{tab:results} and~\ref{tab:quant})
Measuring and comparing the run times (between non-SW query tools and RDF2SOLAP) and development costs are not the only scope of the evaluation since algorithms in the implementation of RDF2SOLAP run in several steps with helper functions for extracting correct data (level members, hierarchy steps), finding the spatial (attribute) values etc., while on the GIS tool and the RDBMS only ``\textit{finding the topological relations}'' part of the algorithms are run. Converting RDF data into GIS and RDBMS native format, loading to these environments, and preparation of batch processes and SQL queries are 1,5-2 days of manual work. Therefore, the overall cost of using offline GIS and RDBMS tools for RDF data is very expensive in terms of developer time compared to our RDF2SOLAP tool. Therefore, we use the comparison baselines for scoping out the accuracy and coverage of number of topological relations found for each algorithm in these three different environments. %We give the number of topological relations in Table~\ref{tab:compbase} and discuss the results in qualitative evaluation section (Section~\ref{subsec:qualeval}).

%However, our comparison baseline is the number of topological relations found for each algorithm in these three different environments, which are given in Table~\ref{tab:compbase}.

		%\resizebox{1.02\columnwidth}{0.3\columnwidth}{

% \usepackage{multirow}
\begin{table}[t]
	\caption{Comparisons of number of topological relations found in each tool (f.s. = farm states, p. = parishes, d.a. = drainage areas)}
	\label{tab:compbase}
	\begin{tabular}{c|l|l|l|l|}
		\cline{3-5}
		\multicolumn{2}{l|}{\multirow{2}{*}{}} & \multicolumn{3}{c|}{\textbf{TOOLS}} \\ \cline{3-5} 
		\multicolumn{2}{l|}{} & \textbf{GIS} & \textbf{RDBMS} & \textbf{RDF2SOLAP} \\ \hline
		\multicolumn{1}{|c|}{\multirow{2}{*}{\textbf{\begin{tabular}[c]{@{}c@{}}Alg. 3: \\ (p. -- d.a.)\end{tabular}}}} & \textit{intersects} & N/A & 1,897 & 636 \\ \cline{2-5} 
		\multicolumn{1}{|c|}{} & \textit{within} & N/A & 785 & 2,046 \\ \hline
		\multicolumn{1}{|c|}{\textbf{\begin{tabular}[c]{@{}c@{}}Alg. 5:\\ (f.s.-- p.)\end{tabular}}} & \textit{within} & N/A & 39,334 & 39,334 \\ \hline
		\multicolumn{1}{|c|}{\multirow{2}{*}{\textbf{\begin{tabular}[c]{@{}c@{}}Alg. 4: \\ (p. -- d.a.)\end{tabular}}}} & \textit{intersects} & 2,556 & 2,802 & 1,088 \\ \cline{2-5} 
		\multicolumn{1}{|c|}{} & \textit{within} & 1,039 & 785 & 3,392 \\ \hline
		\multicolumn{1}{|c|}{\textbf{\begin{tabular}[c]{@{}c@{}}Alg. 6: \\ (f.s. -- p.)\end{tabular}}} & \textit{within} & 39,805 & 39,984 & 39,998 \\ \hline
		\multicolumn{1}{|c|}{\textbf{\begin{tabular}[c]{@{}c@{}}Alg. 6: \\ (f.s. -- d.a.)\end{tabular}}} & \textit{within} & 39,441 & 39,845 & 39,845 \\ \hline
	\end{tabular}
\end{table}

\subsection{Qualitative Evaluation}\label{subsec:qualeval}
 
%We give the number of topological relations in Table~\ref{tab:compbase} and discuss the results in qualitative evaluation section (Section~\ref{subsec:qualeval}).

The number of topological relations found for each enrichment algorithm  in RDF2SOLAP, RDBMS, and GIS, are given in Table~\ref{tab:compbase}.  

%are given in comparison with two other development environments (RDBMS and GIS), where we can get similar results for finding topological relations in Table~\ref{tab:compbase}.
 
As mentioned earlier, the GIS tool does not employ explicit relations between the parent-child members but instead spatial joins. Therefore, Alg.~\ref{algo:detectspatialHS} and Alg.~\ref{algo:detectFactLevel} are denoted with N/A in Table~\ref{tab:compbase} for the GIS tool as these algorithms employ explicit relations between child-parent members. We only tested the finding of (discovering) implicit topological relations (\texttt{discoverSpatialHS} and \texttt{discoverFactLevelRelations}) by utilizing the spatial join functionality of the (GIS data management) tool to emulate the results for Alg.~\ref{algo:discoverspatialHS} and Alg.~\ref{algo:discoverFactLevel}. %, where there are no direct links between child-parent members,  

 %Therefore, on the GIS tool we tested only finding (discovering) implicit topological relations (\texttt{discoverSpatialHS} and \texttt{discoverFactLevelRelations}), where we did not use direct links between farms, parishes, and drainage areas (through referential integrity of defining the explicit relations), but we used the spatial join functionality of the (GIS data management) tool.  

%%The results from the GIS tool are summarized in Fig.~\ref{fig:finalmaps}. The GIS tool gives convincing results, since we can show the results on the map and verify them visually. We can also group the results (parish child members) for intersects relations based on how many intersecting parent members (drainage areas) exist. In Fig.~\ref{fig:finalmaps}, the green scale given in the legend of the parishes represents the intersects relations with 2, 3, 4, and 5 drainage areas. The white symbol for parishes represents the within relations where they can be only within one (1) single drainage area as shown in the legend. 

%In total, there are 1,039 within relations and 2,556 intersects relations found between parishes and drainage areas (Table~\ref{tab:compbase}, Alg.~\ref{algo:discoverspatialHS}). 

%%The red/pink symbol represents the parishes that could not be related to a drainage area with a topological relation, which are 0,6\% of total parishes.  

On the RDBMS, we tested both (detect and discover topological relations), where we queried with joins on the unique IDs if it was present (drainage area foreign key in parishes, parish foreign key in farm states), and with spatial joins by using \texttt{STWithin}, \texttt{STIntersects}, and \texttt{STOverlaps} built-in functions. In the RDBMS, we found fewer \textit{within} relations compared to the GIS tool and more \textit{intersects} relations (Table~\ref{tab:compbase}, Alg.~\ref{algo:discoverspatialHS}). On the contrary, RDF2SOLAP finds more \textit{within} relations and fewer \textit{intersects} relations than found in the GIS tool and the RDBMS. In Alg.~\ref{algo:detectspatialHS} RDF2SOLAP detects 2,046 \textit{within} relations, which is 47\% more than the RDBMS, where 785 \textit{within} relations are detected. Similarly, in Alg.~\ref{algo:discoverspatialHS}, which is also between parishes and drainage areas, RDF2SOLAP finds 47\% more (3,392 \textit{within} relations) than GIS and 54\% more than RDBMS (Table~\ref{tab:compbase}). This is due to generalizing the multi-part POLYGON data as bounding boxes in RDF2SOLAP, where in the GIS tool and in the spatial RDBMS, multi-port POLYGON data is processed in its original format. The GIS tool presents the most accurate results for finding topological relations between parishes and drainage areas (polygon-polygon relations), where in Alg.~\ref{algo:discoverspatialHS}, the GIS tool discovers 2,556 \textit{intersects} relations and the RDBMS discovers similar results (2,802 \textit{intersects} relations) to the GIS tool with 8\% difference, where RDF2SOLAP discovers (1,088 \textit{intersects}) 40\% fewer relations than the GIS tool.

We can see that the results from RDF2SOLAP for finding implicit and explicit relations between POINT-POLYGON data types with Alg.~\ref{algo:detectFactLevel} (farm states-parishes: 39,334) and Alg.~\ref{algo:discoverFactLevel} (farm states-parishes: 39,998 and farm states-drainage areas: 39,845) are very similar to the relations found in RDBMS and the GIS tool, which can be observed in Table~\ref{tab:compbase}.

We found the exact same number of topological relations (\textit{within}) in RDF2SOLAP and the RDBMS for  Alg.\ref{algo:detectFactLevel} (farm states-parishes: 39,334) and Alg.~\ref{algo:discoverFactLevel} (farm states-drainage areas: 39,845). For Alg.~\ref{algo:discoverFactLevel} (farm states-parishes), we found 39,984 \textit{within} relations in RDBMS, and 39,998 within relations in RDF2SOLAP with only 14 difference (0,03\%) (Table~\ref{tab:compbase}). There is a \textit{little} divergence between the number of (\textit{within}) relations found in the GIS tool and in RDF2SOLAP in Alg.~\ref{algo:discoverFactLevel}. There are fewer relations found in the GIS tool than RDF2SOLAP, where the difference is 193 (0,4\%) for farm states-parishes and the difference is 404 (1\%) for farm states-drainage areas. All the tools present very similar results for Alg.~\ref{algo:detectFactLevel} and Alg.~\ref{algo:discoverFactLevel} with less than 1\% divergence (for topological relations between POINT-POLYGON parent-child members), which we find is acceptable. %can be tolerated. %This can be tolerated compared to the discrepancy between the GIS tool and RDF2SOLAP (or RDBMS and RDF2SOLAP) for POLYGON-POLYGON relations. 

\subsection{Technical Lessons}\label{subsec:techless}

The deviation in RDF2SOLAP for POLYGON-POLYGON relations can be prevented by using multi-part POLYGON and MULTIPOLYGON data as its original form instead of generalizing them as bounding boxes.    

% SQL Server STWithin and STIntersects and STOverlaps (MakeValid())

However, in practice, storing the multi-part POLYGON data as MULTIPOLYGONs in a triple store, loading the data to Node.js in JSON format, and applying spatial Boolean functions from Turf.js library was not possible at many levels. We encountered performance and formatting problems while loading MULTIPOLYGON data to Virtuoso, where the debugger was not capable of providing a stack trace, where the error occurred. This led to missing data in the triple store for drainage areas. Even assuming that the MULTIPOLYGON data was successfully loaded, Turf.js could not handle POLYGON-MULTIPOLYGON \textit{within} relations, which is normally possible on the GIS tool or on the RDBMS. %to a degree.
Since keeping the multi-part POLYGON and MULTIPOLYGON data in its original form was not feasible for the Web/Semantic Web technologies (Turf.js API and RDF Store), we had a trade-off between implementing the POLYGON-POLYGON relations in generalized bounding boxes and a slight deviation in the precision of results found in two other non-SW environments.

\subsection{Experimental Summary}\label{subsec:expsum}
%TODO: First conclude on the quality of our implementation compared to the others, then on the performance. Finish off by saying what is needed from triple stores to support us. 

RDF2SOLAP demonstrated accurate results in comparison to two other tools (GIS and RDBMS) in terms of found topological relations between POINT-POLYGON data types. Due to the generalization of multipart POLYGON relations, RDF2SOLAP has a minor divergence on the detected/discovered number of topological relations and did not meet the exact same results as the other two tools for detecting and discovering topological relations between POLYGON-POLYGON data types. Better spatial RDF libraries and support of spatial Boolean operators between multi-polygon geometries are essential in order to get better accuracy in the results without generalizing the multi-part polygon data as bounding boxes.

 %either by using
%the spatial capabilities of a triple store where multi-polygon data and spatial
%Boolean operators are natively supported or by using a third-party library
%that can operate with spatial Boolean operators over multi-polygon data natively
%without generalizing them into bounding boxes.

In terms of productivity and performance, RDF2SOLAP significantly reduces
the development effort to operate with RDF data compared to in-house non-SW proprietary platforms (such as RDBMS and GIS) by 2 to 3 orders of magnitude since it does not require the manual work to prepare and process the data but operates on native RDF/SW data. %data with implicitly and explicitly created test cases. 

%In terms of productivity, RDF2SOLAP is far ahead of in-house GIS and RDBMS tools for enriching spatial RDF data by finding hierarchy steps and fact-level relations since it does not require (overall) manual work to prepare and process the data, but operates on native RDF/SW data with implicitly and explicitly defined test cases. %By providing RDF2SOLAP enrichment tool, user development costs can be significantly reduced from hours of preparation of data, query platforms, scripts, queries, and etc. to one time 1/2 hour one time overall cost by fetching the data sets from the SPARQL endpoint in test cases.  

Comparing and evaluating the technical capabilities (for supporting spatial data handling) of triple stores, APIs, and libraries is beyond the scope of this paper. Improvements of the underlying technologies can provide a better development environment to implement RDF2SOLAP (spatial) enrichment algorithms (Section~\ref{sec:rdf2solap}) with better performance and accuracy. Further improvements to the performance (processing times) of RDF2SOLAP can be achieved either by implementing spatial indexes directly on the RDF data in triple stores or we can build an R-tree in memory on node.js using the Turf.js library.

\section{Related work}
\label{sec:relatedwork}
% !TeX spellcheck = en_US
% !TEX root = ../iosart2x.tex
%

Utilizing DW/OLAP technologies on the Semantic Web with RDF data makes RDF data sources more easily available for interactive analysis. 
As summarized by Abell\'{o} et al.~\cite{abello2015using}, related work has studied OLAP and data warehousing possibilities on the Semantic Web (SW) in general. Our work, however, is centered around spatial OLAP (SOLAP) and spatial data warehouses (SDW) on the Semantic Web, which is not yet a comprehensively studied research topic. We focus on performing spatial OLAP analysis directly over multi-dimensional data published on the Semantic Web. Therefore, we review the related work with relevant approaches classified under the following titles: (1) \emph{data modeling and representation} (on the SW for multi-dimensional and spatial data), (2) \emph{metadata enrichment and MD analysis} (OLAP-like analysis over RDF data). \\

\paragraph{Data modeling and representation.}

The RDF Data Cube (QB) vocabulary~\cite{rdfdatacube} is the W3C recommendation to publish statistical data and its metadata in RDF. Thus, QB is commonly used to publish raw or already aggregated multidimensional data sets. However, QB lacks the underlying metadata for multidimensional models and OLAP operations. The set of MD concepts, such as, hierarchy levels along a cube dimension, semantics of the relationships between levels, semantics and definitions of aggregate functions are missing in QB vocabulary, are essential in an MD schema to enable OLAP analysis. Therefore, K\"ampgen et al. define an OLAP data model on top of QB by using SKOS~\cite{alistair2009skos} extensions\footnote{\url{http://www.w3.org/2011/gld/wiki/ISO_Extensions_to_SKOS}} to support multi-dimensional hierarchies~\cite{kampgen2012interacting, kampgen2013no}. However, the proposed model has some limitations on levels to exists only in one hierarchy. The OLAP operations are made available on the data cubes with the proposed model but restricting the cubes with only one hierarchy per dimension. Etcheverry et al. propose QB4OLAP~\cite{etcheverry2014modeling} as an extension to the QB vocabulary, which supports modeling a complete MD data cube and querying the cube with OLAP operations on the Semantic Web. Modeling of MD data on the Semantic Web motivated the publication of datasets from several domains (e.g., statistical data sets from EuroStat and World Bank data, AirBase air quality data, and many other environmental and governmental open data) as RDF data cubes~\cite{datacubeimplementations}.  

The need of fully multi-dimensional semantic data warehouses (where OLAP operations are enabled in SPARQL) made the QB4OLAP vocabulary prominent. Therefore, RDF data cubes from statistical and environmental domains~\cite{DBLP:conf/icde/VargaEVRPT16, galarraga2017qboairbase, nuref2016solap} are published with an extended QB vocabulary. Moreover, semantic Extract-Transform-Load (ETL) tools automate and ease the process of annotating and publishing open data with QB4OLAP on the Semantic Web~\cite{nath2017setl}. Therefore, we can see more and more multi-dimensional datasets annotated with QB4OLAP on the Semantic Web.

These multi-dimensional semantic modeling approaches and querying with OLAP on the Semantic Web lead us to find ways for modeling, publishing, and querying \textit{spatial} data warehouses in particular since modeling and querying \textit{spatial} data bring new challenges. QB4SOLAP~\cite{nuref2015solap} - a spatial extension to a fully multi-dimensional QB4OLAP vocabulary emerges the need of modeling and publishing geo-semantic data warehouses on the Semantic Web. 

Modeling and publishing (non multi-dimensional) spatial data on the Semantic Web has been a focus by many communities and research groups. Some of the efforts for standardizing and aligning vocabularies to describe spatial data (e.g., locations, geometries, etc.) are GeoSPARQL~\cite{geosparql} by   
the Open Geospatial Consortium (OGC), Basic Geo (WGS84 lat/long) Vocabulary by W3C Semantic Web Interest Group~\cite{wgs84}, NeoGeo Vocabularies by GeoVocab working group~\cite{geovocab}, INSPIRE Directive metadata on the Semantic Web~\cite{Patroumpas:2015:EIS:2869969.2870204}, and GeoNames Ontology~\cite{geonames} among many others.     

These standards have been commonly used in a wide range of projects. Government Linked Data (GLD) working group listed some of these geo-vocabularies as standards to publish governmental linked data sets~\cite{lgdworkinggroup}. Andersen et al. re-use some of these vocabularies for publishing governmental and spatial data on the Semantic Web~\cite{govagribusDenmark2014}. LinkedGeoData is a big contribution to the Semantic Web, which interactively transforms OpenStreetMap data to RDF data~\cite{stadler2012linkedgeodata}. The GeoKnow project focuses on linking geospatial data from heterogeneous sources~\cite{le2014geoknow}. More recent works by  Kyzirakos et al. to transform geospatial data into RDF graphs using R2RML mappings~\cite{kyzirakos2018geotriples} and geo-semantic labelling of open data~\cite{neumaier2018geo} by Neumaier et al. show that spatial data on the Semantic Web will keep growing. However, none of these standards considers the MD aspects of spatial data for geo-semantic data warehouses.   

Large volumes of spatial data on the Semantic Web yield a need for advanced modeling and analysis of such data. As mentioned earlier, QB4SOLAP~\cite{nuref2015solap} remedies this need. Aggregate functions, cardinality relationships, and topological relations are rich sources of knowledge in spatial data cubes in order to query with spatial OLAP operations in SPARQL~\cite{nuref2017qb4solapSWJ}. 

QB4ST~\cite{qb4st} is a recent attempt to define extensions for spatio-temporal components to RDF Data Cube (QB). However, it has the inherent limitations of QB to support OLAP dimensions with hierarchies, levels, and aggregate functions. Lack of OLAP hierarchies and aggregate functions in QB4ST hinders to define and operate with topological relations at hierarchy steps or spatial aggregate functions on spatial measures, which are essential MD concepts for SOLAP operators. These spatial MD concepts in geo-semantic data warehouses are defined together with SOLAP to SPARQL query mappings in~\cite{nuref2017qb4solapSWJ}.

\paragraph{Metadata enrichment and MD analysis.}

Increasing popularity of RDF data cubes and MD OLAP cubes on the Semantic Web raised interest in tools and frameworks that can ease the annotation and querying of MD data on the Semantic Web from existing RDF sources. 

Ibragimov et al. present a framework for exploratory OLAP over Linked Open Data (LOD), where the MD schema of the data cube is annotated with QB4OLAP~\cite{ibragimov2015towards}. Based on this MD schema, they propose a system that is capable of querying data sources, extracting and aggregating data to build OLAP cubes in RDF~\cite{Ibragimov2016}. Similarly, Gallinucci et al. propose an exploratory OLAP approach, namely iMOLD by interactively MD modeling of linked data~\cite{gallinucci2018interactive}. Their approach allows users to enrich RDF cubes with aggregation hierarchies through a user-guided process. 
During this interactive process, the recurring modeling patterns that express roll-up relationships between RDF concepts are recognized in the LOD, then these patterns are translated into aggregation hierarchies to enrich the RDF cube.  Varga et al. enables OLAP analysis with the QB2OLAP tool in~\cite{DBLP:conf/icde/VargaEVRPT16} over statistical data published with QB vocabulary, by applying dimensional enrichment steps described thoroughly in~\cite{varga2016dimensional}. The proposed enrichment steps allow users to enrich a QB dataset with QB4OLAP concepts such as fully-fledged dimension hierarchies. However, none of these frameworks and approaches supports spatial data warehouses and SOLAP operations. 

In this paper, we propose a framework, where OLAP cubes in RDF can be enriched with spatial MD concepts from the \textit{QB4SOLAP} vocabulary by employing RDF2SOLAP enrichment algorithms over QB4OLAP triples. This allows users to query MD cubes with SOLAP operators in SPARQL. Optionally, users can utilize GeoSemOLAP\cite{nuref2017GeoSemOLAP} tool on top of QB4SOLAP data sets, which helps users formulate SOLAP queries in SPARQL.

%-------------------------------------------------------------------------

\section{Conclusion and Future Work}
\label{sec:conclusion}
% !TeX spellcheck = en_US
% !TEX root = ../iosart2x.tex

Motivated by the need to conciliate MD/OLAP RDF data cubes and spatial data on the Semantic Web as geo-semantic data warehouses, we have presented a number of contributions in this paper. As a first attempt to enrich RDF data cubes with spatial concepts, we have shown that the QB4SOLAP vocabulary yields the need for fully-fledged spatial data warehouse concepts (that is built on top of non-spatial QB4OLAP and RDF Data Cube (QB) vocabularies), by demonstrating the running use case examples from real world governmental open data sets from various domains (i.e., environment, farming) with complex geometry types. We introduced running use case examples annotated both with QB4OLAP and QB4SOLAP vocabularies, in RDF triples and formalized the RDF triples as parameters to use in the enrichment algorithms. Second, we have built our conceptual architecture in relation to existing semantic (spatial) OLAP tools (e.g., on top of the QB2OLAPem enrichment module and at the back-end of GeoSemOLAP). Third, we have provided hierarchical enrichment algorithms for two cases that cover finding explicit hierarchy steps with direct links between the level members and finding implicit hierarchy steps (without direct links between the level members) by comparing geometry attributes of the level members. We have defined and deployed the necessary algorithms as spatial helper functions for finding spatial attributes and comparing these attributes to derive topological relations. Fourth, we have presented the factual enrichment phase for both implicit and explicit fact-level relations between the fact and level members. Moreover, we have presented how to re-define the fact schema after the factual enrichment phase in an automated manner. Re-defining the fact schema also includes finding the spatial measures and associating them with spatial aggregate functions. In the end, we have implemented all the algorithms that are designed for both hierarchical enrichment and factual enrichment processes, then we presented the details of our implementation.

Finally, we have evaluated our approach and its accuracy as well as the implementation with the underlying technologies by comparing the number of topological relations found in the RDF2SOLAP framework (between the level members in spatial hierarchies and between the level members and the fact members, respectively, during the hierarchical enrichment phase and the factual enrichment phase) against two different non-SW environments. We have presented the experimental set-up and our comparison baselines and concluded our evaluation with technical lessons learned.

In conclusion, RDF2SOLAP facilitates the spatial enrichment of RDF data cubes and fills an important gap in our vision of SOLAP on the Semantic Web despite of the challenges and restrictions in supporting complex spatial data types with the current state of the most common triple stores~\cite{huang2019assessment, ioannidis2019evaluating}.

Several directions are interesting for future research: creating a comprehensive benchmark by implementing the RDF2SOLAP enrichment algorithms on different platforms and testing on different use cases, deriving spatial hierarchy levels and level member instances from external geo-vocabularies and extending our approach in QB4SOLAP, GeoSemOLAP and RDF2SOLAP to handle highly dynamic spatio-temporal data and multi-dimensional analytical queries~\cite{jakobsen2015optimizing}. Another line of future work would be runtime optimizations for scalable querying of spatial data warehouses~\cite{10.1007/978-3-030-00671-6_32}. Moreover, it is also important to develop query optimization techniques for OLAP queries on semantic DW/RDF data, similar to the ones developed for cubes and XML data~\cite{pedersen2002, pedersen2004integrating, yin2006evaluating}. Furthermore, to achieve scalable querying and runtime optimization, new research directions can be taken with binary serialization of the QB4SOLAP RDF data such as header dictionary triples (HDT), which is a compact data structure that can be compressed and kept in-memory, thus
it enables high performance (and also concurrent) querying.

%-------------------------------------------------------------------------

\begin{acks}
	\footnotesize{This research is partially funded by the European Commission through the Erasmus Mundus Joint Doctorate Information Technologies for Business Intelligence (EM IT4BI-DC), the Danish Council for Independent Research (DFF) under grant agreement no. DFF-4093-00301B, and the Poul Due Jensen Foundation.}
\end{acks}

\bibliographystyle{abbrv}
\bibliography{bibliography}

\end{document}